%% file: main.tex
\documentclass[sigconf]{acmart}

\AtBeginDocument{%
  \providecommand\BibTeX{{%
    \normalfont B\kern-0.5em{\scshape i\kern-0.25em b}\kern-0.8em\TeX}}}

\copyrightyear{2024} 
\acmYear{2024} \setcopyright{acmlicensed}\acmConference[CCS '24]{Proceedings of the 2024 ACM SIGSAC Conference on Computer and Communications Security}{October 14--18, 2024}{Salt Lake City, UT, USA} \acmBooktitle{Proceedings of the 2024 ACM SIGSAC Conference on Computer and Communications Security (CCS '24), October 14--18, 2024, Salt Lake City, UT, USA} \acmDOI{10.1145/3658644.3690295} \acmISBN{979-8-4007-0636-3/24/10}

\usepackage{multirow}
\usepackage{graphicx}
\usepackage{balance}
\usepackage[normalem]{ulem}
\useunder{\uline}{\ul}{}
\usepackage{colortbl}
\usepackage{tcolorbox}
\usepackage{tabularx}

\usepackage{amssymb}
\usepackage{amsmath}
\usepackage{glossaries}
\usepackage{hyperref}
\input{meta/abbreviation}

\newtheorem{assumption}{Assumption}

\begin{document}

\title[Unveiling the Vulnerability of Private Fine-Tuning in Split-Based Frameworks for LLMs]{Unveiling the Vulnerability of Private Fine-Tuning in Split-Based Frameworks for Large Language Models: A Bidirectionally Enhanced Attack}

\include{meta/authors}
\begin{abstract}

Recent advancements in pre-trained large language models (LLMs) have significantly influenced various domains. Adapting these models for specific tasks often involves fine-tuning (FT) with private, domain-specific data. However, privacy concerns keep this data undisclosed, and the computational demands for deploying LLMs pose challenges for resource-limited data holders. This has sparked interest in split learning (SL), a Model-as-a-Service (MaaS) paradigm that divides LLMs into smaller segments for distributed training and deployment, transmitting only intermediate activations instead of raw data. SL has garnered substantial interest in both industry and academia as it aims to balance user data privacy, model ownership, and resource challenges in the private fine-tuning of LLMs. Despite its privacy claims, this paper reveals significant vulnerabilities arising from the combination of SL and LLM-FT: \emph{the Not-too-far property of fine-tuning} and \emph{the auto-regressive nature of LLMs}. Exploiting these vulnerabilities, we propose Bidirectional Semi-white-box Reconstruction (BiSR), the first data reconstruction attack (DRA) designed to target both the forward and backward propagation processes of SL. BiSR utilizes pre-trained weights as prior knowledge, combining a learning-based attack with a bidirectional optimization-based approach for highly effective data reconstruction. Additionally, it incorporates a Noise-adaptive Mixture of Experts (NaMoE) model to enhance reconstruction performance under perturbation. We conducted systematic experiments on various mainstream LLMs and different setups, empirically demonstrating BiSR's state-of-the-art performance. Furthermore, we thoroughly examined three representative defense mechanisms, showcasing our method's capability to reconstruct private data even in the presence of these defenses.





\end{abstract}




\maketitle
\input{sections/1_introduction}
\input{sections/3_preliminaries}

\input{sections/4_method}

\input{sections/5_experiments}

\input{sections/2_relatedwork}
\input{sections/6_conclusion}
\input{sections/10_acknowledgment}

\bibliographystyle{ACM-Reference-Format}
\balance
\bibliography{main}

\input{sections/9_appendix}
\end{document}

%% file: meta/abbreviation.tex
\newacronym{ours}{BiSR}{Bidirectional Semi-white-box Reconstruction}
\newacronym{m1}{SIP}{Semi-white-box Forward Inversion Paradigm}
\newacronym{m2}{BRE}{Bidirectional Reconstruction Enhancement}
\newacronym{m3}{NaMoE}{Noise-adaptive Mixture of Experts}

\newacronym{plm}{PLM}{pre-trained language model}
\newacronym{llm}{LLM}{large language model}
\newacronym{sl}{SL}{split learning}
\newacronym{fl}{FL}{federated learning}
\newacronym{sdl}{SDL}{split-based distributed learning}
\newacronym{dp}{DP}{differential privacy}
\newacronym{dxp}{$d_{\mathcal{X}}P$}{$d_{\mathcal{X}}$-Privacy}
\newacronym{peft}{PEFT}{parameter-efficient fine-tuning}
\newacronym{dlg}{DLG}{Deep Leakage from Gradients}
\newacronym{tag}{TAG}{TAG}
\newacronym{dl}{DL}{distributed learning}
\newacronym{moe}{MoE}{mixture of experts}
\newacronym{gdpr}{GDPR}{EU General Data Protection Regulation}
\newacronym{nlp}{NLP}{natural language processing}
\newacronym{nlg}{NLG}{natural language generation}
\newacronym{fsl}{FSL}{federated split learning}
\newacronym{maas}{MaaS}{Model-as-a-Service}
\newacronym{dra}{DRA}{data reconstruction attack}
\newacronym{gma}{GMA}{gradient matching attack}
\newacronym{sft}{SFT}{supervised fine-tuning}
\newacronym{ft}{FT}{fine-tuning}
\newacronym{clm}{CLM}{causal language modeling}
\newacronym{mlm}{MLM}{masked language modeling}

%% file: meta/authors.tex
\newcommand{\corres}{$\dagger$}

\author{Guanzhong Chen}
\affiliation{%
  \institution{Harbin Institute of Technology, Shenzhen}
  \city{Shenzhen}
  \country{China}
}
\email{muxichenz@outlook.com}

\author{Zhenghan Qin}
\affiliation{%
  \institution{Zhejiang University}
  \city{Hangzhou}
  \country{China}
}
\email{leonine@zju.edu.cn}

\author{Mingxin Yang}
\orcid{0009-0004-0700-8299}
\affiliation{%
  \institution{Huazhong University of Science and Technology}
  \city{Wuhan}
    \country{China}
}
\email{ymx@hust.edu.cn}

\author{Yajie Zhou}
\orcid{0009-0006-5125-5400}
\affiliation{%
 \institution{Zhejiang University}
 \city{Hangzhou}
 \country{China}
 }
 \email{yajiezhou@zju.edu.cn}

\author{Tao Fan}
\affiliation{%
  \institution{Hong Kong University of Science and Technology}
  \city{Hong Kong}
  \country{China}
}
\affiliation{
\institution{Webank}
\city{Shenzhen}
  \country{China}
}
 \email{tfanac@cse.ust.hk}

\author{Tianyu Du}
\authornotemark[1]
\affiliation{%
  \institution{Zhejiang University}
  \city{Hangzhou}
  \country{China}
  }
\email{zjradty@zju.edu.cn}

\author{Zenglin Xu}
\affiliation{%
  \institution{Fudan University; Shanghai Academy of AI for Science}
  \city{Shanghai}
    \country{China}
  }
\affiliation{
\institution{Pengcheng Lab}
\city{Shenzhen}
\country{China}
}
\email{zenglinxu@fudan.edu.cn}
\authornote{Corresponding authors.}

%% file: sections/1_introduction.tex
\section{Introduction}

\Glspl{llm}, such as the GPT series \cite{radford2018improving, radford2019language, DBLP:conf/nips/BrownMRSKDNSSAA20, DBLP:journals/corr/abs-2303-08774} and LLaMA \cite{DBLP:journals/corr/abs-2302-13971, touvron2023llama}, have demonstrated exceptional performance across various domains. This success is largely due to the prevalent \emph{pre-train then fine-tune} paradigm. In this approach, models are first pre-trained on extensive public domain datasets (spanning hundreds of gigabytes) to develop foundational capabilities, and then fine-tuned on smaller, task-specific datasets to adapt the model to particular domains. Given the substantial costs associated with the pre-training phase and the billions of parameters in \glspl{llm}, \gls{ft} a pre-trained foundation model (particularly with \gls{peft}) stands out as the most cost-effective approach for real-world applications. However, despite the reduced costs of fine-tuning, most users still lack the computational resources and technical expertise to independently obtain and fine-tune \glspl{llm}. This has led to the emergence of a new business model known as \gls{maas}. In \gls{maas}, enterprises with sufficient computational resources and technical capabilities (referred to as model vendors) offer \glspl{llm} as cloud services, providing customers with fine-tuning APIs that allow them to customize \glspl{llm} using their own data.

While \gls{maas} offers customers efficient and customizable \gls{llm} services, it also presents significant privacy risks. On one hand, transmitting raw customer data through \gls{ft} APIs to model vendors can result in direct data leakage. On the other hand, sending the entire \gls{llm} to customers for local fine-tuning is impractical, as the large size of \glspl{llm} often exceeds customers' limited capacity, and the weights of \glspl{llm} are generally considered valuable proprietary assets by model vendors and cannot be disclosed. In this context, a distributed machine learning framework known as \gls{sl} \cite{DBLP:journals/jnca/GuptaR18, DBLP:journals/corr/abs-1812-00564} has emerged as a promising solution that effectively balances data privacy with model privacy. \gls{sl} involves partitioning and deploying models between the client-side (customer) and server-side (model vendor), enabling model vendors to utilize their extensive computational resources while keeping the majority of \glspl{llm} undisclosed. At the same time, users need only upload intermediate activations (referred to as \emph{smashed data}) to avoid direct exposure of their sensitive data. Consequently, \gls{sl} for \glspl{llm} has garnered substantial interest in both industry and academia \cite{DBLP:journals/corr/abs-2312-15603, DBLP:journals/corr/abs-2311-14030, DBLP:journals/corr/abs-2306-17465}.

However, prior research indicates that even with the use of intermediate data, adversaries can still execute privacy attacks, such as property inference attacks \cite{melis2019exploiting}, membership inference attacks \cite{shokri2017membership, salem2018ml}, and data reconstruction attacks \cite{DBLP:journals/tbd/ZhangPTEM23, DBLP:journal/iot/he2020attacking, qiu2024evaluating, DBLP:conf/ccs/PasquiniAB21}. Among these, the \gls{dra} is of particular concern, as it constitutes the most severe breach of user privacy by attempting to reconstruct the user's original data. While \glspl{dra} have been extensively studied in traditional \gls{sl} with discriminative models and standard classification tasks, there is a notable absence of research on \gls{sl} involving \glspl{llm} in the existing literature. This paper seeks to address the question: \emph{Are generative \glspl{llm} in \gls{sl} also susceptible to \glspl{dra}?}

Different from conventional \gls{sl}, integrating \glspl{llm} into \gls{sl} introduces two notable features: (\emph{\romannumeral1}) \emph{Reliance on Pre-trained Weights}: Traditional \gls{sl} typically involves simple model architectures with small-scale parameters, where client models are often initialized with random parameters and trained from scratch \cite{DBLP:conf/aaai/ThapaCCS22, abedi2024fedsl, turina2021federated}. In contrast, \gls{llm}-based approaches utilize pre-trained weights, which provide essential language modeling capabilities crucial for effective \gls{ft}. Consequently, existing frameworks \cite{DBLP:journals/corr/abs-2312-15603, DBLP:journals/corr/abs-2306-17465} either require clients to access publicly available \gls{llm} weights or distribute portions of pre-trained \glspl{llm} directly to clients, allowing client-side models to inherit these pre-trained weights. (\emph{\romannumeral2}) \emph{Generation Task}: Whereas conventional \gls{sl} primarily focuses on discriminative models and classification tasks, mainstream \glspl{llm} are typically \gls{clm} models designed for \gls{nlg} tasks.

\textbf{Our work.} Focusing on the scenario involving an \emph{honest-but-curious server} as the adversary, we extend earlier findings and argue that the two new features introduced by the combination of \gls{sl} and \gls{llm} \gls{ft} lead to even higher privacy threats from \glspl{dra}. We identify \emph{two vulnerabilities} that contribute to this amplification but are not captured by existing \glspl{dra}.

\textbf{Vulnerability$_1$ -- The "Not-too-far" Property of \gls{llm} \gls{ft}.} The \gls{ft} process's reliance on pre-trained weights provides adversaries with significant prior knowledge, due to a critical characteristic observed during \gls{llm} \gls{ft}: the “Not-too-far” property. This property suggests that the features embedded in the pre-trained weights, which are crucial for \glspl{dra}, remain largely \emph{unchanged} during \gls{ft}. Consequently, even if the client's model weights diverge from the pre-trained state, attackers can still leverage the pre-trained weights to conduct straightforward yet highly effective \glspl{dra}.

\textbf{Vulnerability$_2$ -- The Auto-regressive Nature of \glspl{llm}}. Mainstream \glspl{llm} are \gls{clm} models that predict the next token based on the preceding sequence. During training, these models exhibit auto-regressive behavior, generating training labels by shifting the input sequence left by one token (teacher-forcing). This creates an overlap in the solution space for reconstructing both the label and the input. Such alignment enables adversaries to exploit both forward smashed data and backward gradients for \emph{bidirectional} attacks. Moreover, in the \gls{sl} setting, the known-forward-activation trait eliminates the sequence order issue in gradient-based \glspl{dra} for language models, allowing adversaries to achieve greater reconstruction accuracy.

To fully exploit these vulnerabilities, we propose a novel \gls{dra} method called \textbf{Bi}directional \textbf{S}emi-white-box \textbf{R}econstruction (BiSR), the first \gls{dra} paradigm specifically targeting \glspl{llm} fine-tuned within the \gls{sl} framework. \acrshort{ours} integrates a learning-based reconstruction, which trains an inversion model to recover the original input from the smashed data, and an optimization-based reconstruction, which approximates the raw input through iterative optimization. To leverage Vulnerability{$_1$}, \acrshort{ours} employs pre-trained weights as an attack prior via \emph{semi-white-box access} (defined in Section \ref{sec:pre-tm}) for both the learning-based and optimization-based reconstruction processes. To exploit Vulnerability{$_2$}, the optimization-based stage of \acrshort{ours} capitalizes on the overlap in the solution space, leveraging both forward (smashed data) and backward (gradient) information to conduct a highly effective \emph{bidirectional} attack.

\textbf{Contributions.} To the best of our knowledge, this work represents the first study on \glspl{dra} within the \gls{sl} framework for \gls{llm} fine-tuning.  Our contributions can be summarized as the following:
\begin{enumerate}
    \item[$ \bullet $] \textbf{New insights.} We present the insight that the combination of \gls{sl} and \glspl{llm} brings about two new crucial vulnerabilities that greatly amplify the threat of \glspl{dra} and are not captured by existing \gls{dra} methods: the "Not-too-far" property of \gls{llm}-\gls{ft} and the auto-regressive nature of \glspl{llm}.

    \item[$ \bullet $] \textbf{Novel Approach.} We introduce \gls{ours}, the first \gls{dra} that integrates learning-based and optimization-based methods to bidirectionally attack both forward and backward transmissions in \gls{sl}. Additionally, we propose the \gls{m3} inversion model to enhance its perturbation adaptability.

    \item[$ \bullet $] \textbf{Systematic Experiments.} We conduct thorough experiments to evaluate various \glspl{dra} across different mainstream \glspl{llm} and \gls{ft} datasets under a range of conditions. We also assess their attack performance against three types of defense mechanisms. Our results reveal the effectiveness and superiority of \gls{ours} compared to baseline methods and its adaptability to perturbation-based defenses, highlighting the significant privacy risks associated with split-based \gls{llm} fine-tuning frameworks. 
    

    \item[$\bullet$] Our code is available at GitHub repository \url{https://github.com/StupidTrees/SplitLLM}. 
\end{enumerate}

%% file: sections/3_preliminaries.tex
\section{Preliminaries}

\subsection{Fine-Tuning of LLMs}
The "pre-train then fine-tune" approach is widely used for training \glspl{llm} and adapting them for practical applications. In contrast to the data-intensive and computationally demanding pre-training phase, the \acrfull{ft} process requires smaller datasets and fewer computational resources but is essential for real-world deployment of \glspl{llm}. 

Current \gls{ft} of \glspl{llm} primarily focuses on \gls{sft} for \gls{nlg} tasks, where the target sentence to be generated serves as the label for supervised training. Unlike encoder-only models such as BERT~\citep{DBLP:conf/naacl/DevlinCLT19}, which are better suited for classification tasks due to their \gls{mlm} characteristics, encoder-decoder models like T5~\citep{DBLP:journals/jmlr/RaffelSRLNMZLL20} and decoder-only models like GPT~\citep{radford2018improving, radford2019language, DBLP:conf/nips/BrownMRSKDNSSAA20} and LLaMA~\citep{DBLP:journals/corr/abs-2302-13971, touvron2023llama} are extensively used in \gls{nlg} tasks due to their \acrfull{clm} properties, where the modeling of each token is based solely on its preceding tokens. These models typically comprise an auto-regressive decoder structure.

During the \gls{sft} of \gls{clm} models, the \emph{teacher forcing}~\citep{DBLP:journals/neco/WilliamsZ89} strategy is employed. In this approach, the tokens preceding a target token in a sentence serve as inputs to train the decoder to predict that token. Let $w_t$ represent the token at position $t$, and $\hat{y}_t[w]$ denote the model's predicted probability for the next token $w$ at that position, i.e., the prediction for position $t+1$. The loss for each time step during the teacher forcing phase is defined as:
\begin{equation}
\mathcal{L}_{CE}(\hat{y}_t,y_t) = - \log \hat{y}_t [w_{t+1}],
\end{equation}
which corresponds to the negative log probability of the model's prediction for the true token at the next position. Consequently, for each sentence $x$ of length $T$ in the \gls{sft} dataset, with the model output denoted as $y=(y_1,y_2,\ldots, y_T)$, the training loss is:
\begin{equation}
\label{eq:shifted_label}
\mathcal{L}_{LM} = \text{CrossEntropy}(y_{1:T-1}, x_{2:T}),
\end{equation}
indicating that \emph{the fine-tuning label is obtained by shifting the input sentence to the left by one token}. This type of loss is also referred to as the \emph{auto-regressive target}.

\subsection{Split Learning}
 \begin{figure}
    \centering
    \includegraphics[width=\linewidth]{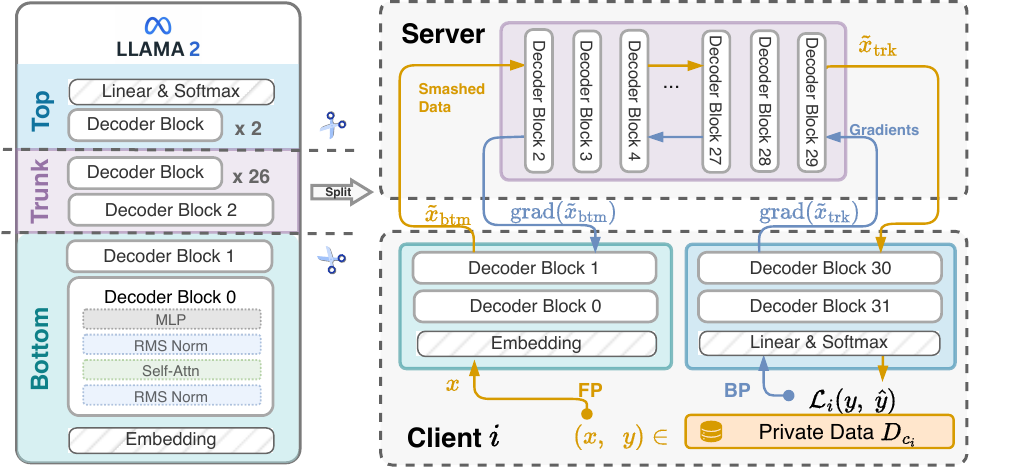}
    \caption{The private-label \gls{sl} framework for fine-tuning LLaMA2-chat-7B, illustrated with a single client. The model's 32 decoder blocks are divided into three segments: \emph{Bottom} (embedding layer and the first two blocks), \emph{Trunk} (middle 28 blocks), and \emph{Top} (final two blocks, normalization layer, and linear layer). The client hosts the Bottom and Top segments, while the server manages the Trunk.}
    \label{fig:sfl-arch}
    \Description{
    The private-label Split Learning framework for fine-tuning LLaMA2-chat-7B, illustrated with a single client. The model's 32 decoder blocks are divided into three segments: Bottom (embedding layer and the first two blocks), Trunk (middle 28 blocks), and Top (final two blocks, normalization layer, and linear layer). The client hosts the Bottom and Top segments, while the server manages the Trunk.
    }
\end{figure}

The \acrfull{sl} architecture typically involves multiple clients ${c_1, c_2, \dots, c_n}$ with private data ${D_{c_1}, D_{c_2}, \dots, D_{c_n}}$ and a central server $S$ that coordinates the learning process.

To collaboratively train a model $f_{\bar{\theta}}$ using private data $D_{c_i}$ across clients, the model is partitioned. Common \gls{sl} frameworks adopt two architectures~\citep{DBLP:conf/ccs/PasquiniAB21}: (\emph{\romannumeral1}) the non-private-label architecture, where clients share labels with the server and the model is divided into two segments, one held by the client and the other by the server, and (\emph{\romannumeral2}) the private-label architecture, where labels remain local and the model is divided into three segments, with the first and last segments residing on the client. We focus on the private-label architecture due to its enhanced privacy and the inherent non-disclosure of labels in \gls{llm}-\gls{ft} scenarios, as depicted in Figure \ref{fig:sfl-arch}. The model $f_{\bar{\theta}}$ is split into three parts: the \emph{Bottom} $f_\text{btm}$ (input end), the \emph{Trunk} $f_\text{trk}$ (middle layers), and the \emph{Top} $f_\text{Top}$ (output end), with the \emph{cut-layers} being the final layers of each segment. The server hosts the parameter-intensive Trunk, while each client holds the lightweight Bottom and Top segments.

Model training in \gls{sl} employs distributed forward and backward propagation~\citep{DBLP:journals/jnca/GuptaR18}. Each step begins with a client encoding a batch $x\in D_{c_i}$ through the model's Bottom part, generating intermediate activation $\tilde{x}_{\text{btm}}=f_{\text{btm}}(x)$, or \emph{smashed data}, which is sent to the server. The server forwards $\tilde{x}_{\text{btm}}$ through the Trunk, creating $\tilde{x}_{\text{trk}}=f_{\text{trk}}(\tilde{x}_{\text{btm}})$, and returns it to the client. The client then finalizes the forward pass through the Top part, yielding output $\hat{y} = f_{\text{Top}}(\tilde{x}_{\text{trk}})$. Following loss computation $\mathcal{L}_i(\hat{y}, y)$ using the private label $y$, backward propagation begins, with gradients flowing through Top-Trunk-Bottom, facilitated by intermediate gradients $\text{grad}(\tilde{x}_{\text{trk}})$ and $\text{grad}(\tilde{x}_{\text{btm}})$ exchanged between client and server.

n a standard scenario, it has been shown that the single-step \gls{sl} process aligns with the centralized forward and backward passes for a client-server pair~\citep{DBLP:journals/jnca/GuptaR18}. In the case of multiple clients, the traditional \gls{sl} paradigm can be applied by sequentially training each client and then passing the parameters to the next~\citep{DBLP:journals/corr/abs-1812-00564}. Alternatively, synergistic methods that combine \gls{sl} and \gls{fl} can be utilized, allowing clients to participate concurrently in split learning with the server, which subsequently aggregates the gradients or model parameters~\citep{DBLP:conf/icoin/JeonK20, zhang2023privacy, DBLP:conf/aaai/ThapaCCS22}. However, the complex processing associated with multiple clients in \gls{sl} is not directly relevant to the privacy data leakage issues addressed in this paper. Therefore, for clarity, we will limit our focus to a scenario involving \emph{a single client}.

After completing \gls{sl}, the model is partitioned between the server and the clients. During inference, a client executes the distributed forward process of \gls{sl} on the input data. In auto-regressive \gls{nlg}, the client feeds the prompt into the lower layers of the model, performs split forward propagation to generate the next token at the Top, appends this token to the partially generated sentence, and repeats the distributed process until the end-of-sentence token is produced.

\subsection{Perturbation-based Privacy Protection methods for Split Learning}
With the feasibility of \gls{sl} validated by numerous studies, it has become standard practice to augment \gls{sl} with additional privacy safeguards. In scenarios involving a curious server, perturbation-based non-cryptographic methods are commonly utilized. These primarily include approaches based on \gls{dp}\citep{DBLP:conf/ccs/AbadiCGMMT016}, which add noise directly to the intermediate data\citep{DBLP:journals/corr/abs-2104-05743, DBLP:journals/tvt/WuCYKYWP24, DBLP:conf/pet/ChatzikokolakisABP13, DBLP:conf/wsdm/FeyisetanBDD20, DBLP:journals/corr/abs-2312-15603}, and NoPeek~\citep{DBLP:conf/icdm/Vepakomma0GR20}, which constrains intermediate representations through regularization terms.

\textbf{Embedding \gls{dxp}}\citep{DBLP:conf/pet/ChatzikokolakisABP13} is a perturbation strategy specifically designed for language models within the \gls{sl} framework. It perturbs the embedding space to satisfy \acrfull{dxp}, a relaxed version of Local \gls{dp}\citep{DBLP:conf/wsdm/FeyisetanBDD20}. Specifically, noise $N$ is added to the output of the embedding layer $\phi_x = \phi(x)$ as follows:
\begin{equation}
\phi_x' = \phi_x + N, ; p(N)(\mathbf{z}) \propto \exp(-\epsilon |\mathbf{z}|),
\end{equation}
where $\epsilon$ is the privacy parameter. Next, within the embedding space, the input embedding closest to the perturbed embedding is identified and used to replace the original embedding:

\begin{equation}
\phi_x = \underset{\phi \in \Phi}{\text{argmin}} |\phi - \phi_x'|.
\end{equation}

\textbf{Smashed-data \gls{dp}} techniques, such as DP-Forward~\citep{DBLP:conf/ccs/DuYC0H023}, add noise directly to the transmitted smashed data $\tilde{x}$ and are widely used in \gls{sl} systems~\citep{DBLP:journals/tvt/WuCYKYWP24, DBLP:journals/corr/abs-2104-05743, DBLP:conf/ccs/DuYC0H023}. Using the Laplace mechanism as an example, the smashed data is first clipped as follows: $\tilde{x} \leftarrow \tilde{x}/\max(1, \frac{|x|{\infty}}{G})$, and then augmented with Laplace noise of a specified scale:
\begin{equation}
\tilde{x} \leftarrow \tilde{x} + \text{Lap}(0, \frac{\Delta f{\text{btm}}}{\epsilon}),
\end{equation}
where $G$ is the clipping threshold, $\text{Lap}(0, \sigma)$ denotes noise following the Laplace distribution with scale $\sigma$, $\epsilon$ is the differential privacy budget, and $\Delta f_{\text{btm}}$ represents the sensitivity of the query function (i.e., the model segment $f_{\text{btm}}$ that generates $\tilde{x}$) with respect to the original data $x$.

\textbf{NoPeek-based perturbation} methods introduce a regularization term into the model's loss function to constrain the correlation between the intermediate smashed data $\tilde{x}$ and the original input $x$, with the goal of minimizing the leakage of original information through the smashed data. The training loss for the model can be expressed as:
\begin{equation}
\mathcal{L}{\text{nopeek}} = \mathcal{L}(\hat{y}, y) + \alpha \text{DCOR}(x, \tilde{x}),
\end{equation}
where $\alpha$ is a scalar weight that adjusts the strength of the regularization, and $\text{DCOR}(x, \tilde{x})$ is a measure of distance correlation~\citep{DBLP:conf/icdm/Vepakomma0GR20}. By incorporating this regularization term, NoPeek encourages the model segment prior to the cut layer (e.g., $f{\text{btm}}$) to adapt such that the smashed data is as dissimilar from the original input as possible, effectively introducing a form of perturbation at the input level.

\subsection{Threat Model}
\label{sec:pre-tm}
\noindent\textbf{Adversary's abilities.} We consider an \emph{honest-but-curious}~\citep{paverd2014modelling} server as the adversary. The “honest” aspect indicates that the adversary adheres to the \gls{sl} protocol, thereby not compromising the efficiency of distributed learning (a particularly realistic assumption for the resource-intensive \gls{llm} training scenario). The “curious” aspect implies that the adversary seeks to access and potentially exploit intermediate data within the system to reconstruct the original private input. We attribute the following capabilities to a curious server: (\emph{\romannumeral1}) Access to all clients' smashed data $\tilde{x}{\text{btm}}$ and the gradients $\text{grad}(\tilde{x}{\text{trk}})$ during the \gls{ft} process, (\emph{\romannumeral2}) White-box access to the model's Trunk part, (\emph{\romannumeral3}) \emph{Semi-white-box access} to the model's Bottom and Top parts (on the client), and (\emph{\romannumeral4}) Access to an auxiliary dataset $D_{a}$, which may (but does not necessarily) resemble the client's private dataset $D_{c_i}$. We define semi-white-box access as follows:

\textbf{Definition: Semi-White-Box Access.} \textit{During an attack on an \gls{llm} undergoing \gls{ft}, the adversary does not have access to the model's exact weights at any training step but is aware of the model's architecture and its pre-trained weights.}

The \emph{semi-white-box} access level lies between white-box and black-box scenarios. In the context of \gls{llm}-\gls{sl}, the semi-white-box assumption is both realistic and reasonable. While the server does not have precise knowledge of the client's updated weights during \gls{sl}, the pre-trained weights of \glspl{llm} are often publicly available or retained by the server as the model vendor. This prior knowledge about the client's model segments can be leveraged when launching attacks. As a result, adversaries can design attack strategies that closely resemble those in a white-box setting, but they must contend with a \emph{fine-tuning gap}—the discrepancy between the pre-trained model and the fine-tuned model.


\vspace{0.2cm}
\noindent\textbf{Adversary's objective.} In the split-based \gls{ft} process, the adversarial server performing \gls{dra} aims to \emph{reconstruct} as many original texts $(x, y) \in D_{c_i}$ from the client's private fine-tuning dataset $D_{c_i}$ as possible.

%% file: sections/4_method.tex
\section{The Proposed B\MakeLowercase{i}SR Method}

 \begin{figure*}
    \centering
    \includegraphics[width=1\textwidth]{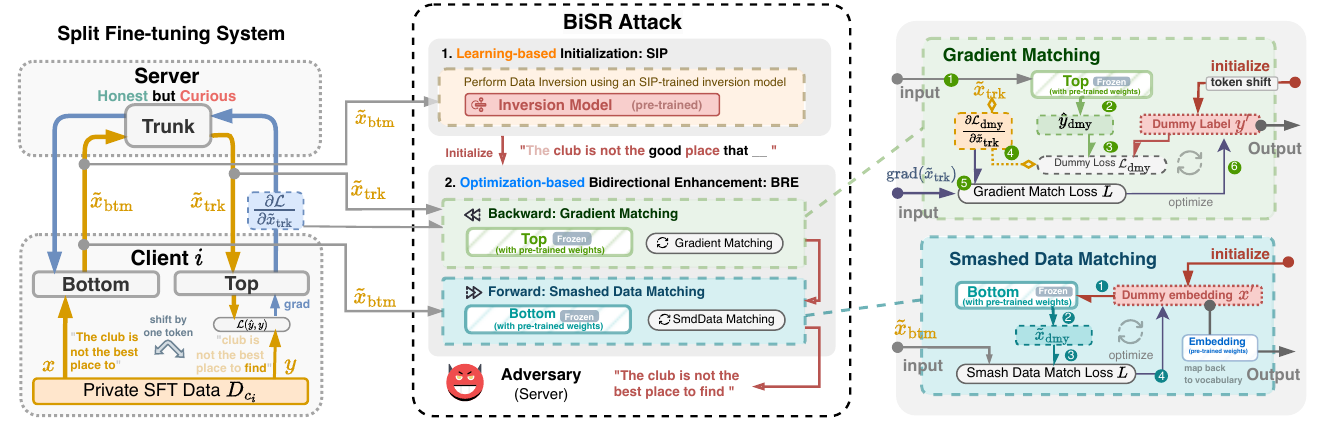}
    \caption{The \gls{ours} attack framework (middle) targeting the split-\gls{llm}-\gls{ft} system (left). This two-stage attack starts with a learning-based approach, where the \gls{m1}-trained attack model (\gls{m3} model) performs direct inversion on smashed data. The second stage builds upon the sentences initially recovered, employing both forward (using smashed data) and backward (using gradients) data for bidirectional, optimization-based enhancement. This process, detailed on the right, results in significantly more comprehensive sentence recovery.}
    \label{fig:oview}
    \Description{
    The \gls{ours} attack framework (middle) targeting the split-\gls{llm}-\gls{ft} system (left). This two-stage attack starts with a learning-based approach, where the \gls{m1}-trained attack model (\gls{m3} model) performs direct inversion on smashed data. The second stage builds upon the sentences initially recovered, employing both forward (using smashed data) and backward (using gradients) data for bidirectional, optimization-based enhancement. This process, detailed on the right, results in significantly more comprehensive sentence recovery.
    }
\end{figure*}

In this section, we introduce our \gls{dra} method targeting \gls{llm} fine-tuned with \gls{sl}, \acrfull{ours}, depicted in Figure \ref{fig:oview}. We begin by establishing the foundational assumption underpinning \gls{ours} for conducting semi-white-box attacks using pre-trained model priors.

\begin{assumption}
\label{as:nottoofar}
\textbf{Not-too-far:} The fine-tuning of \glspl{llm} does not drastically alter the core characteristics established during the pre-training phase, which are adequate for both learning-based and optimization-based \glspl{dra}.
\end{assumption}

This assumption posits that the intrinsic language modeling capabilities of \glspl{llm}, developed during pre-training, remain stable throughout the fine-tuning process for downstream tasks. Consequently, a fine-tuned \gls{llm} can generate intermediate outputs similar to those produced in its original, untuned state, thereby making semi-white-box access closely resemble white-box access. It is important to note that regardless of how clients personalize their model components (e.g., deviating in parameters or structure from pre-trained parts), as long as the intermediate results they produce (smashed data, gradients) are compatible with the pre-trained Trunk part on the server, they fall within the scope of the Not-too-far assumption. We empirically demonstrate that the Not-too-far assumption holds for \glspl{llm} in practice, even when clients adopt model components that are structurally or parametrically inconsistent with the pre-trained parts. Additionally, in Appendix \ref{app:vis}, we highlight through a comparison with image models that this assumption may be unique to language models.

Based on this assumption (utilizing Vulnerability$_1$) and within the semi-white-box setting, \gls{ours} implements a two-stage attack strategy, as depicted in Figure \ref{fig:oview}. The first stage employs \acrfull{m1}, a learning-based approach that trains an inversion model to recover input sentences from the smashed data, as detailed in subsection \ref{subsec:m1}. Following this initial sentence recovery, a second-stage attack is initiated using \acrfull{m2}. This stage leverages Vulnerability$_2$ by utilizing both forward (smashed data) and backward (gradients) information for bidirectional optimization-based refinement, aiming to enhance reconstruction across various domain datasets, as discussed in subsection \ref{subsec:m2}. Additionally, to strengthen the resilience of the \acrshort{m1}-trained attack model against perturbation-based defense mechanisms, we introduce \acrfull{m3}, an advanced inversion model for learning-based attacks, which is described in subsection \ref{subsec:m3}.

\subsection{Learning-based Initialization: SIP}
\label{subsec:m1}

\begin{figure}
    \centering
    \includegraphics[width=\linewidth]{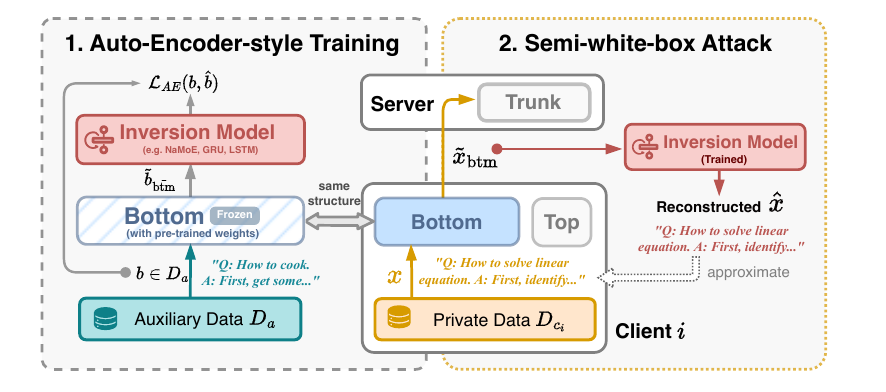}
    \caption{The proposed \acrfull{m1}. During the training phase (left), the encoder mimicking the target model's Bottom segment but with pre-trained parameters is frozen, while the decoder is trained on a dataset $D_a$ similar to $D_{c_i}$, with original inputs $b \in D_{a}$ as labels. In the attack stage (right), the inversion model receives the real model's Bottom output $\tilde{x}_{\text{btm}}$, the smashed data, and directly decodes it to obtain the attack results $\hat{x}$.}
    \label{fig:dra-paradigm}
    \Description{
    The proposed \acrfull{m1}. During the training phase (left), the encoder mimicking the target model's Bottom segment but with pre-trained parameters is frozen, while the decoder is trained on a dataset $D_a$ similar to $D_{c_i}$, with original inputs $b \in D_{a}$ as labels. In the attack stage (right), the inversion model receives the real model's Bottom output $\tilde{x}_{\text{btm}}$, the smashed data, and directly decodes it to obtain the attack results $\hat{x}$.
    }
\end{figure}

Considering Vulnerability$_1$, the first stage of \gls{ours} involves a learning-based attack using an inversion model on smashed data. This model, trained on auxiliary data $D_a$, is designed to reconstruct the original data from the intermediate outputs of the \gls{llm}. The training and attack process of \gls{m1} is illustrated in Figure \ref{fig:dra-paradigm}.

Viewing the Bottom part of the \gls{llm} as an encoder, \gls{m1} aims to train a malicious inversion model as a decoder, thereby forming an auto-encoder system. We design an encoder $e$ (mimicking the attacked \gls{llm}‘s Bottom segment) and a decoder $d$ to be trained on an auxiliary dataset $D_a$, which may or may not be similar to the attacked data $D_{c_i}$. Our objective is to minimize the auto-encoder loss $\mathbb{E}_{x\sim D_a}[\mathcal{L}_{\text{AE}}(d(e(x)), x)]$, enabling $d$ to function as the inversion model. For $e$, we select a model structurally identical to the Bottom segment of the attacked model but with \emph{pre-trained weights} (i.e., semi-white-box access to the client's Bottom model), referred to as $f_{\bar{\text{btm}}}$, to replicate $f_{\text{btm}}$. During the auto-encoder-style training, we freeze $e$ and train only the decoder.

Assuming the target \gls{llm} has a vocabulary size of $V$ and a hidden size of $H$, each batch of text (represented as token IDs with a unified sequence length $S$) from the auxiliary dataset $b^{B\times S\times 1} \in D_a$ is encoded by $e=f_{\bar{\text{btm}}}$, resulting in $f_{\bar{\text{btm}}}(b) = \tilde{b}_{\bar{\text{btm}}}^{B\times S \times H}$. The decoder $d$ (which could be a linear model, RNN, etc.) then decodes $\tilde{b}_{\bar{\text{btm}}}$ to produce $d(\tilde{b}_{\bar{\text{btm}}}) = \hat{b}^{B\times S\times V}$. We use cross-entropy as the auto-encoder loss, defined as $\mathcal{L}_{\text{AE}}(\hat{b}, b) = \text{CrossEntropy}(\hat{b}, b)$. After training, the adversary conducts an attack by decoding the observed smashed data $\tilde{x}{\text{btm}}$ using the decoder $d$ to recover $\bar{x} = d(\tilde{x}_{\text{btm}})$, which approximates the original private data $x$.

The effectiveness of the auto-encoder-trained inversion model is anticipated to rely on Assumption \ref{as:nottoofar}, the degree of similarity between $D_a$ and the target dataset $D_{c_i}$, and the precise alignment of the encoder $e=f_{\bar{\text{btm}}}$ with the Bottom segment of the target \gls{llm}. In Section \ref{sec:exp-m1}, we will validate Assumption \ref{as:nottoofar} and demonstrate that the inversion model trained using \gls{m1} exhibits \emph{strong versatility}, attributed to its capacity for dual learning of both structural and semantic information. This model proves effective in cross-dataset scenarios (training \gls{m1} on one dataset while attacking \gls{sl} on another with a different data distribution) and cross-layer scenarios (training \gls{dra} with $e$ having a different number of layers than the emulated segment). Moreover, it demonstrates robustness to discrepancies between the weights of $e$ and the emulated segment and shows potential adaptability to forward noise addition.

\begin{figure}
    \centering
    \includegraphics[width=1\linewidth]{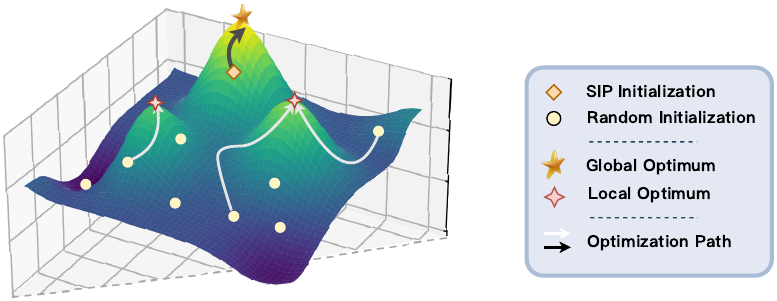}
    \caption{Demonstration of the impact of \gls{ours}'s learning-based initialization (\gls{m1}) on optimization-based process: providing an effective starting point for achieving global optima.}
    \label{fig:opt}
    \Description{
    Demonstration of the impact of \gls{ours}'s learning-based initialization (\gls{m1}) on optimization-based process: providing an effective starting point for achieving global optima.
    }
\end{figure}

In practice, adversaries can apply the \gls{m1}-trained inversion model to smashed data from various layers of the \gls{llm}. This process produces a substantially reconstructed version of the original input, serving as a strong initialization point for the optimization-based \gls{dra} process discussed in the following section. Such initialization effectively mitigates the local optima issue in non-convex optimization~\citep{DBLP:conf/nips/li2024gan}, thereby significantly enhancing the final accuracy, as illustrated in Figure \ref{fig:opt}.

\subsection{Optimization-based Enhancement: BRE}
\label{subsec:m2}

The \gls{m1}-trained inverter can initially generate high-quality data reconstructions, but it still faces performance bottlenecks due to the \emph{fine-tuning gap} in semi-white-box assumption and the disparity between $D_a$ and real data. To overcome these challenges, \gls{ours} progresses from initial reconstruction to a second stage, leveraging both vulnerability{$_1$} and vulnerability{$_2$}. It employs an \emph{optimization-based} approach, \gls{m2}, which attacks in \emph{two directions} to utilize the \emph{posterior information} observed during the \gls{sl} process to mitigate the \emph{fine-tuning gap} for higher-precision reconstruction.

\subsubsection{Forward Enhancement by Smashed-Data Matching}

In the forward direction, existing \glspl{dra} assume white-box access, utilizing the smashed data $\tilde{x}_{\text{btm}}$ to perform relaxed optimization in a discrete vocabulary space for attacks \citep{DBLP:conf/ccs/song2020information}. However, in \gls{sl}, the adversary lacks white-box access, and within the \gls{llm} context, relaxed optimization becomes increasingly challenging and unstable due to the larger vocabulary space and model parameters (see Section \ref{sec:exp}). To address these challenges, \gls{ours} shifts to continuous optimization in the embedding space under a semi-white-box setting.

Specifically, we construct a dummy embedding $e'$ and iteratively optimize it to make the model's output smashed data $\tilde{x}_{\text{btm}}'$ approximate the observed $\tilde{x}_{\text{btm}}$, expecting $e'$ to approach the real embedding. Unlike previous work, the adversary has only semi-white-box access to the Bottom part of the model, meaning they can access \( f_{\bar{\text{btm}}} \) but not \( f_{\text{btm}} \). The optimization process is as follows:
\begin{equation}
e^* = \underset{e'}{\text{argmin}} \mathcal{D} (f'_{\bar{\text{btm}}}(e'),  \tilde{x}_{\text{btm}}),
\end{equation}
where \( \mathcal{D} \) is a distance function. Unlike \cite{DBLP:conf/ccs/song2020information}, we use cosine similarity rather than Euclidean distance as the loss function. This method effectively addresses the challenges associated with high-dimensional embeddings in \glspl{llm} (details in Appendix \ref{sec:diff-loss}), leading to significantly improved empirical reconstruction performance.

After optimization, using the embedding layer \( \text{emb}() \) of the pre-trained model, we can identify the sentence \( x^* \) in the vocabulary space whose embedding is closest to \( e^* \), which is then output as the reconstruction result:

\begin{equation}
x^* =  \underset{x'}{\text{argmin}} 
\; \| \text{emb}(x') -  e^* \|.
\end{equation}
In reverse, to initiate this optimization process with a previously reconstructed text, we simply pass the sentence through \( \text{emb}() \) to obtain its embedding, then proceed directly with optimization.

\subsubsection{Backward Enhancement by Gradient Matching}

The combination of \gls{sl} and \gls{llm} fine-tuning enables the reconstruction of original data \emph{in the backward direction} (Vulnerability{$_2$}). To achieve this, \gls{m2} employs a \gls{gma}-based approach. Previous \glspl{gma}~\citep{DBLP:conf/nips/ZhuLH19, DBLP:conf/nips/BalunovicD0V22, DBLP:conf/emnlp/DengWLWSLRD21} have primarily targeted \gls{fl} systems, framing the attack as an optimization problem. Specifically, for a network \( f_{\theta} \) and a loss function \( \mathcal{L} \), given a private training sample \( x \) with label \( y \), an attacker with white-box access to \( f_{\theta} \) and the true gradient \( \nabla_{\theta}\mathcal{L}_{\theta}(x,y) \) can generate a randomly initialized dummy sample \( (x^{'}, y^{'}) \), compute a dummy gradient through \( f_{\theta} \), and optimize the gradient-matching loss~\citep{DBLP:conf/nips/ZhuLH19} to minimize the discrepancy between the true gradient \( \nabla_{\theta}\mathcal{L}_{\theta}(x,y) \) and the dummy gradient \( \nabla_{\theta}\mathcal{L}_{\theta}(x^{'},y^{'}) \), thereby making \( x^{'} \) and \( y^{'} \) approximate the original private data.

\gls{m2} adapts \gls{gma} to the private-label \gls{sl} scenario, where \( f_\theta \) corresponds to the model's Top segment \( f_{\text{top}} \), and the input \( x \) becomes \( \tilde{x}_{\text{trk}} \). The adversary has only semi-white-box access to \( f_\theta \), meaning they can only access the mimic network \( f_{\bar{\text{top}}} \). After completing forward propagation, the client sends the gradient \( \text{grad}(\tilde{x}_{\text{trk}}) = \nabla_{\tilde{x}_{\text{trk}}} \mathcal{L}_{\text{top}}(\tilde{x}_{\text{trk}}, y) \) to the curious server, enabling a \gls{gma}. This process includes generating a dummy ground truth \( y^{'} \) (which must be a probability distribution over the vocabulary), passing \( \tilde{x}_{\text{trk}} \) through the mimic network \( f_{\bar{\text{top}}} \), calculating the dummy loss \( \mathcal{L}_{\text{dmy}}(\tilde{x}_{\text{trk}}, y^{'}) \), and minimizing the discrepancy between the gradient of the dummy loss and the real gradient represented by:

\begin{equation}
\begin{split}
L(y^{'}) & = \beta ||\nabla_{\tilde{x}_{\text{trk}}}\mathcal{L}_{\text{dmy}}(\tilde{x}_{\text{trk}},y^{'})-\nabla_{\tilde{x}_{\text{trk}}}\mathcal{L}(\tilde{x}_{\text{trk}},y)||_{2} \\
& +(1-\beta)||\nabla_{\tilde{x}_{\text{trk}}}\mathcal{L}_{\text{dmy}}(\tilde{x}_{\text{trk}},y^{'})-\nabla_{\tilde{x}_{\text{trk}}}\mathcal{L}(\tilde{x}_{\text{trk}},y)||_{1},
\end{split}
\end{equation}
where a combined distance of L2 norm and L1 norm is adopted, similar to TAG~\citep{DBLP:conf/emnlp/DengWLWSLRD21}, to enhance the attack performance on language models, with \( \beta \) serving as a coefficient parameter. It is important to note that \glspl{gma} in \gls{fl} settings often encounter the \emph{sequence order issue}, where the reconstructed input \(x'\) may be token-wise permuted. However, \gls{clm}'s input-label alignment (detailed later) and \gls{sl}‘s known-forward-activation trait eliminate this issue, allowing adversaries to obtain a correctly ordered sequence by simply optimizing $y'$.

What makes \gls{gma} effective for input-data reconstruction in \gls{sl} is \emph{\gls{llm}'s auto-regressive \gls{ft} objective}, which ensures alignment between the reconstruction of label and the input. Specifically, the $y$ to be recovered is a left-shifted version of the original input $x$, corresponding to $\tilde{x}_{\text{trk}}$. This alignment enables the \emph{collaborative use} of \gls{m2}'s forward and backward reconstruction processes, thereby significantly enhancing attack performance.

In practice, the forward and backward enhancements of \gls{m2} can be utilized in two ways: (1) sequentially, where gradient matching is performed first, followed by smashed-data matching. For the output $\hat{x}_{\text{SIP}}^{B\times T\times V}$ of \gls{m1}, softmax is applied along its last dimension to initialize dummy labels $y'$ for gradient matching, then after several rounds of gradient matching, the resulting $\hat{x}_{\text{GM}}$ is processed with a pre-trained embedding layer to obtain initial $e'$ for smashed-data matching, resulting in $\hat{x}_{\text{SM}}$. This method is denoted as \gls{ours} (b+f) in the experimental section. (2) Independently, where gradient matching and smashed-data matching are performed separately on the results of \gls{m1}. This is denoted as \gls{ours} (b) and \gls{ours} (f) in the experimental section, respectively. Adversaries can modularly employ \gls{ours} to adapt to different scenarios of forward and backward perturbations.

\subsection{Adapting to Noise: NaMoE Model}
\label{subsec:m3}

\begin{figure}
    \centering
    \includegraphics[width=1\linewidth]{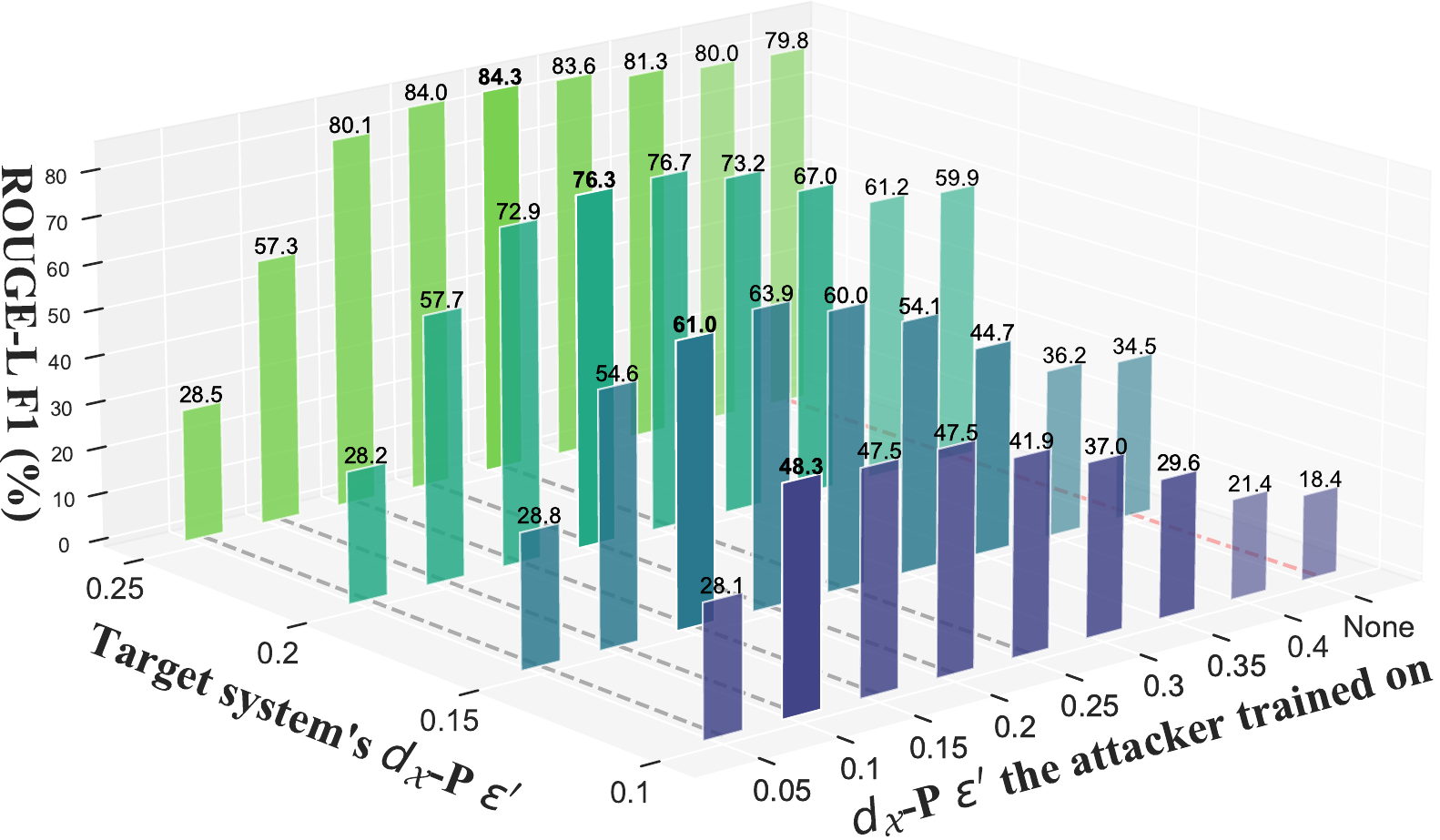}
    \caption{Performance boost brought by noise-aware training, demonstrated by attack performance (ROUGE-L F1 Score \%) of inverters trained with embedding-\gls{dxp}-awareness on split-\gls{ft} systems using embedding-gls{dxp} with varying noise scales, evaluated on GPT2-large and PIQA datasets.}
    \label{fig:noise-adaption}
    \Description{
    Performance boost brought by noise-aware training, demonstrated by attack performance (Rouge-L-F) of inverters trained with embedding-\gls{dxp}-awareness on split-\gls{ft} systems using embedding-gls{dxp} with varying noise scales, evaluated on GPT2-large and PIQA datasets.
    }
\end{figure}

\begin{figure}
    \centering
    \includegraphics[width=1\linewidth]{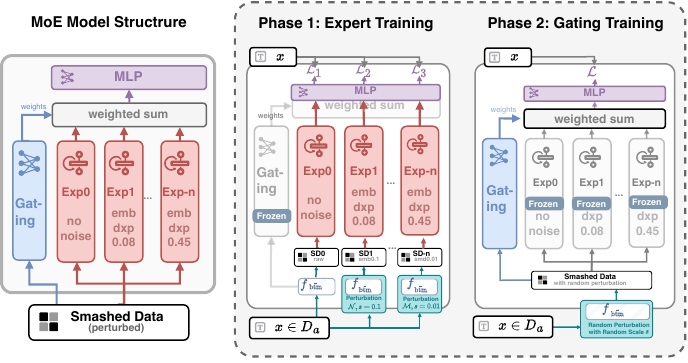}
    \caption{Proposed \gls{m3} model structure (left) and its training process (right). Each "Exp" represents an Expert handling a specific noise scale under a particular perturbation mechanism. The "emb" denotes Embedding-\gls{dxp}.}
    \label{fig:moe}
    \Description{
    Proposed \gls{m3} model structure (left) and its training process (right). Each "Exp" represents an Expert handling a specific noise scale under a particular perturbation mechanism. The "emb" denotes Embedding-\gls{dxp}.
    }
    
\end{figure}

Another goal of \gls{ours} is to provide adaptability against privacy-preserving methods that perturb intermediate outputs in the \emph{forward} direction. In the curious server scenario, perturbation serves as the primary non-cryptographic technique for privatizing \gls{sl}'s smashed data. Methods include adding noise directly to the smashed data~\citep{DBLP:journals/tvt/WuCYKYWP24, DBLP:journals/corr/abs-2104-05743} or to the output of the embedding layer~\citep{DBLP:conf/wsdm/FeyisetanBDD20, DBLP:journals/corr/abs-2312-15603}. These techniques not only degrade the performance of \gls{dra}, particularly optimization-based \gls{dra}, but also negatively impact model training performance, requiring more iterations to converge. Nevertheless, the ability of models to converge despite noise demonstrates their adaptability to forward perturbations under supervision.

This raises an intriguing question: \emph{Can learning-based inversion models also adapt to forward perturbations similar to \glspl{llm}}? To explore this, we first highlight the similarity between the split-\gls{ft} system (Bottom-Trunk-Top) and \gls{m1}‘s training system (Bottom-inversion model): (1) Both systems share a Bottom \gls{llm} segment with weights that are \emph{Not-too-far} from the pre-trained state. (2) They follow similar training paradigms, with the former's \gls{ft} label being the input sentence shifted by one token, while the latter's training label is the original input sentence.

Consequently, when perturbations impact the forward path, the noise in the smashed data necessitates adjustments in the \gls{sl} system's Trunk and Top layers via the \gls{ft} label to correct these disturbances and maintain functionality. Similarly, if noise is introduced during \gls{m1}'s training phase (\emph{noise-aware training}), the inversion model should be guided by the auto-encoder training label to develop perturbation correction capabilities. This leads us to the following observation:


\textbf{Observation}: \textit{Noise-aware-trained inversion models adapt more effectively to perturbations and achieve superior attack performance compared to their noise-unaware counterparts}.

More specifically, consider a model $f$ with a perturbation strategy (noise scale $s$) represented as $\mathcal{N}_s(f)$. If an adversary knows the perturbation strategy $\mathcal{N}$ used by the target and uses $e=\mathcal{N}_{s'}(f_{\bar{\text{btm}}})$ instead of $e=f_{\bar{\text{btm}}}$ during \gls{m1}'s auto-encoder-style training phase, the noise-aware trained inversion model can significantly improve performance when attacking an \gls{sl} system using the same perturbation strategy $\mathcal{N}_{s}$. Furthermore, the closer the noise scale $s'$ used by the attacker is to the real noise scale $s$, the higher the attack accuracy, as illustrated in Figure \ref{fig:noise-adaption}.

Building on this observation, a straightforward attack enhancement strategy could involve the adversary pre-training multiple \gls{dra} models on various noise scales and then deploying these models concurrently during the attack phase, choosing the highest-quality attack outcomes (evaluated by perplexity, for instance). However, this approach has the significant limitation that the adversary must anticipate the target's potential noise scales in advance and can only select from a restricted set of candidates with uncertain quality. To address this limitation, this paper introduces the \acrfull{m3} model, which incorporates the adaptability of different expert models to various noise scales. This enables adversaries to conduct \gls{dra} effectively even without knowledge of the target's specific privacy strategies.

As depicted in Fig. \ref{fig:moe} (left), the \gls{m3} model primarily comprises several expert networks and a gating network. Each expert network simultaneously decodes smashed data inputs, with the results weighted and combined based on the gating network's weights and processed by the output layer to yield the final attack outcome.

The \gls{m3} model utilizes a two-stage training approach, illustrated in Fig. \ref{fig:moe} (right). In the first stage, with the gating network frozen, each attack dataset sample $x \in D_a$ undergoes forward propagation with noise added per the perturbation mechanism $\mathcal{N}$ and noise scale $s'$ for each expert, creating personalized smashed data $\tilde{x} = \mathcal{N}_{s'}(f_{\bar{\text{btm}}})(x)$. The decoding capabilities of each expert are then independently trained on these smashed data. In the second stage, the expert networks are frozen, and each sample $x \in D_a$ is forward propagated with a random perturbation mechanism and noise scale to obtain smashed data $\tilde{x}$, which trains the gating network's ability to select the appropriate experts for each input type.

The \gls{m3} model trained in this manner can effectively mitigate performance degradation caused by various perturbation strategies without requiring the adversary to make excessive assumptions about the target. Its adaptability to different forward perturbations will be demonstrated in Section \ref{sec:exp}. In practice, the adversary can establish multiple expert networks that cover the range of noise levels acceptable for effective \gls{llm} training (from minor perturbations to levels that completely destroy fine-tuning utility), thereby training a highly adaptable attack model.

%% file: sections/5_experiments.tex
\section{Experiments}
\label{sec:exp}
\input{images/tables/hps}
\begin{figure*}
    \centering
    \includegraphics[width=\linewidth]{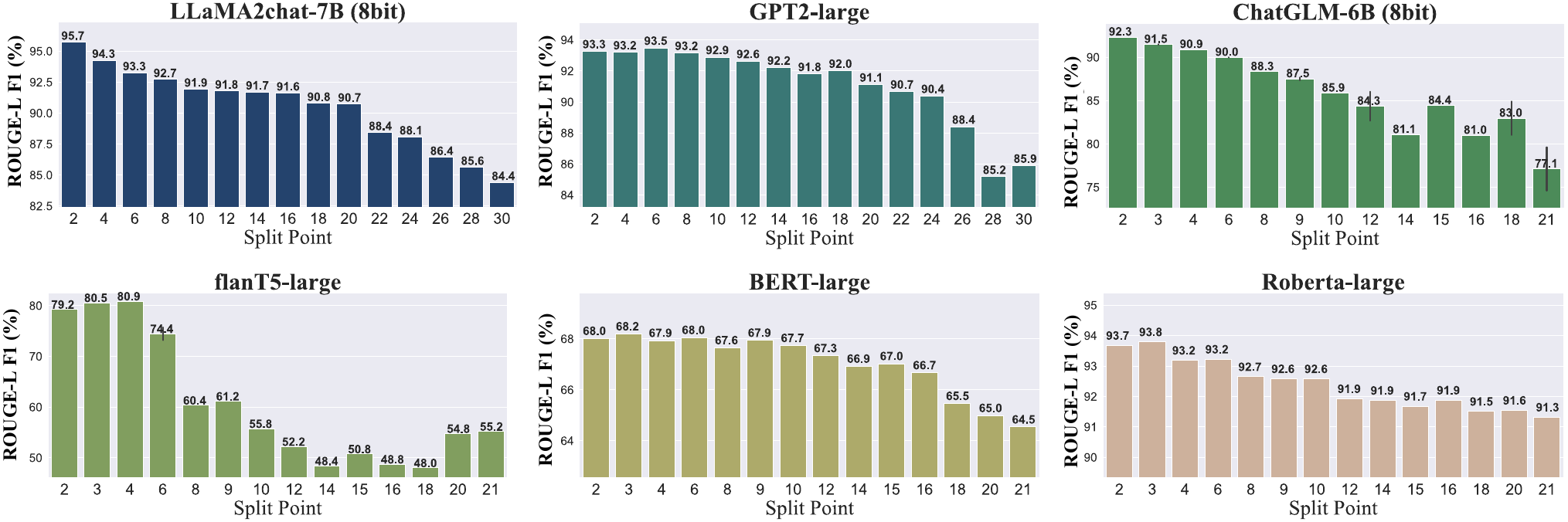}
    \caption{Attack performance (ROUGE-L F1 \%) of the proposed \gls{m1} on different \glspl{llm} using varying split points (e.g., split point=3 means blocks 0-2 are Bottom, and the output of the 3rd block is the attacked smashed data). For the encoder-decoder model flanT5, we consider the split point on the decoder, training the inverter to simultaneously recover the original inputs of the encoder and decoder from the decoder's smashed data.}
    \label{fig:diff-split}
    \Description{
    Attack performance (ROUGE-L F1 \%) of the proposed \gls{m1} on different \glspl{llm} using varying split points (e.g., split point=3 means blocks 0-2 are Bottom, and the output of the 3rd block is the attacked smashed data). For the encoder-decoder model flanT5, we consider the split point on the decoder, training the inverter to simultaneously recover the original inputs of the encoder and decoder from the decoder's smashed data.
    }
\end{figure*}

\subsection{Setup}

\textbf{Models.} Our experiments primarily focus on mainstream \glspl{llm} with decoder-only architectures, including GPT2-large~\citep{radford2019language}, LLaMA2-chat-7B~\citep{touvron2023llama}, and ChatGLM3-6B~\citep{DBLP:conf/acl/DuQLDQY022}. Due to hardware constraints, we employ 8-bit quantization during both the inference and fine-tuning of LLaMA2-chat-7B and ChatGLM3-6B, which, as shown in Appendix~\ref{sec:quant}, does not significantly impact attack performance. Additionally, to evaluate \gls{m1}, we conducted experiments on encoder-only models (BERT-large~\citep{DBLP:conf/naacl/DevlinCLT19} and RoBERTa-large~\citep{DBLP:journals/corr/abs-1907-11692}) as well as an encoder-decoder model (FLAN-T5~\citep{https://doi.org/10.48550/arxiv.2210.11416}).

\noindent\textbf{Datasets.} To assess the vulnerability of split-\gls{ft} in \glspl{llm} across various tasks, we utilize multiple datasets: QA datasets PIQA~\citep{Bisk2020} and GSM8K~\citep{cobbe2021gsm8k}, encyclopedic text dataset WikiText2-v1~\citep{merity2016pointer}, and the code dataset CodeAlpaca-20K~\citep{codealpaca} (details in Appendix \ref{sec:ds}). To further investigate the attack performance of DRA methods on \emph{sensitive information}, we modified the CNN-DailyMail News Text Summarization dataset~\citep{see-etal-2017-get} (details in Appendix \ref{sec:ss}) to create the Sensitive Datasets (\emph{Sensi-Series}): (1) SensiMarked, which annotates sensitive entities (names, dates, etc.) and focuses attack performance on them during evaluation; (2) SensiReplaced, created by substituting sensitive entities in SensiMarked with random, similar words to simulate a large-scale desensitized dataset; (3) SensiMasked, produced by masking all sensitive entities in SensiMarked with entity name tags for comparison. \emph{In all experiments, SensiReplaced is used by default as $D_a$}.

\noindent\textbf{Evaluation Metrics.} To thoroughly evaluate data reconstruction attacks, we employ two main metrics. (1) ROUGE-L~\citep{rouge2004package}, our most stringent metric, measures sentence similarity based on the longest common subsequence, requiring strict adherence to both order and tokens. Scores range from 0 to 1, with 1 indicating perfect attack performance. (2) Meteor~\citep{DBLP:conf/acl/BanerjeeL05} assesses similarity using n-gram matching and incorporates WordNet for synonym recognition, allowing for more flexibility in word order and synonym use, making it our most lenient metric. Additional metrics and corresponding results are detailed in Appendix \ref{sec:other-metrics}.

\noindent\textbf{Evaluation Settings.} By default, a private-label \gls{sl} setting with 6 Bottom blocks and 21 Trunk blocks (expressed as 6-21) is used for the three models. Attack performance was evaluated across steps of this split-\gls{ft} system, capped at 600 steps. Attacks were conducted on 5 batches every 200 steps during \gls{ft} to assess average performance. \gls{ft} effectiveness was gauged by the perplexity of the \gls{llm} on the test dataset. Due to hardware constraints, a batch size of 2 was used, and LoRA was applied for \gls{peft}. For \gls{m1}'s inversion model and \gls{m3}'s experts, a single-layer GRU (hidden size = 256, dropout = 0.1) was utilized. \gls{m2}'s optimization process used the best empirical hyperparameters for each \gls{llm}, as listed in Table \ref{tab:hps}. Experiments were conducted on a single machine (PyTorch 2.0, using the Huggingface Transformers library) to simulate multi-party \gls{sl} scenarios, all on Tesla V100-SXM2 GPUs (32GB). Main results were averaged over three random seeds to ensure reproducibility.

\subsection{Evaluation on SIP}
\label{sec:exp-m1}

\vspace{\parskip}
\begin{tcolorbox}[colback=gray!20!white,colframe=black,left=2mm, right=2mm, top=1mm, bottom=1mm, boxsep=0pt]
\textbf{Question 1.} \textit{Is \gls{m1} effective across various models? How does the split point impact attack performance?}
\end{tcolorbox}
\vspace{\parskip}

\gls{m1} is applicable to all \glspl{llm} architectures. We evaluated its performance, measured by the ROUGE-L F1 score, on various \gls{llm} architectures to assess attack effectiveness on smashed data across different layers. These \glspl{llm} include decoder-only models (LLaMA2-chat-7B~\citep{touvron2023llama}, GPT2-large~\citep{radford2019language}, and ChatGLM3-6B~\citep{glm2024chatglm}), encoder-only models (BERT-large~\citep{DBLP:conf/naacl/DevlinCLT19} and Roberta-large~\citep{DBLP:journals/corr/abs-1907-11692}), and an encoder-decoder model (FLAN-T5~\citep{https://doi.org/10.48550/arxiv.2210.11416}). Using the PIQA dataset, which supports \gls{sft} tasks across these models (QA generation for encoder-decoder and decoder-only models, binary classification for encoder-only models), we trained the \gls{m1} inversion model on the validation set and evaluated its performance on an \gls{llm} fine-tuned with the PIQA training set. Note that for encoder-decoder models like T5, where the decoder depends on encoder outputs for cross-attention, the standard private-label \gls{sl} architecture is unsuitable. Therefore, for comparison, we set the split point \emph{only on the decoder}, using a concatenation of the encoder's last hidden state and the decoder's smashed data as input for the inversion model.

Figure \ref{fig:diff-split} illustrates the attack performance of \gls{m1} on different \glspl{llm} at various split depths. \gls{m1} achieves high reconstruction performance, above 0.9, in shallow layers, especially in large-scale models like LLaMA2 and ChatGLM3. Deeper splits generally increase the difficulty of attacks. Compared to decoder-only models, masked language models such as Roberta and BERT show less intermediate output leakage.

\vspace{\parskip}
\begin{tcolorbox}[colback=gray!20!white,colframe=black,left=2mm, right=2mm, top=1mm, bottom=1mm, boxsep=0pt]
\textbf{Question 2.} \textit{What if the attack dataset $D_a$ and the target dataset $D_{c_i}$ are not sufficiently similar, i.e., if there is a distribution shift?}
\end{tcolorbox}
\vspace{\parskip}

\input{images/tables/cross_dataset}
To assess the generalizability of \gls{m1}, we evaluated its performance in cross-dataset attack tasks. Table \ref{tab:cross-dataset} presents the performance of \gls{m1} in cross-dataset scenarios, where the inversion model is trained on one dataset and used to attack an \gls{llm} splitly fine-tuned on another dataset. We observed performance across various datasets: CodeAlpaca (CodeAl), Gsm8k, PIQA, WikiText-20K (WikiT), SensiMarked (SsMkd), and SensiReplaced (SsRpl). The results indicate that \gls{m1} demonstrates strong cross-dataset attack performance (with ROUGE-L scores consistently above 0.5). However, the gap between $D_a$ and $D_{c_i}$ does affect its effectiveness; for instance, performance tends to be lower on the CodeAlpaca dataset, which significantly differs from the others.  

To further investigate \gls{m1}'s inversion patterns, we tested \gls{m1} inversion models trained on SensiReplaced and SensiMasked against the SensiMarked dataset (see the right part of Table \ref{tab:cross-dataset}). The SensiReplaced-trained attacker, despite its training data having sensitive words replaced with similar entities, successfully recovers most sensitive entities in SensiMarked (0.95+). In contrast, the SensiMasked-trained attacker, trained with sensitive entities masked, shows much lower recovery performance (0.74). This underscores the importance of using complete sentences and sufficient entity words in the training data for effective sensitive entity attacks. It also indicates that \gls{m1} may prioritize learning specific token representations over decoding (inversion) rules for the \gls{llm}'s intermediate layer outputs. For further discussion on \gls{m1}'s inversion mechanism, see Appendix \ref{sec:sip-insp}.

\vspace{\parskip}
\begin{tcolorbox}[colback=gray!20!white,colframe=black,left=2mm, right=2mm, top=1mm, bottom=1mm, boxsep=0pt]
\textbf{Question 3.} \textit{What if the simulated $f_{\bar{\text{btm}}}$ by the attacker is structurally inconsistent with the real $f_{\text{btm}}$?}
\end{tcolorbox}
\vspace{\parskip}

\begin{figure}
    \centering
    \includegraphics[width=1\linewidth]{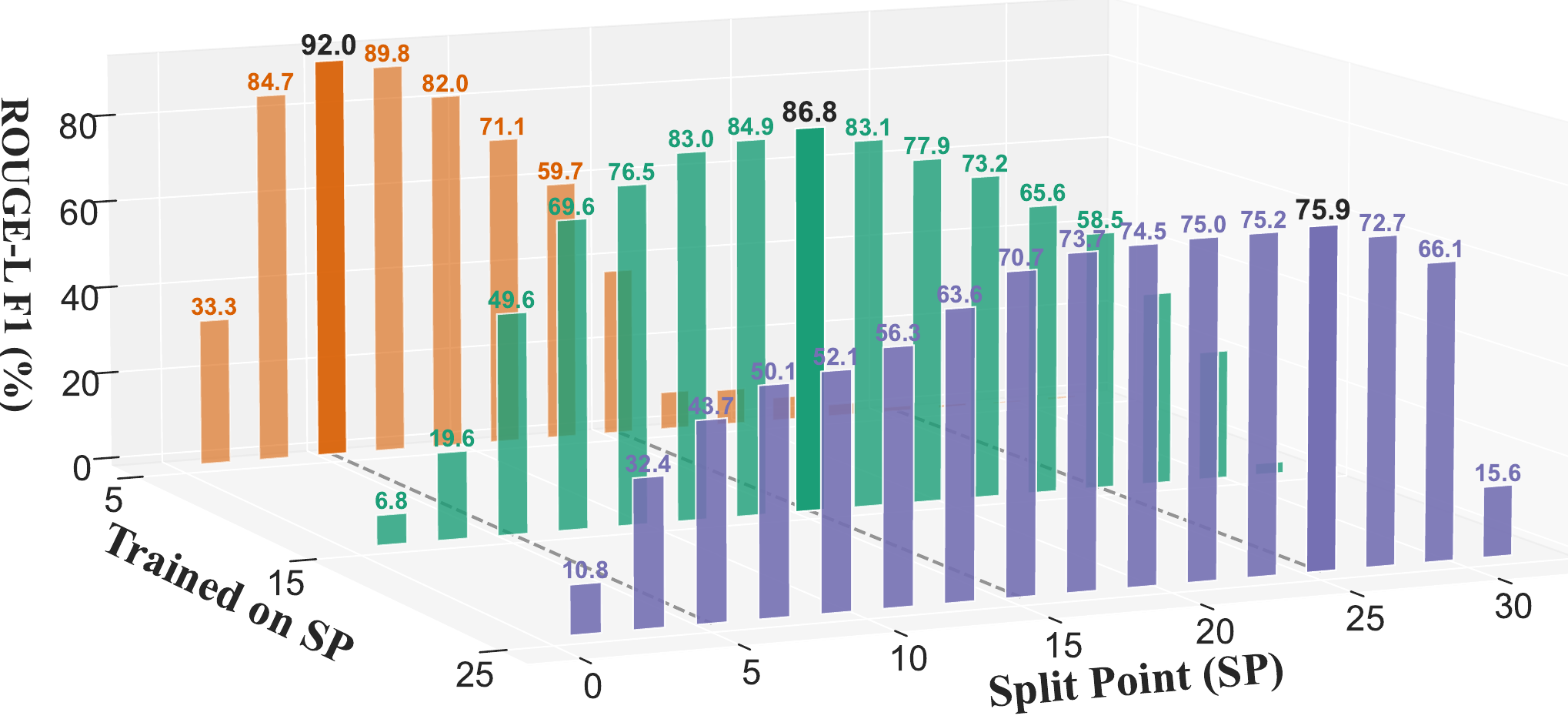}
    \caption{Attack performance of \gls{m1} in cross-split-point attacks. The three sets of bar graphs represent ROUGE-L F1 scores (\%) of \gls{m1} inversion models trained on the outputs of layers 5, 10, and 15 when attacking different layers of LLaMA2-chat-7B, with their performance at the split point they trained on emphasized. }
    \label{fig:cross-layer}
    \Description{
    Attack performance of \gls{m1} in cross-split-point attacks. The three sets of bar graphs represent ROUGE-L F1 scores (\%) of \gls{m1} inversion models trained on the outputs of layers 5, 10, and 15 when attacking different layers of LLaMA2-chat-7B, with their performance at the split point they trained on emphasized. 
    }
    
\end{figure}

\input{images/tables/big_table}

Figure \ref{fig:cross-layer} illustrates the cross-split-point attack performance of \gls{m1}, where the model is trained on one split setting (e.g., Split Point (SP) = 5, meaning the inversion model is trained on the output of the 5th block) and then used to attack a split-\gls{ft} system with a different setting (e.g., SP = 10, meaning the Bottom segment has 10 blocks). This indicates that the attack model's $f_{\bar{\text{btm}}}$ and the actual $f_\text{btm}$ have varying layer counts. Results show that \gls{m1} possesses robust cross-layer attack capabilities, performing better on closer layers, and models trained on deeper layers exhibit better generalization across different model layers.

\subsection{Evaluation on BiSR}
\label{sec:exp-m2}

\vspace{2mm}
\begin{tcolorbox}[colback=gray!20!white,colframe=black,left=2mm, right=2mm, top=1mm, bottom=1mm, boxsep=0pt]
\textbf{Question 4.} \textit{What are the advantages of \gls{ours} in performance relative to other \gls{dra} methods across different \glspl{llm}? }
\end{tcolorbox}
\vspace{\parskip}

Table \ref{tab:cross-dataset-bisr} compares \gls{ours}‘s performance with various \gls{dra} methods attacking different \glspl{llm} fine-tuned on distinct datasets. We evaluated forward-\glspl{dra} methods based on smashed data, including AE~\citep{DBLP:journals/tbd/ZhangPTEM23}, which uses an autoencoder with an encoder structurally identical to the model's Bottom but without pre-trained weights, and Embedding Inversion Attack (EIA)\citep{DBLP:conf/acsac/HeZL19}, which performs smashed-data matching through relaxed optimization on vocabulary and uses Euclidean Distance as the target. We also evaluated backward-\glspl{dra} methods based on gradient matching, including TAG\citep{DBLP:conf/emnlp/DengWLWSLRD21} and LAMP~\citep{DBLP:conf/nips/BalunovicD0V22}.

It's worth noting that, most of these baseline \glspl{dra} (EIA, DLG, TAG, and LAMP) were designed for \emph{white-box} scenarios. For comparison, we adapted them to the \emph{semi-white-box} \gls{sl} setting by replacing the white-box model segments they access with the same model but with \emph{pre-trained weights}, and marked their results with \textbf{*}. For \gls{m1} and AE, the inversion model (GRU) was trained on SensiReplaced. For EIA, we followed \citet{DBLP:conf/acsac/HeZL19}, pretraining a Mapper network on SensiReplaced to map the targeted \gls{llm} hidden states back to a shallow layer (the first layer) before performing the attack. A two-layer MLP Mapper produced the best results for the \glspl{llm} considered. For gradient-based attacks (TAG and LAMP), we adapted them to the \gls{sl} setting by reconstructing only the label $y$, given that the network input is known. All hyperparameters for each method and dataset were meticulously fine-tuned.


We present the outputs of \gls{ours} at various stages: \gls{m1}-only showcases the initial output from the learning-based inversion model's direct attack, while \gls{ours} (b) and \gls{ours} (f) respectively represent the results enhanced with gradient matching and smashed-data matching based on \gls{m1}. \gls{ours} (b+f) indicates the sequentially enhanced results using \gls{m2}, with backward enhancement followed by forward enhancement. As the table illustrates, for larger models, purely optimization-based methods like TAG and EIA are limited by extensive search spaces, resulting in a lower reconstruction performance ceiling. It is worth noting that in the considered setup, LAMP did not achieve an advantage over TAG. This is because when applying gradient attacks in \gls{sl} scenarios, there is no issue of sentence order. As a result, the token reorder operation in LAMP does not provide a substantial benefit. The learning-based \gls{m1}-only approach produces high-quality recovered sentences without any iteration steps but suffers performance drawbacks when significant discrepancies exist between $D_a$ and the target data (e.g., CodeAlpaca). \gls{ours}, integrating learning-based and optimization-based approaches, achieves significantly optimal recovery performance. This effectiveness stems from \gls{m1} providing an excellent starting point for optimization, thereby facilitating the bidirectional optimization process to more effectively find the global optimum.

\input{images/tables/pre_ft}

\vspace{\parskip}
\begin{tcolorbox}[colback=gray!20!white,colframe=black,left=2mm, right=2mm, top=1mm, bottom=1mm, boxsep=0pt]
\textbf{Question 5.} \textit{Does the Not-too-far Assumption hold in practice?}
\end{tcolorbox}
\vspace{\parskip}

To validate the Not-too-far Assumption, we assessed various \glspl{dra} in scenarios where the client does not initiate \gls{ft} from the pre-trained weights provided by the server. We simulated this by: (1) Attacking a LLaMA2-chat-7B model that was splitly fine-tuned on PIQA, with the client performing \gls{ft} on the Bottom and Top parts for several steps on CodeAlpaca-20K beforehand (pre-\gls{ft}), diverging them from the pre-trained state; (2) Attacking a well-fine-tuned model, specifically the LLaMA2-based Vicuna-v1.5-7B~\citep{zheng2023judging}. As shown in Table \ref{tab:pre-ft}, when the client's model portions diverge further from the pre-trained states, with the L2-norm of their tunable parameters (LoRA Adapters) increasing from 19.59 to 32.89, the attack performance of \gls{ours} and other \glspl{dra} does not exhibit a decline. Even when attacking Vicuna while semi-white-box accessing LLaMA2's weights, the performance remains unaffected. Thus, Assumption \ref{as:nottoofar} holds empirically. Furthermore, we compared \gls{dra} for language and vision models in Appendix \ref{app:vis}, suggesting that The Not-too-far may be uniquely applicable to language models.

\subsection{Evaluation on BiSR's Noise Adaptability}

\vspace{\parskip}
\begin{tcolorbox}[colback=gray!20!white,colframe=black,left=2mm, right=2mm, top=1mm, bottom=1mm, boxsep=0pt]
\textbf{Question 6.} \textit{How does \gls{ours} adapt to various forward perturbation mechanisms?}
\end{tcolorbox}
\vspace{\parskip}


\begin{figure}
    \centering
    \includegraphics[width=\linewidth]{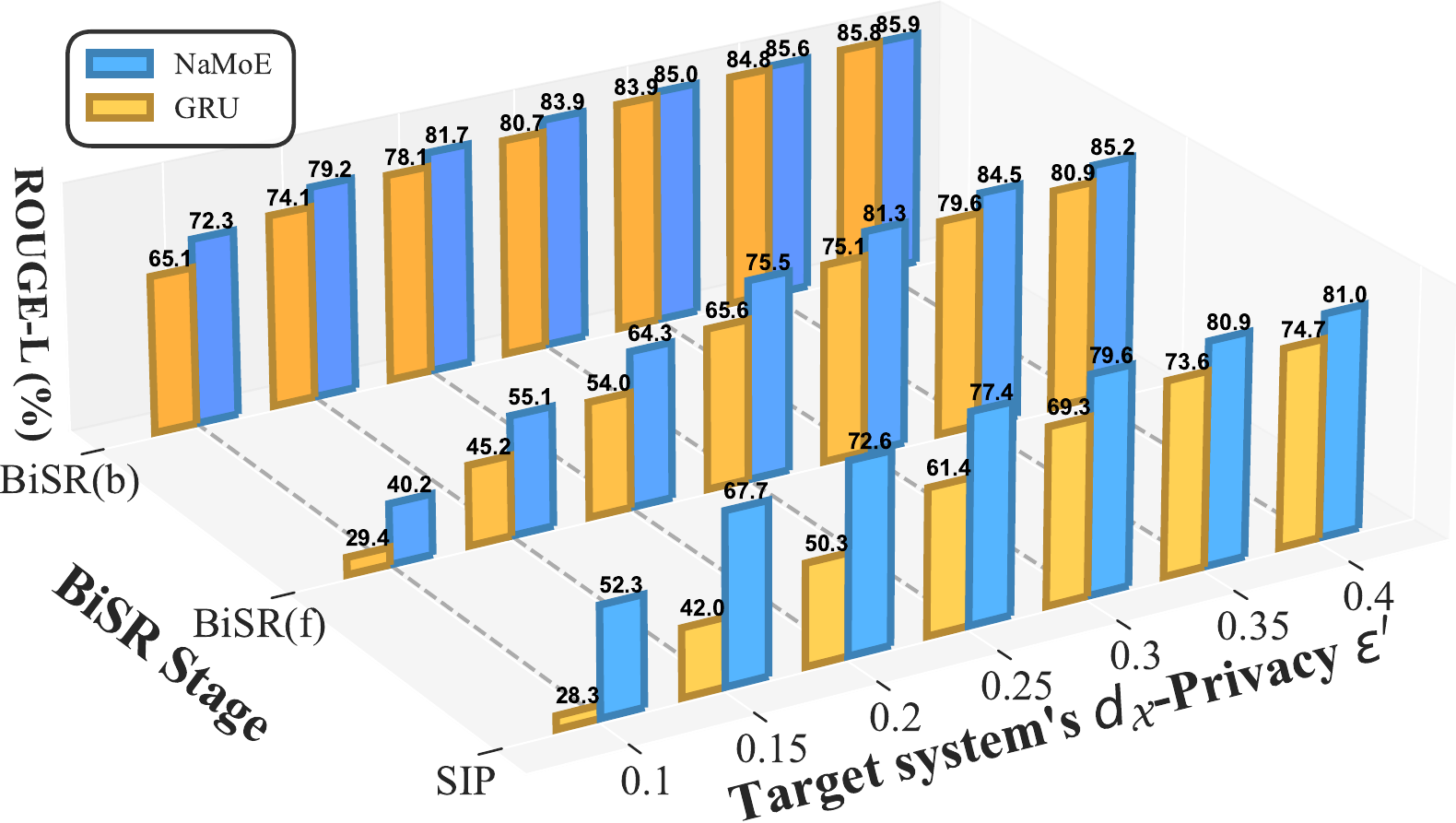}
    \caption{Comparison of attack performance at different stages of \gls{ours} when using \gls{m3} versus a standard GRU-based \gls{m1} for initial reconstruction under varying scales of \gls{dxp} noise. Each set of bars represents the ROUGE-L F1 Score (\%) of the attack method when the target system applies a specific scale of \gls{dxp} perturbation.}
    \label{fig:namoe-comp}
    \Description{
    Comparison of attack performance at different stages of \gls{ours} when using \gls{m3} versus a standard GRU-based \gls{m1} for initial reconstruction under varying scales of \gls{dxp} noise. Each set of bars represents the ROUGE-L F1 Score (\%) of the attack method when the target system applies a specific scale of \gls{dxp} perturbation.
    }
\end{figure}

We evaluated three representative non-cryptographic privacy preservation techniques for \gls{sl}, which mitigate privacy leakage by perturbing intermediate results during forward propagation. The selected mechanisms include: Embedding-\gls{dxp}\citep{DBLP:conf/pet/ChatzikokolakisABP13}, which perturbs sentence embeddings in language models; Smashed-Data-\gls{dp}\citep{DBLP:conf/ccs/DuYC0H023, DBLP:journals/tvt/WuCYKYWP24, DBLP:journals/corr/abs-2104-05743}, which injects noise directly into smashed data; and NoPeek~\citep{DBLP:conf/icdm/Vepakomma0GR20}, which restricts the relevance of intermediate data to the input.

\input{images/tables/dxp_table}

\vspace{0.3cm}
\noindent \textbf{Embedding-\gls{dxp}}. Table \ref{tab:dxp} presents the results of embedding-\gls{dxp}~\citep{DBLP:conf/pet/ChatzikokolakisABP13} tests conducted using the LLaMA2-chat-7B model. The model was fine-tuned on PIQA with \gls{dxp} noise applied to the Bottom part's embedding layer. The noise scale is defined by the relative parameter $\epsilon' = \epsilon / n$, where $n$ is the dimensionality of the model's hidden states and $\epsilon$ is the original $d_{\mathcal{X}}$-Privacy parameter. A smaller $\epsilon'$ indicates stronger perturbations. In this experiment, the \gls{m3} model comprised four experts with $\epsilon'$ values of $+\infty$, 0.08, 0.38, and 0.21, each using a GRU with 256 hidden units. We trained the experts for 15 epochs and the gating network for 10 epochs on SensiReplaced. As shown in Table \ref{tab:dxp}, the \gls{m3} inversion model consistently outperforms a standard, noise-unaware GRU across various noise scales, even without strictly following the noise scales used by the expert models. Notably, even when model performance collapses (test perplexity > 100) at $\epsilon'=0.1$, \gls{m3} still achieves a recovery performance above 0.5. The advantages of \gls{m3} over GRU and its impact on subsequent optimization-based attacks are more clearly illustrated in Figure \ref{fig:namoe-comp}. It is also important to note that optimization-based methods relying on smashed data matching, such as EIA and \gls{ours}(b+f), are more sensitive to noise than the learning-based \gls{m1}. Under high noise levels, the (f) phase of \gls{m2} can degrade \gls{m1} performance, while gradient-based \glspl{dra} remain largely unaffected.

\input{images/tables/gaussian_table}

\vspace{0.3cm}
\noindent \textbf{Smashed-data-\gls{dp}}. Table \ref{tab:gaussian} displays the performance of \gls{ours} compared to other \gls{dra} methods across various scales of smashed-data-\gls{dp}~\citep{DBLP:journals/tvt/WuCYKYWP24, DBLP:journals/corr/abs-2104-05743, DBLP:conf/ccs/DuYC0H023}. We define $\epsilon^*=\epsilon / G$ as the relative \gls{dp} parameter, where $G$ is the clipping threshold (fixed at 2000 for this experiment), and we evaluated the performance of \gls{ours} on LLaMA2-chat-7B model and the PIQA dataset. The \gls{m3} model here also includes three experts, corresponding to $\epsilon^*=+\infty$, $\epsilon^*=3.0$, $\epsilon^*=5.0$, and $\epsilon^*=8.0$. It is important to note that directly training the inversion model on heavily perturbed smashed data is challenging. To address this, we employed an \emph{attack-deeper strategy}: the curious server targets hidden states a few layers deeper within the Trunk instead of at the cut layer. This approach leverages the \gls{llm}'s noise adaptability—during noisy training, the \gls{llm} reduces perturbation impact, leading to decreased noise in hidden states after several \gls{llm} blocks. Thus, although the cut layer was set at layer 6 in LLaMA2 (where noise is added), we trained the \gls{m3} model on layer 9, significantly reducing training difficulty. As shown in Table \ref{tab:gaussian}, \gls{m3} exhibits stronger adaptability than the GRU, maintaining recovery performance above 0.2 even under extreme conditions. However, forward optimization (f) becomes ineffective when noise is directly added to smashed data.

\input{images/tables/nopeek}


\vspace{0.3cm}
\noindent \textbf{NoPeek}. Since NoPeek's perturbation method optimizes regularization terms rather than directly adding random noise, the evaluation approach and \gls{m3} model training differ significantly from the previous methods. We employed GPT2-large with an increased batch size (=6) to ensure distance correlation was effective. Instead of attacking 5 batches every 200 steps, we reduced the attack frequency to 5 batches every 1200 steps to allow the regularization term to stabilize—posing a greater challenge to the attacker, as earlier attacks typically yield better results.A simulation strategy was employed to train the noise-aware \gls{m3} model (see Appendix \ref{sec:namoe-np}). Additionally, the \emph{attack-deeper} approach was applied, targeting layer 8 of GPT2-large. As shown in Table \ref{tab:nopeek}, \gls{m3} demonstrates adaptability to NoPeek perturbations, outperforming the GRU. Notably, the purely gradient-based \gls{dra} method, TAG, also experienced performance degradation under NoPeek perturbations.

\subsection{Impact of Batch Size}

\begin{figure}
    \centering
    \includegraphics[width=1\linewidth]{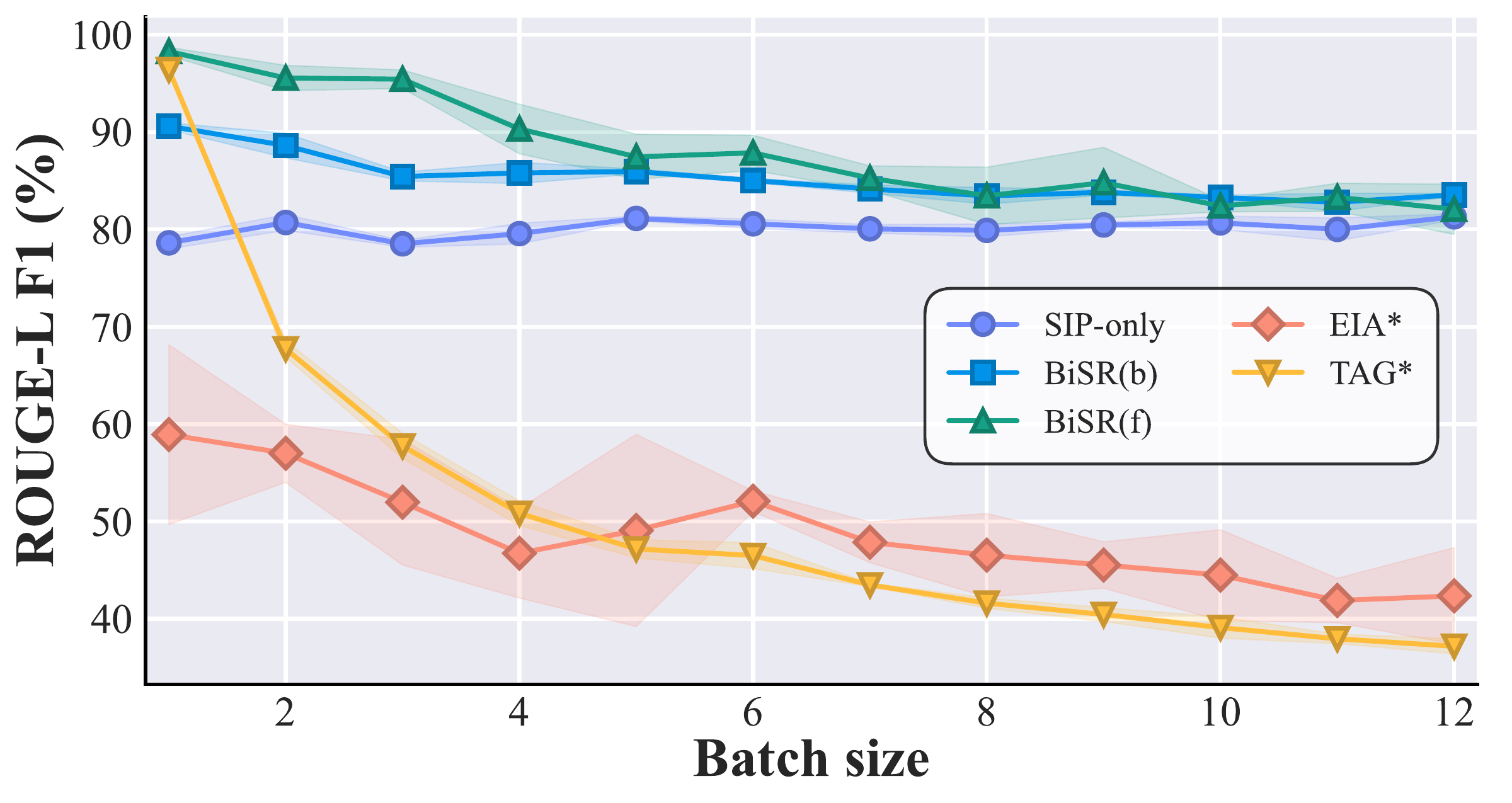}
    \caption{The impact of batch size on the attack performance of \gls{ours} and other \glspl{dra}, tested on the GPT2-large and the PIQA dataset (with the maximum sentence length truncated to 128). Error bars are plotted as shaded regions.}
    \label{fig:diff-bs}
    \Description{
    The impact of batch size on the attack performance of \gls{ours} and other \glspl{dra}, tested on the GPT2-large and the PIQA dataset (with the maximum sentence length truncated to 128). Error bars are plotted as shaded regions.
    }
\end{figure}

The minibatch size is a critical factor frequently examined in \gls{dra} literature, particularly in gradient-based \glspl{dra}, as gradient averaging across batches can complicate the success of attacks. The optimization-based procedures in \gls{m2} are similarly impacted by this factor. Although batch sizes in \gls{llm} scenarios are generally limited by resource constraints—especially in the \gls{sl} scenario discussed in this paper—we evaluated the performance of \gls{ours} and various baselines as the batch size increased. These experiments were conducted using the GPT2-large model and the PIQA dataset, with a maximum sentence length of 128 to allow for larger batch sizes while keeping the experimental setup consistent.

As illustrated in Fig. \ref{fig:diff-bs}, increasing the batch size significantly diminishes the effectiveness of both optimization-based attacks, i.e., EIA and TAG, while the learning-based \gls{m1}-only approach remains unaffected. Due to the inclusion of forward and backward optimizations in \gls{m2}, \gls{ours} also experiences reduced performance at larger batch sizes. However, under the hyperparameter settings detailed in Table \ref{tab:hps}, both \gls{ours}(b) and \gls{ours}(f) still outperform \gls{m1} even with larger batch sizes (though a higher number of \gls{m2} epochs may adversely affect performance in this context). This is because \gls{ours} benefits from \gls{m1}'s initialization, leveraging its learning-based strengths to handle larger batches more effectively. This results in lower search difficulty during optimization-based attacks, thereby providing greater robustness to larger batch sizes compared to other baselines.

Additionally, it is important to note that while increasing the batch size can effectively reduce \gls{dra} risks, it also demands higher computational resources, leading to a trade-off between privacy protection and resource expenditure. Besides batch size, other factors (e.g., quantization, split points) also affect \gls{dra} performance. For additional experimental results, please refer to Appendix \ref{sec:detail-exp}.

%% file: images/tables/hps.tex
\begin{table}
\caption{Empirical hyperparameters for \gls{ours}'s gradient-matching (GM) and smashed-data-matching (SM) enhancements targeting different \glspl{llm}.}
\label{tab:hps}
\resizebox{\columnwidth}{!}{%
\begin{tabular}{l|ll|lllllll}
\hline
\multirow{2}{*}{Model} & \multicolumn{2}{l|}{\multirow{2}{*}{Optimizer}} & \multicolumn{7}{c}{Hyper Parameters} \\ \cline{4-10} 
 & \multicolumn{2}{l|}{} & \multicolumn{1}{c}{GM-epc} & \multicolumn{1}{c}{GM-lr} & \multicolumn{1}{c}{GM-$\beta$} & \multicolumn{1}{c}{GM-$\tau$} & \multicolumn{1}{c}{SM-epc} & \multicolumn{1}{c}{SM-lr} & \multicolumn{1}{c}{SM-wd} \\ \hline
LLaMA2-7B & \multicolumn{2}{l|}{\multirow{3}{*}{\begin{tabular}[c]{@{}l@{}}AdamW\\ (0.9, \\ 0.999)\end{tabular}}} & 18 & 0.09 & 0.85 & 1.2 & 800 & 0.005 & 0.02 \\
GPT2-large & \multicolumn{2}{l|}{} & 32 & 0.09 & 0.85 & 1.2 & 800 & 0.01 & 0.01 \\
ChatGLM3-6B & \multicolumn{2}{l|}{} & 18 & 0.09 & 0.85 & 1.2 & 800 & 0.005 & 0.01 \\ \hline
\end{tabular}%
}
\end{table}

%% file: images/tables/cross_dataset.tex
\begin{table}
\caption{Attack performance of \gls{m1} in cross-dataset attacks. Each row represents the ROUGE-L F1 score (\%) of \gls{m1} attacks on LLaMA2 split-fine-tuned (cutlayer=6) on a specific dataset (with the inverter model trained on different datasets). }
\label{tab:cross-dataset}
\resizebox{\columnwidth}{!}{%
\begin{tabular}{l|ccccccc}
\hline
\multirow{2}{*}{Dataset} & \multicolumn{7}{c}{Inversion Model's Training Dataset} \\ \cline{2-8} 
 & CodeAl & GSM8K & PIQA & WikiT & \multicolumn{1}{c|}{SsMkd} & SsRpl & SsMsk \\ \hline
CodeAl & \begin{tabular}[c]{@{}c@{}}95.78\\ $\pm$0.15\end{tabular} & \begin{tabular}[c]{@{}c@{}}59.25\\ $\pm$0.81\end{tabular} & \begin{tabular}[c]{@{}c@{}}61.13\\ $\pm$1.43\end{tabular} & \begin{tabular}[c]{@{}c@{}}55.18\\ $\pm$1.36\end{tabular} & \multicolumn{1}{c|}{\begin{tabular}[c]{@{}c@{}}67.54\\ $\pm$0.85\end{tabular}} & \begin{tabular}[c]{@{}c@{}}65.02\\ $\pm$1.31\end{tabular} & \begin{tabular}[c]{@{}c@{}}67.11\\ $\pm$1.13\end{tabular} \\
GSM8K & \begin{tabular}[c]{@{}c@{}}70.71\\ $\pm$0.43\end{tabular} & \begin{tabular}[c]{@{}c@{}}93.85\\ $\pm$0.33\end{tabular} & \begin{tabular}[c]{@{}c@{}}74.53\\ $\pm$1.86\end{tabular} & \begin{tabular}[c]{@{}c@{}}91.41\\ $\pm$0.42\end{tabular} & \multicolumn{1}{c|}{\begin{tabular}[c]{@{}c@{}}88.05\\ $\pm$0.33\end{tabular}} & \begin{tabular}[c]{@{}c@{}}85.27\\ $\pm$0.51\end{tabular} & \begin{tabular}[c]{@{}c@{}}83.73\\ $\pm$1.14\end{tabular} \\
PIQA & \begin{tabular}[c]{@{}c@{}}66.92\\ $\pm$0.92\end{tabular} & \begin{tabular}[c]{@{}c@{}}79.13\\ $\pm$1.42\end{tabular} & \begin{tabular}[c]{@{}c@{}}94.97\\ $\pm$1.22\end{tabular} & \begin{tabular}[c]{@{}c@{}}61.91\\ $\pm$0.80\end{tabular} & \multicolumn{1}{c|}{\begin{tabular}[c]{@{}c@{}}83.87\\ $\pm$0.95\end{tabular}} & \begin{tabular}[c]{@{}c@{}}82.91\\ $\pm$0.72\end{tabular} & \begin{tabular}[c]{@{}c@{}}83.80\\ $\pm$1.03\end{tabular} \\
WikiT & \begin{tabular}[c]{@{}c@{}}50.42\\ $\pm$1.22\end{tabular} & \begin{tabular}[c]{@{}c@{}}60.20\\ $\pm$1.00\end{tabular} & \begin{tabular}[c]{@{}c@{}}56.19\\ $\pm$1.07\end{tabular} & \begin{tabular}[c]{@{}c@{}}91.41\\ $\pm$0.42\end{tabular} & \multicolumn{1}{c|}{\begin{tabular}[c]{@{}c@{}}96.96\\ $\pm$0.36\end{tabular}} & \begin{tabular}[c]{@{}c@{}}95.81\\ $\pm$0.31\end{tabular} & \begin{tabular}[c]{@{}c@{}}88.15\\ $\pm$2.32\end{tabular} \\
SsMkd & \begin{tabular}[c]{@{}c@{}}39.07\\ $\pm$1.59\end{tabular} & \begin{tabular}[c]{@{}c@{}}52.63\\ $\pm$1.91\end{tabular} & \begin{tabular}[c]{@{}c@{}}42.62\\ $\pm$1.98\end{tabular} & \begin{tabular}[c]{@{}c@{}}75.36\\ $\pm$1.84\end{tabular} & \multicolumn{1}{c|}{\begin{tabular}[c]{@{}c@{}}97.54\\ $\pm$0.80\end{tabular}} & \begin{tabular}[c]{@{}c@{}}95.57\\ $\pm$1.15\end{tabular} & \begin{tabular}[c]{@{}c@{}}74.17\\ $\pm$2.78\end{tabular} \\ \hline
\end{tabular}%
}
\end{table}

%% file: images/tables/big_table.tex

\begin{table*}
\caption{Performance of \gls{ours} on various \glspl{llm} fine-tuned on different datasets (ROUGE-L F1 Score \%) compared to alternative methods. Experiments utilized split points set at 6-27, with SensiReplaced serving as the auxiliary dataset $D_a$. The best results for each configuration are highlighted in \textbf{bold}, with the second best results \underline{underlined}.}
\label{tab:cross-dataset-bisr}
\resizebox{\textwidth}{!}{
\begin{tabular}{l|l|ccccllllll}
\hline
\multicolumn{1}{c|}{} & \multicolumn{1}{c|}{} & \multicolumn{10}{c}{Split Fine-tuning Datasets} \\ \cline{3-12} 
\multicolumn{1}{c|}{} & \multicolumn{1}{c|}{} & \multicolumn{2}{c}{SensiMkd} & \multicolumn{2}{c}{Codealpaca} & \multicolumn{2}{c}{Gsm8k} & \multicolumn{2}{c}{PIQA} & \multicolumn{2}{c}{WikiText} \\ \cline{3-12} 
\multicolumn{1}{c|}{\multirow{-3}{*}{Model}} & \multicolumn{1}{c|}{\multirow{-3}{*}{Methods}} & RougeL-F & Meteor & RougeL-F & Meteor & \multicolumn{1}{c}{RougeL-F} & \multicolumn{1}{c}{Meteor} & \multicolumn{1}{c}{RougeL-F} & \multicolumn{1}{c}{Meteor} & \multicolumn{1}{c}{RougeL-F} & \multicolumn{1}{c}{Meteor} \\ \hline
 & AE & {\color[HTML]{333333} 49.99$\pm$0.34} & {\color[HTML]{333333} 46.99$\pm$1.62} & {\color[HTML]{333333} 60.31$\pm$0.48} & {\color[HTML]{333333} 62.09$\pm$0.74} & {\color[HTML]{333333} 42.32$\pm$0.33} & {\color[HTML]{333333} 46.41$\pm$0.83} & {\color[HTML]{333333} 32.63$\pm$1.29} & {\color[HTML]{333333} 29.78$\pm$1.71} & {\color[HTML]{333333} 47.15$\pm$0.67} & {\color[HTML]{333333} 41.05$\pm$0.92} \\
 & EIA* & {\color[HTML]{333333} 83.06$\pm$4.92} & {\color[HTML]{333333} 77.98$\pm$5.35} & {\color[HTML]{333333} 54.00$\pm$2.90} & {\color[HTML]{333333} 49.83$\pm$3.31} & \multicolumn{1}{c}{{\color[HTML]{333333} 59.97$\pm$3.78}} & \multicolumn{1}{c}{{\color[HTML]{333333} 62.22$\pm$3.19}} & \multicolumn{1}{c}{{\color[HTML]{333333} 57.86$\pm$0.46}} & \multicolumn{1}{c}{{\color[HTML]{333333} 65.93$\pm$1.45}} & \multicolumn{1}{c}{{\color[HTML]{333333} 83.42$\pm$1.25}} & \multicolumn{1}{c}{{\color[HTML]{333333} 81.50$\pm$0.22}} \\
 & TAG* & {\color[HTML]{333333} 80.22$\pm$1.47} & {\color[HTML]{333333} 79.61$\pm$1.25} & {\color[HTML]{333333} 84.94$\pm$0.25} & {\color[HTML]{333333} 85.34$\pm$0.89} & \multicolumn{1}{c}{{\color[HTML]{333333} 82.40$\pm$1.93}} & \multicolumn{1}{c}{{\color[HTML]{333333} 84.45$\pm$1.01}} & \multicolumn{1}{c}{{\color[HTML]{333333} 77.05$\pm$2.30}} & \multicolumn{1}{c}{{\color[HTML]{333333} 78.10$\pm$1.75}} & \multicolumn{1}{c}{{\color[HTML]{333333} 74.36$\pm$1.16}} & \multicolumn{1}{c}{{\color[HTML]{333333} 72.68$\pm$0.75}} \\
 & LAMP* & {\color[HTML]{333333} 79.13$\pm$0.66} & {\color[HTML]{333333} 78.55$\pm$0.26} & {\color[HTML]{333333} 83.51$\pm$0.13} & {\color[HTML]{333333} 85.11$\pm$0.17} & \multicolumn{1}{c}{{\color[HTML]{333333} 83.23$\pm$0.14}} & \multicolumn{1}{c}{{\color[HTML]{333333} 83.82$\pm$0.31}} & \multicolumn{1}{c}{{\color[HTML]{333333} 76.50$\pm$1.03}} & \multicolumn{1}{c}{{\color[HTML]{333333} 76.22$\pm$1.09}} & \multicolumn{1}{c}{{\color[HTML]{333333} 74.23$\pm$0.89}} & \multicolumn{1}{c}{{\color[HTML]{333333} 72.40$\pm$0.86}} \\ \cline{2-12} 
 & SIP-only & {\color[HTML]{333333} 95.66$\pm$0.98} & 96.37$\pm$1.18 & {\color[HTML]{333333} 65.08$\pm$1.12} & 66.67$\pm$0.89 & \multicolumn{1}{c}{{\color[HTML]{333333} 85.37$\pm$0.75}} & 88.05$\pm$0.25 & \multicolumn{1}{c}{{\color[HTML]{333333} 82.83$\pm$0.64}} & 96.57$\pm$0.26 & \multicolumn{1}{c}{{\color[HTML]{333333} 95.90$\pm$0.41}} & 89.06$\pm$0.54 \\
 & BiSR(b) & {\color[HTML]{333333} 96.80$\pm$0.78} & 97.42$\pm$0.76 & {\color[HTML]{333333} 83.02$\pm$1.12} & 85.65$\pm$0.83 & \multicolumn{1}{c}{{\color[HTML]{333333} 90.82$\pm$0.69}} & 93.80$\pm$1.03 & \multicolumn{1}{c}{{\color[HTML]{333333} 89.34$\pm$0.67}} & 97.10$\pm$0.32 & \multicolumn{1}{c}{{\color[HTML]{333333} 96.31$\pm$0.38}} & 94.65$\pm$0.32 \\
 & BiSR(f) & {\color[HTML]{333333} {\ul 98.70$\pm$0.58}} & {\ul 98.94$\pm$0.34} & {\color[HTML]{333333} {\ul 86.93$\pm$0.88}} & {\ul 94.39$\pm$1.22} & \multicolumn{1}{c}{{\color[HTML]{333333} {\ul 94.59$\pm$0.40}}} & {\ul 95.96$\pm$1.05} & \multicolumn{1}{c}{{\color[HTML]{333333} {\ul 92.30$\pm$1.14}}} & {\ul 99.01$\pm$0.29} & \multicolumn{1}{c}{{\color[HTML]{333333} {\ul 98.09$\pm$0.33}}} & {\ul 97.70$\pm$0.19} \\
\multirow{-8}{*}{\begin{tabular}[c]{@{}l@{}}LLa-\\ MA2\end{tabular}} & BiSR & {\color[HTML]{333333} \textbf{99.64$\pm$0.26}} & \textbf{99.47$\pm$0.42} & {\color[HTML]{333333} \textbf{92.62$\pm$0.77}} & \textbf{96.58$\pm$0.14} & \multicolumn{1}{c}{{\color[HTML]{333333} \textbf{95.98$\pm$0.18}}} & \textbf{97.71$\pm$0.06} & \multicolumn{1}{c}{{\color[HTML]{333333} \textbf{95.84$\pm$0.15}}} & \textbf{99.46$\pm$0.14} & \multicolumn{1}{c}{{\color[HTML]{333333} \textbf{98.49$\pm$0.27}}} & \textbf{98.03$\pm$0.17} \\ \hline
 & AE & {\color[HTML]{333333} 75.47$\pm$0.31} & 78.04$\pm$1.21 & {\color[HTML]{333333} 85.66$\pm$0.36} & 90.23$\pm$0.28 & {\color[HTML]{333333} 73.20$\pm$1.02} & 88.00$\pm$0.52 & {\color[HTML]{333333} 64.11$\pm$0.82} & 69.32$\pm$1.55 & {\color[HTML]{333333} 82.55$\pm$4.96} & 63.04$\pm$2.36 \\
 & EIA* & {\color[HTML]{333333} 85.15$\pm$3.11} & 61.35$\pm$6.70 & {\color[HTML]{333333} 50.54$\pm$7.12} & 39.76$\pm$4.66 & {\color[HTML]{333333} 65.75$\pm$4.84} & 57.29$\pm$7.51 & {\color[HTML]{333333} 60.59$\pm$2.10} & 57.45$\pm$2.64 & {\color[HTML]{333333} 81.50$\pm$0.56} & 71.17$\pm$2.23 \\
 & TAG* & {\color[HTML]{333333} 62.32$\pm$5.48} & 45.99$\pm$4.74 & {\color[HTML]{333333} 77.62$\pm$1.08} & 74.82$\pm$1.48 & {\color[HTML]{333333} 78.63$\pm$1.19} & 75.11$\pm$1.40 & {\color[HTML]{333333} 71.04$\pm$1.92} & 67.71$\pm$0.82 & {\color[HTML]{333333} 68.47$\pm$0.86} & 67.67$\pm$0.29 \\
 & LAMP* & {\color[HTML]{333333} 60.29$\pm$0.24} & 48.33$\pm$0.60 & {\color[HTML]{333333} 76.48$\pm$0.18} & 73.01$\pm$0.14 & {\color[HTML]{333333} 75.16$\pm$0.27} & 67.33$\pm$0.31 & {\color[HTML]{333333} 68.59$\pm$0.44} & 64.48$\pm$0.24 & {\color[HTML]{333333} 67.57$\pm$0.27} & 66.34$\pm$0.11 \\ \cline{2-12} 
 & SIP-only & {\color[HTML]{333333} 80.84$\pm$0.75} & 90.69$\pm$0.27 & {\color[HTML]{333333} 88.77$\pm$0.32} & 93.22$\pm$0.38 & {\color[HTML]{333333} 93.19$\pm$0.17} & 92.85$\pm$0.86 & {\color[HTML]{333333} 78.41$\pm$1.25} & 84.84$\pm$1.04 & {\color[HTML]{333333} 93.53$\pm$0.34} & 78.04$\pm$4.68 \\
 & BiSR(b) & {\color[HTML]{333333} 88.16$\pm$0.74} & 92.87$\pm$0.24 & {\color[HTML]{333333} 94.81$\pm$0.35} & 96.48$\pm$0.06 & {\color[HTML]{333333} 95.76$\pm$0.21} & 95.12$\pm$0.71 & {\color[HTML]{333333} 91.14$\pm$0.34} & 93.41$\pm$0.15 & {\color[HTML]{333333} 97.28$\pm$0.75} & 82.26$\pm$3.66 \\
 & BiSR(f) & {\color[HTML]{333333} {\ul 95.07$\pm$0.60}} & {\ul 97.43$\pm$0.24} & {\color[HTML]{333333} \textbf{97.22$\pm$0.81}} & {\ul 99.05$\pm$0.28} & {\color[HTML]{333333} {\ul 99.06$\pm$0.29}} & {\ul 99.30$\pm$0.13} & {\color[HTML]{333333} {\ul 92.30$\pm$1.12}} & {\ul 96.17$\pm$0.63} & {\color[HTML]{333333} {\ul 99.71$\pm$0.26}} & {\ul 83.91$\pm$3.71} \\
\multirow{-8}{*}{\begin{tabular}[c]{@{}l@{}}GPT2\\ -large\end{tabular}} & BiSR & {\color[HTML]{333333} \textbf{95.20$\pm$0.83}} & \textbf{97.46$\pm$0.43} & {\color[HTML]{333333} {\ul 97.16$\pm$0.55}} & \textbf{99.11$\pm$0.11} & {\color[HTML]{333333} \textbf{99.14$\pm$0.11}} & \textbf{99.38$\pm$0.01} & {\color[HTML]{333333} \textbf{92.64$\pm$0.70}} & \textbf{96.20$\pm$0.45} & {\color[HTML]{333333} \textbf{99.79$\pm$0.29}} & \textbf{84.06$\pm$3.89} \\ \hline
 & AE & 66.36$\pm$1.12 & 62.08$\pm$0.67 & 68.89$\pm$0.57 & 64.22$\pm$0.69 & \multicolumn{1}{c}{62.61$\pm$0.91} & 64.86$\pm$0.80 & \multicolumn{1}{c}{48.94$\pm$0.44} & 41.83$\pm$1.82 & 65.85$\pm$1.08 & 59.06$\pm$2.21 \\
 & EIA* & 61.53$\pm$13.47 & 62.42$\pm$15.69 & 12.14$\pm$1.33 & 10.43$\pm$1.66 & \multicolumn{1}{c}{11.41$\pm$0.25} & 12.80$\pm$1.04 & \multicolumn{1}{c}{29.50$\pm$6.14} & 27.84$\pm$6.35 & 26.65$\pm$0.15 & 26.76$\pm$1.61 \\
 & TAG* & 60.86$\pm$1.97 & 59.22$\pm$2.04 & 77.55$\pm$0.98 & 81.75$\pm$2.49 & \multicolumn{1}{c}{79.23$\pm$1.64} & 83.41$\pm$3.09 & \multicolumn{1}{c}{67.30$\pm$0.52} & 68.30$\pm$1.02 & \multicolumn{1}{c}{66.04$\pm$1.17} & 65.42$\pm$0.82 \\
 & LAMP* & 39.20$\pm$0.47 & 36.01$\pm$0.13 & 76.27$\pm$0.40 & 78.28$\pm$0.13 & 77.60$\pm$0.26 & 81.34$\pm$0.33 & 66.95$\pm$0.47 & 67.22$\pm$0.28 & \multicolumn{1}{c}{67.18$\pm$0.52} & 65.59$\pm$0.21 \\ \cline{2-12} 
 & SIP-only & 93.78$\pm$1.14 & 94.75$\pm$1.20 & 63.81$\pm$1.40 & 61.92$\pm$2.45 & 77.00$\pm$0.88 & 79.74$\pm$0.49 & 92.57$\pm$0.30 & 92.37$\pm$0.60 & 82.34$\pm$0.69 & 86.13$\pm$0.37 \\
 & BiSR(b) & 94.96$\pm$1.29 & 95.54$\pm$1.02 & \textbf{86.96$\pm$0.42} & 83.13$\pm$0.74 & {\ul 90.59$\pm$0.63} & {\ul 90.97$\pm$0.42} & 94.45$\pm$0.23 & 94.68$\pm$0.17 & \textbf{93.11$\pm$0.82} & 93.43$\pm$0.74 \\
 & BiSR(f) & {\ul 100.00$\pm$0.00} & {\ul 99.99$\pm$0.01} & 86.64$\pm$1.48 & {\ul 90.66$\pm$1.05} & 89.92$\pm$1.07 & 90.13$\pm$0.68 & \textbf{99.72$\pm$0.10} & \textbf{99.76$\pm$0.04} & 86.12$\pm$1.72 & {\ul 95.74$\pm$0.34} \\
\multirow{-8}{*}{\begin{tabular}[c]{@{}l@{}}Chat-\\ GLM3\end{tabular}} & BiSR & \textbf{100.00$\pm$0.00} & \textbf{99.99$\pm$0.01} & {\ul 86.64$\pm$1.55} & \textbf{93.37$\pm$0.61} & \multicolumn{1}{c}{\textbf{90.90$\pm$0.31}} & \textbf{96.09$\pm$0.14} & \multicolumn{1}{c}{{\ul 99.71$\pm$0.09}} & {\ul 99.74$\pm$0.05} & {\ul 86.52$\pm$1.97} & \textbf{96.38$\pm$0.38} \\ \hline
\end{tabular}%
}
\end{table*}

%% file: images/tables/pre_ft.tex
\begin{table*}
\caption{Validation of the \emph{Not-too-far Assumption}: comparing the attack performance (ROUGE-L F1 Score \%) of different \glspl{dra} on LLaMA2-chat-7B with varying pre-fine-tuning steps, and on a well-fine-tuned model Vicuna-v1.5-7B.}
\label{tab:pre-ft}
\resizebox{\textwidth}{!}{%
\begin{tabular}{l|cccccc|c}
\hline
 & \multicolumn{6}{c|}{Pre-fine-tuning Steps} & Cross-model \\ \cline{2-8} 
\multirow{-2}{*}{DRA Methods} & 0 & 4800 & 9600 & 14400 & 19200 & 24000 & Vicuna \\ \hline
EIA & {\color[HTML]{333333} $49.15\pm2.27$} & {\color[HTML]{333333} $50.89\pm4.32$} & {\color[HTML]{333333} $57.49\pm6.50$} & {\color[HTML]{333333} $53.50\pm6.99$} & {\color[HTML]{333333} $62.96\pm3.53$} & $62.17\pm4.40$ & $48.59\pm15.51$ \\
TAG & $75.36\pm4.24$ & $69.62\pm2.13$ & $70.14\pm1.42$ & $70.46\pm0.75$ & $70.20\pm1.94$ & $70.04\pm2.63$ & $73.24\pm1.67$ \\ \hline
SIP & $80.50\pm1.78$ & $77.99\pm2.05$ & $81.23\pm2.75$ & $81.14\pm1.89$ & $82.38\pm1.56$ & $82.63\pm1.80$ & $82.27\pm2.67$ \\
BiSR(b) & $88.26\pm2.81$ & $89.09\pm1.66$ & $88.70\pm2.04$ & $88.14\pm0.62$ & $89.58\pm2.50$ & $89.19\pm1.07$ & $89.39\pm1.74$ \\
BiSR(f) & $82.91\pm2.10$ & $79.29\pm1.89$ & $81.17\pm2.98$ & $79.42\pm1.95$ & $80.06\pm2.26$ & $80.15\pm2.12$ & $95.66\pm0.77$ \\
BiSR(b+f) & $93.22\pm1.37$ & $91.73\pm3.30$ & $92.02\pm4.07$ & $92.11\pm2.86$ & $95.56\pm0.63$ & $95.05\pm0.40$ & $96.49\pm0.55$ \\ \hline
\rowcolor[HTML]{EFEFEF} 
LoRA L2-Norm & $19.59\pm0.01$ & $26.29\pm0.08$ & $28.18\pm0.18$ & $29.89\pm0.21$ & $31.44\pm0.23$ & $32.89\pm0.29$ & - \\ \hline
\end{tabular}%
}
\end{table*}

%% file: images/tables/dxp_table.tex
\begin{table}
\caption{Performance comparison (ROUGE-L F1 Score \% $\uparrow$) of \gls{ours} against various \glspl{dra} under embedding-\gls{dxp} perturbations, alongside a comparison between the \gls{m3} model and a standard noise-unaware GRU model. \gls{llm} utility was assessed using test perplexity ($\downarrow$). Experiments were conducted with the LLaMA2-chat-7B model and the PIQA dataset, with SensiReplaced serving as the auxiliary dataset $D_a$.}
\label{tab:dxp}
\resizebox{\columnwidth}{!}{%
\begin{tabular}{l|ccccccc}
\hline
 & \multicolumn{7}{c}{\gls{dxp}-$\epsilon'$} \\ \cline{2-8} 
\multirow{-2}{*}{Methods} & 0.4 & 0.35 & 0.30 & 0.25 & 0.20 & 0.15 & \multicolumn{1}{l}{0.10} \\ \hline
EIA* & \begin{tabular}[c]{@{}c@{}}51.98\\ $\pm$2.72\end{tabular} & \begin{tabular}[c]{@{}c@{}}54.45\\ $\pm$2.72\end{tabular} & \begin{tabular}[c]{@{}c@{}}51.80\\ $\pm$2.84\end{tabular} & \begin{tabular}[c]{@{}c@{}}43.66\\ $\pm$0.35\end{tabular} & \begin{tabular}[c]{@{}c@{}}42.33\\ $\pm$0.45\end{tabular} & \begin{tabular}[c]{@{}c@{}}32.25\\ $\pm$3.83\end{tabular} & \begin{tabular}[c]{@{}c@{}}17.56\\ $\pm$0.23\end{tabular} \\
TAG* & \begin{tabular}[c]{@{}c@{}}73.95\\ $\pm$0.65\end{tabular} & \begin{tabular}[c]{@{}c@{}}73.26\\ $\pm$0.25\end{tabular} & \begin{tabular}[c]{@{}c@{}}71.86\\ $\pm$0.35\end{tabular} & \begin{tabular}[c]{@{}c@{}}69.57\\ $\pm$0.96\end{tabular} & \begin{tabular}[c]{@{}c@{}}68.56\\ $\pm$0.97\end{tabular} & \begin{tabular}[c]{@{}c@{}}68.36\\ $\pm$1.24\end{tabular} & \begin{tabular}[c]{@{}c@{}}63.19\\ $\pm$1.16\end{tabular} \\ \hline
SIP (GRU) & \begin{tabular}[c]{@{}c@{}}74.73\\ $\pm$0.56\end{tabular} & \begin{tabular}[c]{@{}c@{}}73.59\\ $\pm$1.16\end{tabular} & \begin{tabular}[c]{@{}c@{}}69.26\\ $\pm$1.89\end{tabular} & \begin{tabular}[c]{@{}c@{}}61.40\\ $\pm$2.16\end{tabular} & \begin{tabular}[c]{@{}c@{}}50.29\\ $\pm$1.61\end{tabular} & \begin{tabular}[c]{@{}c@{}}41.98\\ $\pm$1.21\end{tabular} & \begin{tabular}[c]{@{}c@{}}28.34\\ $\pm$2.08\end{tabular} \\
SIP (NaMoE) & \begin{tabular}[c]{@{}c@{}}80.96\\ $\pm$0.89\end{tabular} & \begin{tabular}[c]{@{}c@{}}80.93\\ $\pm$0.80\end{tabular} & \begin{tabular}[c]{@{}c@{}}79.62\\ $\pm$0.23\end{tabular} & \begin{tabular}[c]{@{}c@{}}77.44\\ $\pm$0.77\end{tabular} & \begin{tabular}[c]{@{}c@{}}72.56\\ $\pm$0.54\end{tabular} & \begin{tabular}[c]{@{}c@{}}67.65\\ $\pm$0.64\end{tabular} & \begin{tabular}[c]{@{}c@{}}52.31\\ $\pm$1.57\end{tabular} \\
BiSR(b) & \begin{tabular}[c]{@{}c@{}}85.95\\ $\pm$1.42\end{tabular} & \begin{tabular}[c]{@{}c@{}}85.58\\ $\pm$1.82\end{tabular} & \begin{tabular}[c]{@{}c@{}}85.01\\ $\pm$1.00\end{tabular} & \begin{tabular}[c]{@{}c@{}}83.86\\ $\pm$0.93\end{tabular} & \begin{tabular}[c]{@{}c@{}}81.71\\ $\pm$0.50\end{tabular} & \begin{tabular}[c]{@{}c@{}}79.20\\ $\pm$0.37\end{tabular} & \begin{tabular}[c]{@{}c@{}}72.34\\ $\pm$0.70\end{tabular} \\
BiSR(b+f) & \begin{tabular}[c]{@{}c@{}}91.06\\ $\pm$0.16\end{tabular} & \begin{tabular}[c]{@{}c@{}}89.82\\ $\pm$0.40\end{tabular} & \begin{tabular}[c]{@{}c@{}}86.74\\ $\pm$0.36\end{tabular} & \begin{tabular}[c]{@{}c@{}}79.78\\ $\pm$0.58\end{tabular} & \begin{tabular}[c]{@{}c@{}}67.65\\ $\pm$0.95\end{tabular} & \begin{tabular}[c]{@{}c@{}}58.64\\ $\pm$0.73\end{tabular} & \begin{tabular}[c]{@{}c@{}}42.26\\ $\pm$1.45\end{tabular} \\ \hline
\rowcolor[HTML]{EFEFEF} 
Test PPL & \begin{tabular}[c]{@{}c@{}}31.82\\ $\pm$0.63\end{tabular} & \begin{tabular}[c]{@{}c@{}}37.71\\ $\pm$2.33\end{tabular} & \begin{tabular}[c]{@{}c@{}}57.00\\ $\pm$1.51\end{tabular} & \begin{tabular}[c]{@{}c@{}}61.86\\ $\pm$2.34\end{tabular} & \begin{tabular}[c]{@{}c@{}}48.44\\ $\pm$1.48\end{tabular} & \begin{tabular}[c]{@{}c@{}}47.51\\ $\pm$1.88\end{tabular} & \begin{tabular}[c]{@{}c@{}}138.86\\ $\pm$4.31\end{tabular} \\ \hline
\end{tabular}%
}
\end{table}

%% file: images/tables/gaussian_table.tex
\begin{table}[]
\caption{Performance comparison (ROUGE-L F1 Score \% $\uparrow$) of \gls{ours} versus various \glspl{dra} and \gls{llm} utility (test perplexity $\downarrow$) under smashed-data-\gls{dp} perturbations. The experiments used the LLaMA2-chat-7B model with split points 6-27 and the PIQA dataset. The \emph{attack-deeper} strategy was employed, with the \gls{m3} model trained on SensiReplaced and targeting the model’s 9th block’s output.}
\label{tab:gaussian}

\resizebox{\columnwidth}{!}{%
\begin{tabular}{l|ccccccc}
\hline
 & \multicolumn{7}{c}{\gls{dp}-$\epsilon'$} \\ \cline{2-8} 
\multirow{-2}{*}{Methods} & 7.0 & 6.5 & 6.0 & 5.5 & 5.0 & 4.5 & 4.0 \\ \hline
EIA* & \begin{tabular}[c]{@{}c@{}}34.27\\ $\pm$1.49\end{tabular} & \begin{tabular}[c]{@{}c@{}}31.71\\ $\pm$1.60\end{tabular} & \begin{tabular}[c]{@{}c@{}}27.00\\ $\pm$0.60\end{tabular} & \begin{tabular}[c]{@{}c@{}}23.23\\ $\pm$0.63\end{tabular} & \begin{tabular}[c]{@{}c@{}}18.27\\ $\pm$0.59\end{tabular} & \begin{tabular}[c]{@{}c@{}}15.83\\ $\pm$0.46\end{tabular} & \begin{tabular}[c]{@{}c@{}}12.62\\ $\pm$0.62\end{tabular} \\
TAG* & \begin{tabular}[c]{@{}c@{}}67.05\\ $\pm$0.85\end{tabular} & \begin{tabular}[c]{@{}c@{}}66.66\\ $\pm$0.51\end{tabular} & \begin{tabular}[c]{@{}c@{}}65.72\\ $\pm$0.56\end{tabular} & \begin{tabular}[c]{@{}c@{}}63.01\\ $\pm$0.32\end{tabular} & \begin{tabular}[c]{@{}c@{}}61.12\\ $\pm$0.88\end{tabular} & \begin{tabular}[c]{@{}c@{}}60.78\\ $\pm$0.71\end{tabular} & \begin{tabular}[c]{@{}c@{}}60.78\\ $\pm$0.71\end{tabular} \\ \hline
SIP (GRU) & \begin{tabular}[c]{@{}c@{}}31.58\\ $\pm$1.18\end{tabular} & \begin{tabular}[c]{@{}c@{}}27.23\\ $\pm$0.96\end{tabular} & \begin{tabular}[c]{@{}c@{}}22.73\\ $\pm$2.24\end{tabular} & \begin{tabular}[c]{@{}c@{}}18.97\\ $\pm$0.35\end{tabular} & \begin{tabular}[c]{@{}c@{}}14.90\\ $\pm$0.61\end{tabular} & \begin{tabular}[c]{@{}c@{}}9.78\\ $\pm$1.06\end{tabular} & \begin{tabular}[c]{@{}c@{}}7.64\\ $\pm$0.50\end{tabular} \\
SIP (NaMoE) & \begin{tabular}[c]{@{}c@{}}55.13\\ $\pm$1.05\end{tabular} & \begin{tabular}[c]{@{}c@{}}52.07\\ $\pm$1.39\end{tabular} & \begin{tabular}[c]{@{}c@{}}49.11\\ $\pm$1.27\end{tabular} & \begin{tabular}[c]{@{}c@{}}45.75\\ $\pm$1.32\end{tabular} & \begin{tabular}[c]{@{}c@{}}39.85\\ $\pm$0.86\end{tabular} & \begin{tabular}[c]{@{}c@{}}33.59\\ $\pm$0.77\end{tabular} & \begin{tabular}[c]{@{}c@{}}26.67\\ $\pm$1.27\end{tabular} \\
BiSR(b) & \begin{tabular}[c]{@{}c@{}}76.29\\ $\pm$1.65\end{tabular} & \begin{tabular}[c]{@{}c@{}}74.39\\ $\pm$1.18\end{tabular} & \begin{tabular}[c]{@{}c@{}}72.27\\ $\pm$0.75\end{tabular} & \begin{tabular}[c]{@{}c@{}}68.48\\ $\pm$0.87\end{tabular} & \begin{tabular}[c]{@{}c@{}}64.95\\ $\pm$0.95\end{tabular} & \begin{tabular}[c]{@{}c@{}}61.43\\ $\pm$0.64\end{tabular} & \begin{tabular}[c]{@{}c@{}}56.41\\ $\pm$1.63\end{tabular} \\
BiSR(b+f) & \begin{tabular}[c]{@{}c@{}}67.90\\ $\pm$2.13\end{tabular} & \begin{tabular}[c]{@{}c@{}}65.85\\ $\pm$1.76\end{tabular} & \begin{tabular}[c]{@{}c@{}}63.41\\ $\pm$1.57\end{tabular} & \begin{tabular}[c]{@{}c@{}}58.88\\ $\pm$1.66\end{tabular} & \begin{tabular}[c]{@{}c@{}}55.60\\ $\pm$2.02\end{tabular} & \begin{tabular}[c]{@{}c@{}}52.37\\ $\pm$1.63\end{tabular} & \begin{tabular}[c]{@{}c@{}}48.09\\ $\pm$2.69\end{tabular} \\ \hline
\rowcolor[HTML]{EFEFEF} 
Test PPL & \begin{tabular}[c]{@{}c@{}}31.80\\ $\pm$0.66\end{tabular} & \begin{tabular}[c]{@{}c@{}}35.42\\ $\pm$0.48\end{tabular} & \begin{tabular}[c]{@{}c@{}}40.18\\ $\pm$0.35\end{tabular} & \begin{tabular}[c]{@{}c@{}}49.43\\ $\pm$0.82\end{tabular} & \begin{tabular}[c]{@{}c@{}}68.18\\ $\pm$1.01\end{tabular} & \begin{tabular}[c]{@{}c@{}}112.71\\ $\pm$2.29\end{tabular} & \begin{tabular}[c]{@{}c@{}}269.10\\ $\pm$67.23\end{tabular} \\ \hline
\end{tabular}%
}
\end{table}

%% file: images/tables/nopeek.tex
\begin{table}[]
\caption{Adaptability of \gls{ours} to NoPeek-based perturbations. Experiments were conducted on the GPT2-large model and the PIQA dataset, with the \gls{m3} inversion model trained on SensiReplaced.}
\label{tab:nopeek}
\resizebox{\columnwidth}{!}{%
\begin{tabular}{l|ccccccc}
\hline
 & \multicolumn{7}{c}{NoPeek-$\alpha 2$} \\ \cline{2-8} 
\multirow{-2}{*}{Methods} & 10 & 30 & 50 & 70 & 90 & 110 & 130 \\ \hline
EIA* & \begin{tabular}[c]{@{}c@{}}14.49\\ $\pm$1.16\end{tabular} & \begin{tabular}[c]{@{}c@{}}13.17\\ $\pm$0.51\end{tabular} & \begin{tabular}[c]{@{}c@{}}12.57\\ $\pm$1.33\end{tabular} & \begin{tabular}[c]{@{}c@{}}12.46\\ $\pm$0.96\end{tabular} & \begin{tabular}[c]{@{}c@{}}12.02\\ $\pm$0.96\end{tabular} & \begin{tabular}[c]{@{}c@{}}11.43\\ $\pm$0.9\end{tabular} & \begin{tabular}[c]{@{}c@{}}10.90\\ $\pm$1.40\end{tabular} \\
TAG* & \begin{tabular}[c]{@{}c@{}}66.03\\ $\pm$0.86\end{tabular} & \begin{tabular}[c]{@{}c@{}}63.92\\ $\pm$1.11\end{tabular} & \begin{tabular}[c]{@{}c@{}}63.02\\ $\pm$1.21\end{tabular} & \begin{tabular}[c]{@{}c@{}}62.83\\ $\pm$0.36\end{tabular} & \begin{tabular}[c]{@{}c@{}}53.05\\ $\pm$1.10\end{tabular} & \begin{tabular}[c]{@{}c@{}}53.16\\ $\pm$1.55\end{tabular} & \begin{tabular}[c]{@{}c@{}}54.76\\ $\pm$2.16\end{tabular} \\ \hline
SIP (GRU) & \begin{tabular}[c]{@{}c@{}}59.45\\ $\pm$0.03\end{tabular} & \begin{tabular}[c]{@{}c@{}}54.07\\ $\pm$1.20\end{tabular} & \begin{tabular}[c]{@{}c@{}}38.68\\ $\pm$0.40\end{tabular} & \begin{tabular}[c]{@{}c@{}}32.88\\ $\pm$4.90\end{tabular} & \begin{tabular}[c]{@{}c@{}}17.43\\ $\pm$3.62\end{tabular} & \begin{tabular}[c]{@{}c@{}}11.38\\ $\pm$2.14\end{tabular} & \begin{tabular}[c]{@{}c@{}}7.19\\ $\pm$2.36\end{tabular} \\
SIP (NaMoE) & \begin{tabular}[c]{@{}c@{}}63.74\\ $\pm$1.35\end{tabular} & \begin{tabular}[c]{@{}c@{}}56.32\\ $\pm$1.03\end{tabular} & \begin{tabular}[c]{@{}c@{}}48.08\\ $\pm$2.39\end{tabular} & \begin{tabular}[c]{@{}c@{}}39.53\\ $\pm$0.01\end{tabular} & \begin{tabular}[c]{@{}c@{}}28.94\\ $\pm$2.89\end{tabular} & \begin{tabular}[c]{@{}c@{}}20.56\\ $\pm$3.50\end{tabular} & \begin{tabular}[c]{@{}c@{}}17.79\\ $\pm$3.69\end{tabular} \\
BiSR(b) & \begin{tabular}[c]{@{}c@{}}81.43\\ $\pm$0.11\end{tabular} & \begin{tabular}[c]{@{}c@{}}77.47\\ $\pm$2.59\end{tabular} & \begin{tabular}[c]{@{}c@{}}64.62\\ $\pm$2.34\end{tabular} & \begin{tabular}[c]{@{}c@{}}55.06\\ $\pm$3.34\end{tabular} & \begin{tabular}[c]{@{}c@{}}41.72\\ $\pm$3.27\end{tabular} & \begin{tabular}[c]{@{}c@{}}35.61\\ $\pm$0.80\end{tabular} & \begin{tabular}[c]{@{}c@{}}33.52\\ $\pm$0.75\end{tabular} \\
BiSR(b+f) & \begin{tabular}[c]{@{}c@{}}81.38\\ $\pm$0.15\end{tabular} & \begin{tabular}[c]{@{}c@{}}75.28\\ $\pm$2.95\end{tabular} & \begin{tabular}[c]{@{}c@{}}61.29\\ $\pm$3.61\end{tabular} & \begin{tabular}[c]{@{}c@{}}56.34\\ $\pm$0.34\end{tabular} & \begin{tabular}[c]{@{}c@{}}40.86\\ $\pm$3.41\end{tabular} & \begin{tabular}[c]{@{}c@{}}34.03\\ $\pm$0.20\end{tabular} & \begin{tabular}[c]{@{}c@{}}32.68\\ $\pm$0.79\end{tabular} \\ \hline
\rowcolor[HTML]{EFEFEF} 
Test PPL & \begin{tabular}[c]{@{}c@{}}5.26\\ $\pm$0.08\end{tabular} & \begin{tabular}[c]{@{}c@{}}9.89\\ $\pm$2.74\end{tabular} & \begin{tabular}[c]{@{}c@{}}30.31\\ $\pm$1.37\end{tabular} & \begin{tabular}[c]{@{}c@{}}51.42\\ $\pm$24.16\end{tabular} & \begin{tabular}[c]{@{}c@{}}60.01\\ $\pm$11.35\end{tabular} & \begin{tabular}[c]{@{}c@{}}82.46\\ $\pm$18.06\end{tabular} & \begin{tabular}[c]{@{}c@{}}93.07\\ $\pm$7.47\end{tabular} \\ \hline
\end{tabular}%
}
\end{table}

%% file: sections/2_relatedwork.tex
\section{Related Work}

\textbf{Split Learning for \glspl{llm}.} \Gls{sl} is a distributed learning technique that partitions an entire model into consecutive chunks of layers among various parties \cite{DBLP:journals/jnca/GuptaR18, DBLP:journals/corr/abs-1812-00564}. By selecting appropriate split points, the model’s architecture and weights can be partially concealed from each party. Leveraging \gls{sl}, FedBERT~\cite{DBLP:journals/tist/TianWLYJS22} partitions the BERT model for split-based pretraining, while PrivateLoRA~\cite{DBLP:journals/corr/abs-2311-14030} splits \glspl{llm} and low-rank adapter matrices for split-based model fine-tuning. SAP~\citep{DBLP:journals/corr/abs-2312-15603} proposed an \gls{sl} framework for fine-tuning \glspl{llm}, adding perturbations to the smashed data to enhance privacy.



\noindent\textbf{Data Reconstruction Attacks on \gls{sl}.} Considering the two phases of intermediate data transmission in \gls{sl}, \glspl{dra} on \gls{sl} can be divided into \emph{forward \glspl{dra}}~\cite{DBLP:journals/tbd/ZhangPTEM23, DBLP:journal/iot/he2020attacking} and \emph{backward \glspl{dra}}~\citep{DBLP:conf/nips/ZhuLH19, DBLP:conf/emnlp/DengWLWSLRD21}, targeting smashed data during forward propagation and gradients from backward propagation, respectively. \glspl{dra} can be further categorized into \emph{learning-based \glspl{dra}} and \emph{optimization-based \glspl{dra}} based on the attack techniques. Learning-based \glspl{dra} involve training an inversion model on auxiliary data to map intermediate representations to the original 
input~\citep{he2019model,DBLP:journals/tbd/ZhangPTEM23}. Optimization-based \glspl{dra} treat data reconstruction as non-convex optimization problems~\citep{jain2017non}. EIA~\cite{DBLP:conf/ccs/song2020information} trains a mapper to project deep embeddings to shallow ones, then optimizes the input to match the smashed data with the shallow output. Gradient attack methods like DLG~\citep{DBLP:conf/nips/ZhuLH19} and TAG~\citep{DBLP:conf/emnlp/DengWLWSLRD21} construct dummy data to match observed gradients, while LAMP~\citep{DBLP:conf/nips/BalunovicD0V22} adds token reordering and language priors to the search, though these are largely ineffective in \gls{sl} scenarios. All these methods attack from a single direction.

%% file: sections/6_conclusion.tex
\section{Conclusion}
This paper highlights the vulnerabilities of split-based \gls{ft} for \gls{llm}s and proposes an attack scheme, \gls{ours}, tailored to the characteristics of this scenario. \gls{m1} is designed for attackers aware of pre-trained weights, and experiments demonstrate the recoverability of intermediate outputs during the \gls{ft} process of mainstream generative \gls{llm}s. \gls{m2} combines forward \gls{m1} with backward gradient matching attacks, significantly enhancing attack performance in scenarios such as cross-dataset attacks and forward noise addition. Additionally, based on observations of the language model's adaptability to forward random noise, \gls{m3} is proposed to enable \gls{dra} attackers to better adapt to unknown forward noise, further strengthening attack performance under perturbations.

Therefore, the direct application of \gls{sl} to \gls{llm}-\gls{ft} carries significant security risks. We aim to spur future research to either address these vulnerabilities or develop protective measures. 

%% file: sections/10_acknowledgment.tex
\section{Acknowledgement}
We thank the anonymous reviewers for their valuable comments. We would like to extend our gratitude to Xuhong Zhang, Qiang Yang for their insightful suggestions. This research is partially supported by the Major Key Project of PCL (PCL2023A09), the NSFC under No. 62402418, and the Ningbo Key Research and Development Program under No. 2024Z115.

%% file: sections/9_appendix.tex
\newpage
\appendix

\section{Split Learning Simulation}

In this work, the experiments for \gls{sl} were conducted in a single-machine simulation environment. Using PyTorch and Huggingface Transformers APIs, we developed an easy-to-use \gls{sl} simulation framework. By leveraging the equivalence between \gls{sl} and centralized forward and backward propagation, the simulation was performed in a non-intrusive manner. Specifically, during training, we recorded the intermediate values of each block's forward propagation and the corresponding gradients without altering the model's forward and backward propagation processes.

We simulated the Client-Server architecture of \gls{sl} systems in a serial manner. Specifically, we maintained the complete \gls{llm} in GPU memory while storing the parameters of different model segments (Bottom, Trunk, Top) from each client and the server in the main memory. In each round of \gls{sl}, we loaded the Bottom and Top parameters of client1 and the Trunk parameters into the corresponding parts of the model in GPU memory. We then fine-tuned the \gls{llm} on the GPU using client1's data. After fine-tuning, we saved the Bottom, Top, and Trunk parameters back to the main memory. This process was repeated for client2 and subsequent clients. Once all clients had completed, we aggregated the Trunk parameters for that round. This serial simulation approach required hardware resources comparable to fine-tuning a single \gls{llm} on a single machine, allowing us to trade time for computational resources.

\section{Dataset Detail}

\input{images/tables/dataset_detail}

\subsubsection{Open-source Datasets}
\label{sec:ds}

To evaluate the overall attack performance of different \gls{dra} methods, we selected four commonly used open-source \gls{llm} instruction-tuning datasets to assess the data privacy vulnerabilities of \glspl{llm} during \gls{sft} under the \gls{sl} framework. The datasets considered in the experiments include the natural language commonsense question-answering dataset PIQA~\citep{Bisk2020} (21k question-answer pairs), the encyclopedia text dataset WikiText2-v1~\citep{merity2016pointer} (44.8k rows), the instructed code generation dataset CodeAlpaca-20K~\citep{codealpaca} (20k code generation instructions and corresponding code), and the math problem question-answering dataset GSM8K~\citep{cobbe2021gsm8k} (17.6k math problem question-answer pairs). Details of these datasets are provided in Table \ref{tab:dataset_detail}. Due to hardware limitations, all datasets were truncated to a maximum input length of 512.

\subsubsection{Customized Datasets: Sensi-Series}
\label{sec:ss}

\begin{figure}
    \centering
    \includegraphics[width=\linewidth]{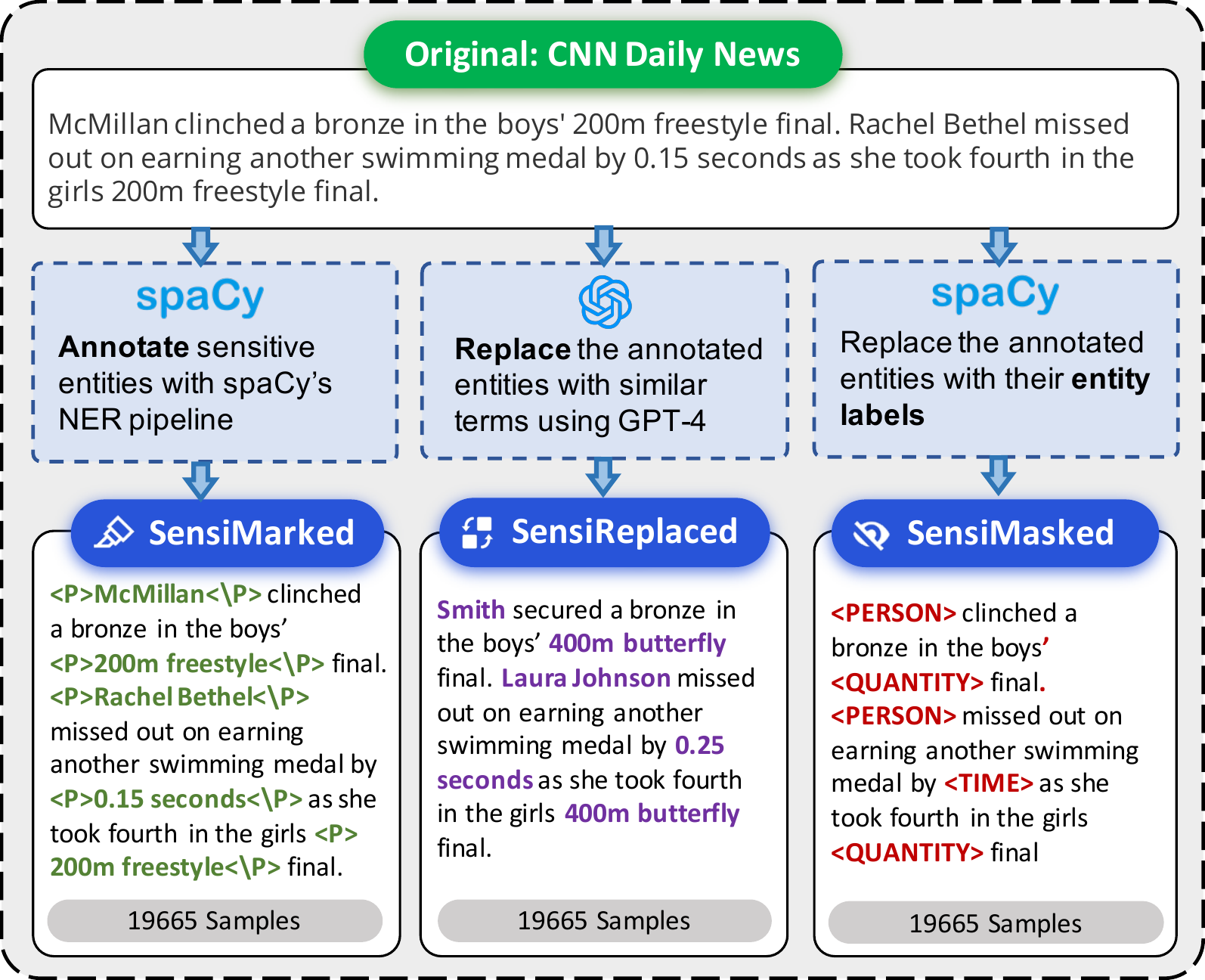}
    \caption{A detailed overview of the three Sensi-Series datasets derived from CNN DailyMail News, including processing methods and examples.}
    \label{fig:sensi-oview}
    \Description{
    A detailed overview of the three Sensi-Series datasets derived from CNN DailyMail News, including processing methods and examples.
    }
\end{figure}
\input{images/tables/label_type}

To further investigate the attack performance of \gls{dra} methods on \emph{sensitive information} and explore the attack mechanisms of learning-based approaches, we modified the open-source CNN-DailyMail News Text Summarization dataset~\citep{see-etal-2017-get} to manually construct a series of Sensitive datasets (hereafter referred to as \emph{Sensi-Series}), as illustrated in Figure \ref{fig:sensi-oview}.

First, we created the \emph{SensiMarked} dataset (19.7k rows) by annotating sensitive entities in the CNN-DailyMail News dataset using the NER pipeline of spaCy~\citep{spacy2}. During testing on \emph{SensiMarked}, we focused solely on the attack performance on these sensitive entities, thus eliminating the interference of frequently occurring common words on attack performance. The twelve types of sensitive entities are listed in Table \ref{tab:label_type}. Next, leveraging these annotations, we used the GPT-4 API to randomly replace the pre-marked sensitive entities within the \emph{SensiMarked} dataset, simulating a high-quality public dataset named \emph{SensiReplaced} that attackers might obtain. The prompt template is as follows:

\begin{tcolorbox}[colbacktitle=white, coltitle=black, colback=white, colframe=black, fonttitle=\bfseries, title={\textbf{Prompt Template of \emph{SensiReplaced} Dataset}}, leftupper=0.5em, rightupper=0.5em, boxrule=1.0pt, halign title=center]
    Replace given entities in the text with other random words.

    \textbf{Text}: [Text $T$]

    \textbf{Given entities}: [Entities List $\{E_i\}$]

    \textbf{Replaced text}:
\end{tcolorbox}


Additionally, we anonymize sensitive entities in the \emph{SensiMarked} dataset using labels corresponding to named entity types to create the \emph{SensiMasked} dataset. The aim is to sanitize sensitive information within the dataset for further revealing the learning capabilities of attackers.

\section{Attack Examples}

\subsection{Example of BiSR Attack}
\begin{figure*}
    \centering
    \includegraphics[width=1\linewidth]{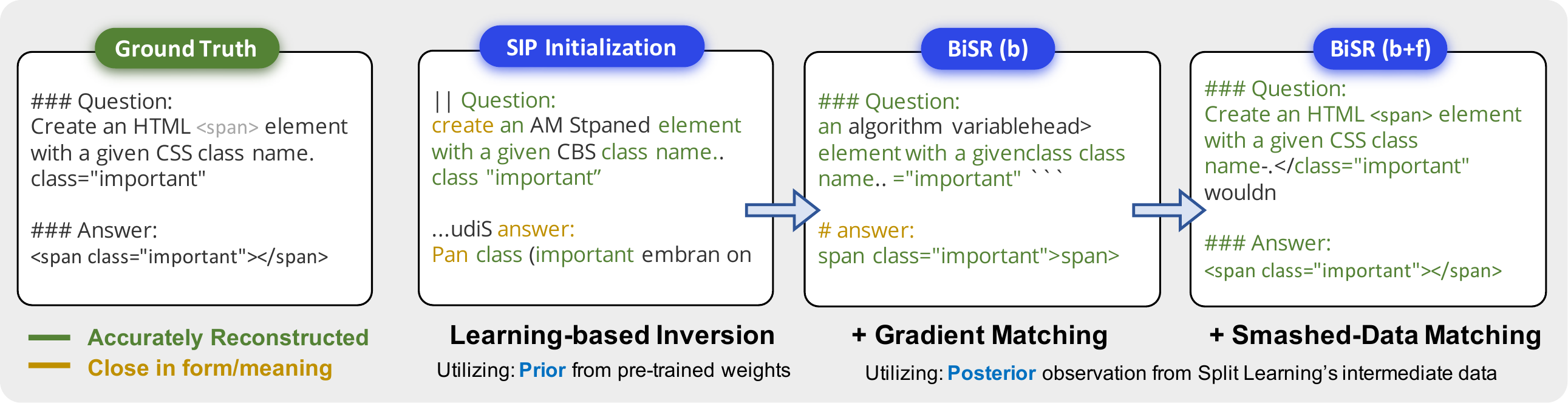}
    \caption{The outputs of \gls{ours}' three-stage attack on a sample from CodeAlpaca. The portions of the sentence that were fully or nearly accurately recovered are highlighted in green and yellow, respectively.}
    \label{fig:example-ours}
    \Description{
    The outputs of \gls{ours}' three-stage attack on a sample from CodeAlpaca. The portions of the sentence that were fully or nearly accurately recovered are highlighted in green and yellow, respectively.
    }
\end{figure*}

Figure \ref{fig:example-ours} shows the multi-stage attack results of \gls{ours} on a sample from CodeAlpaca. In this case, the \gls{m1} model was trained on SensiReplaced, a dataset that differs significantly from CodeAlpaca. As a result, while the initial attack by \gls{m1} successfully recovers a large number of tokens, it fails to accurately recover domain-specific tokens, such as code, due to this gap. This is where the optimization-based process of \gls{m2} comes into play, utilizing posterior information observed during \gls{sl} (i.e., smashed data and gradients) to enhance the attack's accuracy. As shown in the figure, after gradient matching and smashed-data matching, the code tokens are successfully recovered. This demonstrates the significant advantage of \gls{ours} in combining learning-based and optimization-based attack methods.

\subsection{Example of NaMOE's Inversion Attack Under $d_{\mathcal{X}}P$ perturbation}

\begin{figure*}
    \centering
    \includegraphics[width=1\linewidth]{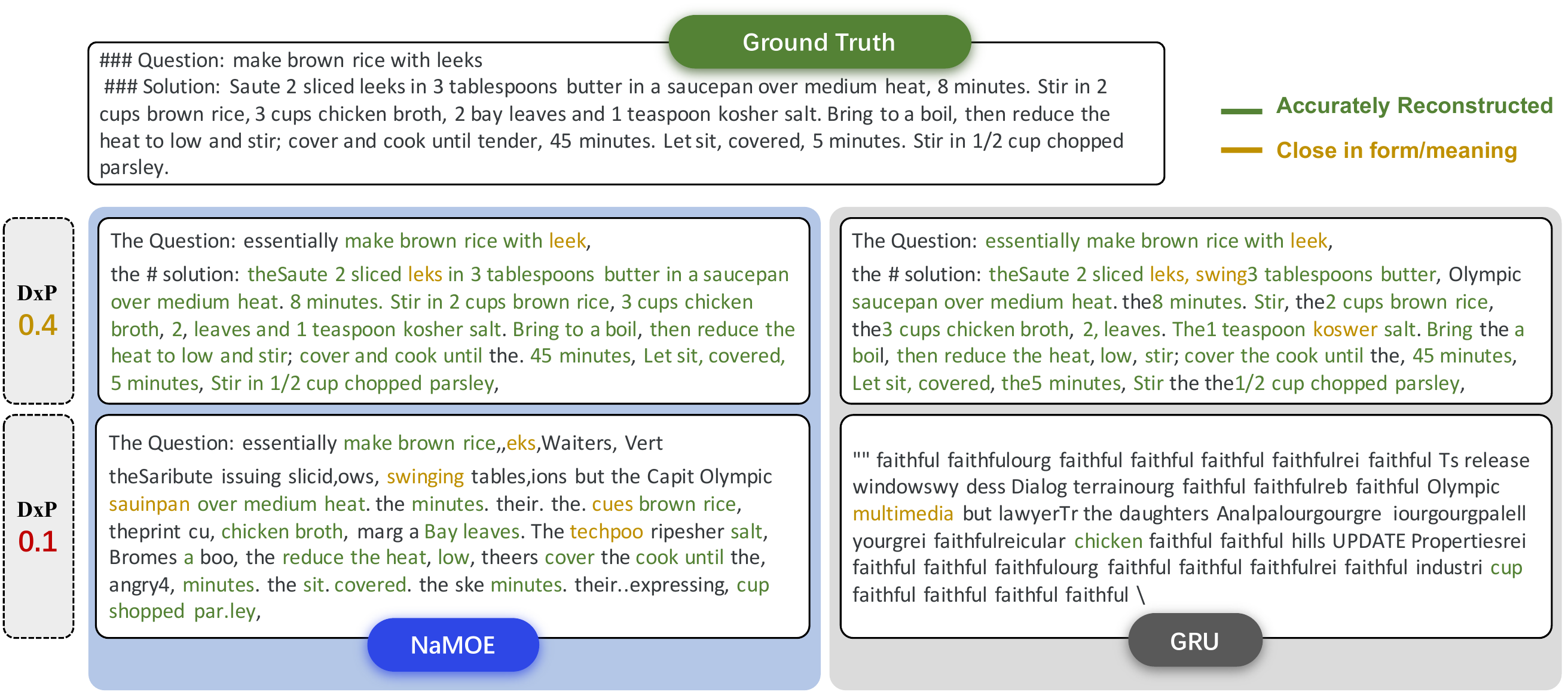}
    \caption{Comparison of attack results on a sample from PIQA between \gls{m3} and a standard GRU under varying levels of \gls{dxp} noise. The test \gls{llm} used is LLaMA2-chat-7B.}
    \label{fig:example-dxp}
    \Description{Comparison of attack results on a sample from PIQA between \gls{m3} and a standard GRU under varying levels of \gls{dxp} noise. The test \gls{llm} used is LLaMA2-chat-7B.}
\end{figure*}

Figure \ref{fig:example-dxp} presents the attack results on a PIQA sample using the \gls{m3} model trained on SensiReplaced and a standard GRU model, under different scales of \gls{dxp} perturbation applied to the target system. As shown, when the noise level is low ($\epsilon' = 0.4$), both models can recover most of the information. However, as the noise level increases ($\epsilon' = 0.1$, where the model's fine-tuning loses utility), the GRU model struggles to recover any data, while \gls{m3} still manages to recover a substantial number of critical tokens. This demonstrates the effectiveness of noise-aware training in enhancing the noise resilience of inversion models.

\section{Detailed Experiment Result}
\label{sec:detail-exp}

\subsection{Attack Performance on Other metrics}
\label{sec:other-metrics}

\input{images/tables/big_table_left}

In addition to ROUGE-L and METEOR, we assessed attack performance using two additional metrics: (1) ROUGE-1~\citep{rouge2004package}, which measures the overlap of unigrams (single words) between the generated and reference texts, disregarding word order, and (2) Token Recovery Rate (TRR), which calculates the proportion of tokens in the reconstructed sentences that match those in the original sentences, also ignoring word order. Both metrics evaluate token-level reconstruction performance. The results for ROUGE-1 and TRR, corresponding to Table \ref{tab:cross-dataset-bisr}, are presented in Table \ref{tab:big-table-left}. These results reflect a consistent pattern with those in Table \ref{tab:cross-dataset-bisr}, demonstrating that \glspl{dra} can effectively reconstruct private information at both the token and sentence levels in the given setting.

\subsection{NaMoE for NoPeek}
\label{sec:namoe-np}

For noise-aware training of the \gls{m3} model, we employed a \emph{simulated fine-tuning} strategy. To produce NoPeek-perturbed intermediate results with parameter $\alpha$, we first fine-tuned the \gls{llm} on SensiReplaced using NoPeek noise with $\alpha$ (600 steps in our experiment). The fine-tuned \gls{llm} then generated the intermediate results for \gls{m3} training. Given the high cost of fine-tuning, this strategy also required modifying the random noise approach during \gls{m3}’s gating training phase. Consequently, we pre-fine-tuned 20 \gls{llm} models (LoRA adapters) with random-scaled NoPeek perturbations for \gls{m3}’s gating training. During each gating training step, one of these 20 models was selected to generate the simulated random-scaled NoPeek-perturbed smashed data.

\subsection{Impact of Split Points}

\subsubsection{Bottom-Trunk Split Point}
\begin{figure}
    \centering
    \includegraphics[width=1\linewidth]{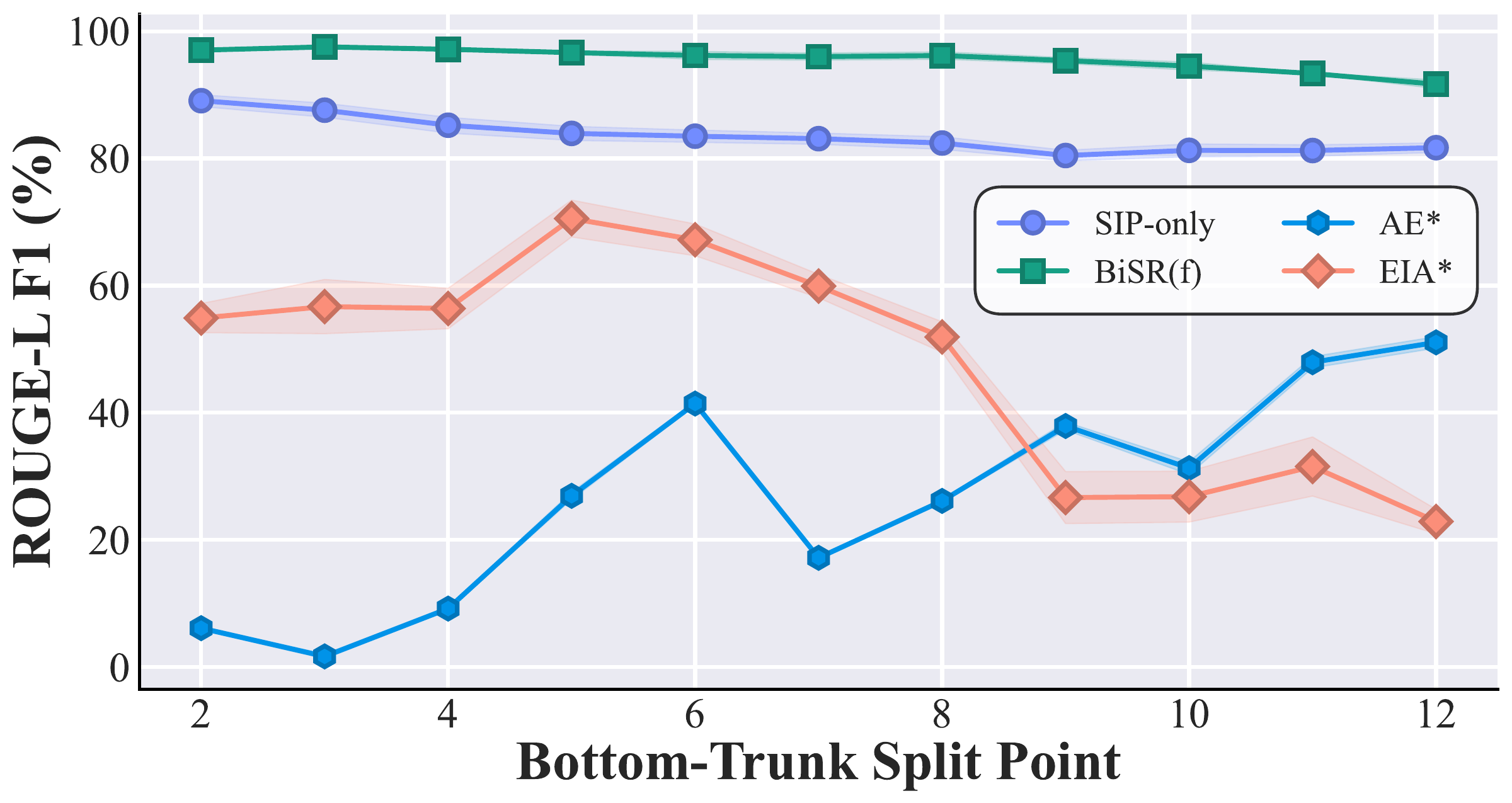}
\caption{The impact of Bottom-Trunk split point on the attack performance (ROUGE-L F1 Score \%) of \gls{ours}, in comparison with other forward \glspl{dra}, tested on the LLaMA2-chat-7B fine-tuned on the PIQA dataset.}
    \label{fig:diff-b2tr-sp}
    \Description{
    The impact of Bottom-Trunk split point on the attack performance (ROUGE-L F1 Score \%) of \gls{ours}, in comparison with other forward \glspl{dra}, tested on the LLaMA2-chat-7B fine-tuned on the PIQA dataset.
    }
\end{figure}

The choice of split point in the Bottom-Trunk interface significantly impacts the performance of \glspl{dra} based on smashed data (forward \glspl{dra}), as intermediate representations that are farther from the input may contain less information about the original input, and a larger Bottom can increase the difficulty of search-based attacks. We tested the effects of different shallow split points on the performance of forward \glspl{dra} using the LLaMA2-7B-chat model and the PIQA dataset. These \glspl{dra} include the learning-based \gls{m1} and AE, as well as the optimization-based EIA and \gls{ours}(f). As shown in Figure \ref{fig:diff-b2tr-sp}, deeper split points significantly degrade the performance of the optimization-based method EIA*, with its effectiveness dropping from over 60 to 20. While \gls{ours}(f) also experiences a slight performance decline at deeper split points, it still outperforms \gls{m1}, demonstrating that forward optimization can still offer advantages. The learning-based \gls{m1} also shows a similar, slight decline in performance. Interestingly, AE's performance improves as the depth increases.

\subsubsection{Trunk-Top Split Point}

\begin{figure}
    \centering
    \includegraphics[width=1\linewidth]{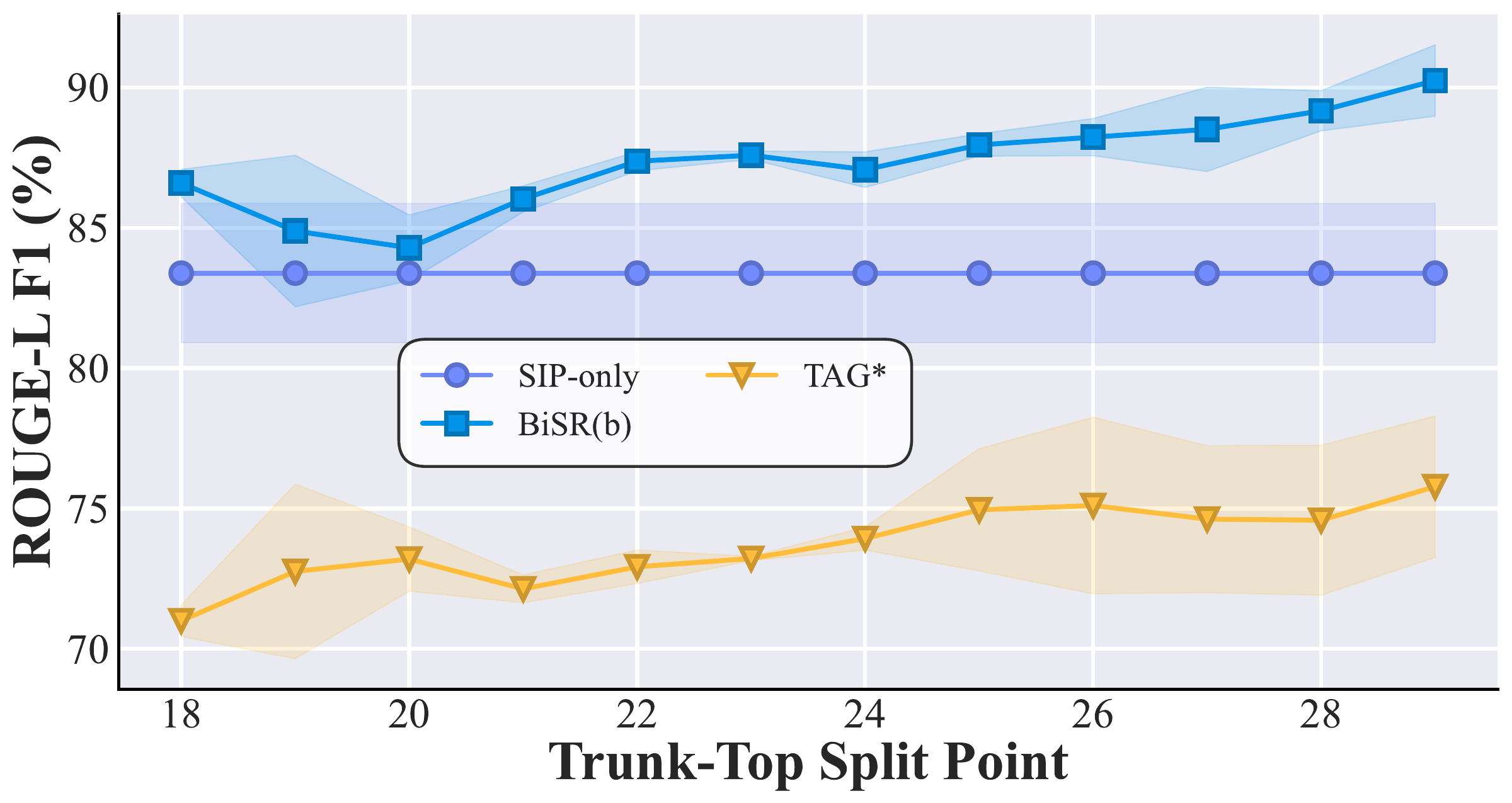}
    \caption{The impact of Trunk-Top split point on the attack performance of backward \glspl{dra}, tested on the LLaMA2-chat-7B and the PIQA dataset.}
    \label{fig:diff-tr2t-sp}
    \Description{The impact of Trunk-Top split point on the attack performance of backward \glspl{dra}, tested on the LLaMA2-chat-7B and the PIQA dataset.}
\end{figure}

Similarly, the number of model layers in the Top part affects the performance of backward \glspl{dra}, i.e., gradient-matching-based attacks represented by TAG and \gls{ours}(b). Therefore, we observed the impact of different Trunk-Top split points on attack performance (for example, for LLaMA2-7B with 32 blocks, a Trunk-Top split point of 28 means the last five decoder blocks are treated as the Top segment), as shown in Figure \ref{fig:diff-tr2t-sp}. Overall, the fewer the number of Top layers, the higher the performance of the gradient attack.

\subsection{Impact of Quantization}
\label{sec:quant}
\input{images/tables/quant}
In practical \gls{llm} fine-tuning scenarios, quantization is a commonly used technique. Quantizing \glspl{llm} to lower precision types significantly reduces memory usage and accelerates computation time. In our experiments, due to resource limitations, we used 8-bit quantized versions for billion-parameter \glspl{llm} by default. To investigate the impact of quantization on \gls{ours} in more detail, we fine-tuned different quantized versions (8-bit and 4-bit) of LLaMA2-chat-7B on PIQA, measuring fine-tuning performance using Test-Perplexity (TestPPL) and conducting attacks. As shown in Table \ref{tab:quant}, although \gls{llm} performance decreases with higher quantization levels (indicated by increased perplexity), quantization does not significantly affect the attack performance of \gls{ours}.

\subsection{Impact of BRE's Optimization Target}
\label{sec:diff-loss}
\begin{figure}
    \centering
    \includegraphics[width=\linewidth]{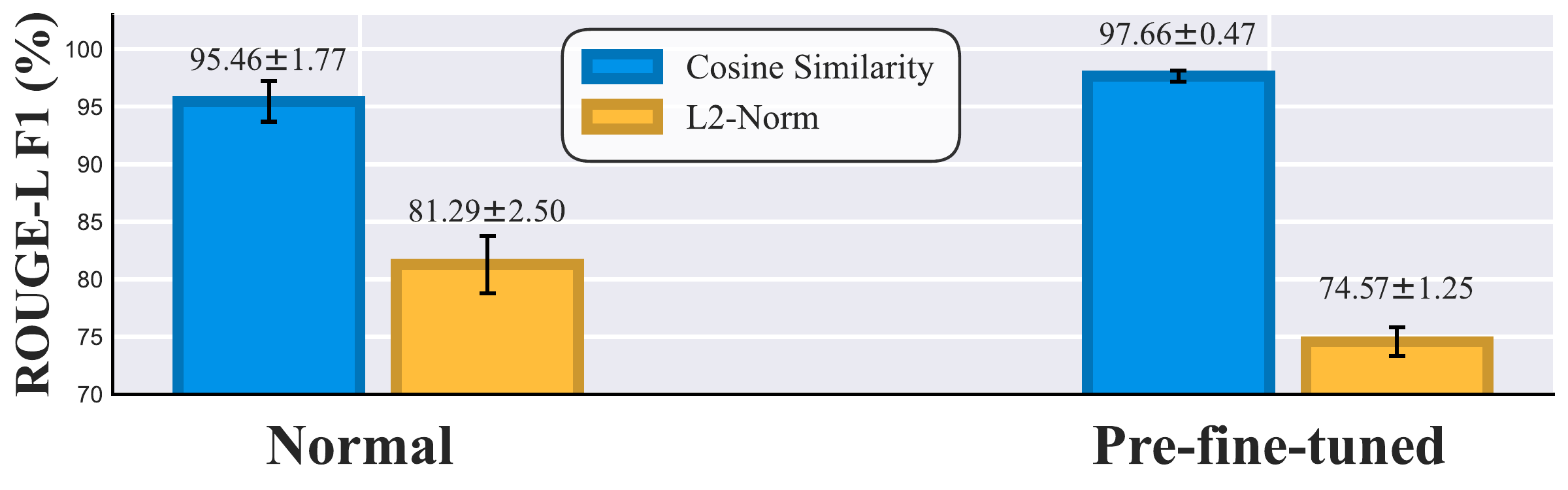}
    \caption{Comparison of the attack performance (ROUGE-L F1 Score\%) and sensitivity to model parameter changes for the two objective functions used in smashed data matching in \gls{m2}. “Normal” refers to attacking a pre-trained LLaMA2 model, while “Pre-fine-tuned” refers to attacking a LLaMA2 model fine-tuned on CodeAlpaca.}
    \label{fig:diff-loss}
    \Description{
    Comparison of the attack performance (ROUGE-L F1 Score\%) and sensitivity to model parameter changes for the two objective functions used in smashed data matching in \gls{m2}. “Normal” refers to attacking a pre-trained LLaMA2 model, while “Pre-fine-tuned” refers to attacking a LLaMA2 model fine-tuned on CodeAlpaca.
    }
\end{figure}

The forward optimization process in \gls{m2}, specifically smashed-data-matching, employs a continuous optimization strategy in the embedding space (rather than in the discrete vocabulary space). Beyond hyperparameters, the choice of the objective function is also critical. We compared two loss functions—Euclidean distance (L2-Norm) and cosine similarity—during the forward attack (BiSR(f)) on LLaMA2-chat-7B. As shown in Figure \ref{fig:diff-loss}, “Normal” refers to directly attacking an \gls{sl} system fine-tuned on PIQA, while “Pre-fine-tuned” refers to attacking a system with parameters slightly deviated from the pre-trained state (i.e., pre-fine-tuned on CodeAlpaca). The results indicate that cosine similarity significantly outperforms L2-Norm, achieving better performance and greater robustness to parameter changes. In contrast, L2-Norm exhibits performance degradation as the model parameters vary, likely due to cosine similarity's superior handling of the curse of dimensionality.

\section{A More Detailed Inspection of SIP Inversion Model}
\label{sec:sip-insp}

We have conducted extensive experiments demonstrating the strong recovery performance of \gls{m1} across different intermediate layers of various model architectures. However, this leads to a curious question: \emph{How much of the original input information is actually retained in the intermediate outputs of \gls{llm}? And on what basis does the \gls{m1} model perform its recovery?} To explore these questions, we conducted a series of observational experiments.

\subsection{Comparison of Different Inversion Models}
\input{images/tables/diff_model}
We explored several commonly used models for the structure of \gls{m1}, including a simple linear model (Linear), LSTM, a single Self-Attention layer (SelfAttn), GRU, and a bidirectional GRU model (GRU(bi)). We compared their performance in inversion attacks on the output of the 6th layer of LLaMA2-chat-7B using the PIQA dataset. As shown in Table \ref{tab:diff_models}, even the simple linear model achieves notably high attack performance, highlighting the significant privacy information contained in the intermediate results. The unidirectional GRU achieved the best attack performance, making it the default \gls{m1} model in our experiments and a component of \gls{m3}.

\subsection{Visualization of SIP Inversion Attack}
\begin{figure}
    \centering
    \includegraphics[width=1.0\columnwidth]{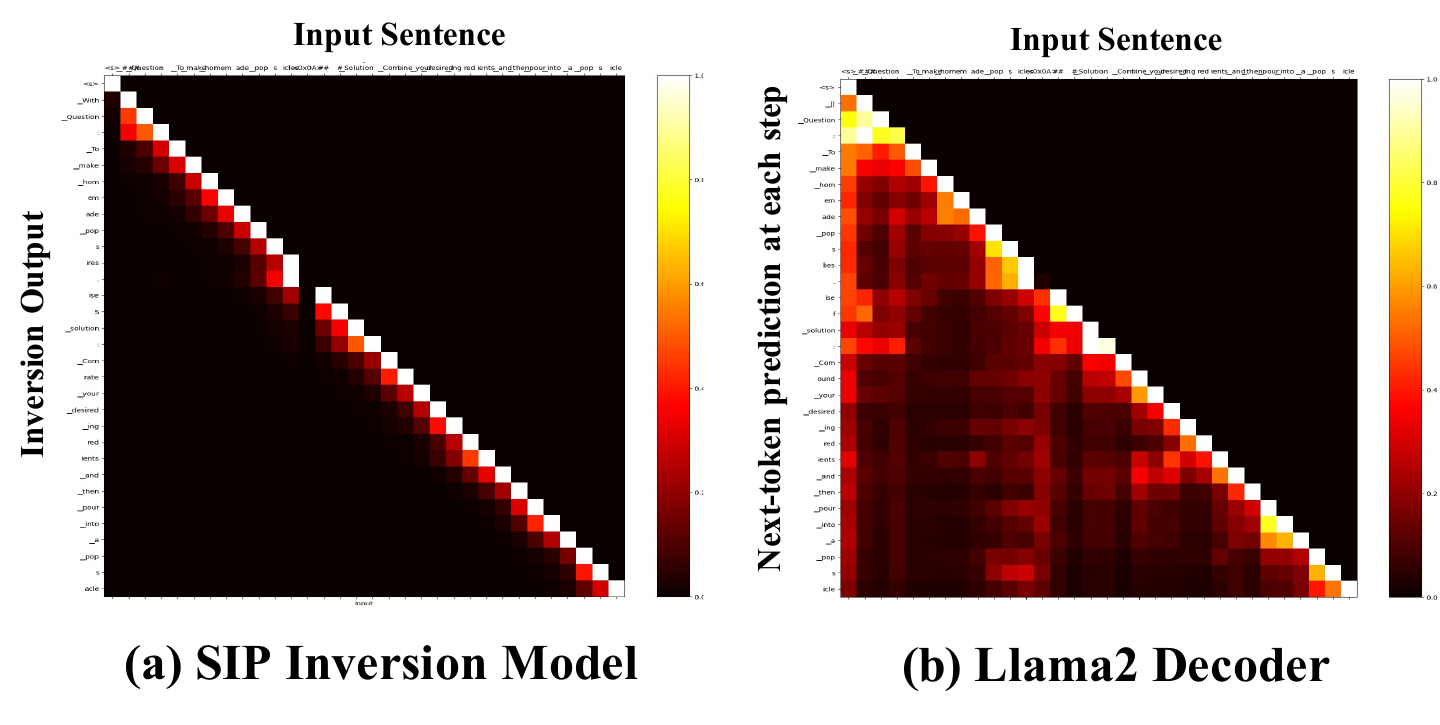}
    \caption{Heatmaps for an example from PIQA showing the contribution of each token position in the intermediate outputs of LLaMA2-chat-7B's 6th layer to each position of the model output: (a) the reconstructed sentence by the inversion model, and (b) LLaMA2's next-token prediction at each position. These heatmaps illustrate the model's attention during inference.}
    \label{fig:sal}
    \Description{
    Heatmaps for an example from PIQA showing the contribution of each token position in the intermediate outputs of LLaMA2-chat-7B's 6th layer to each position of the model output: (a) the reconstructed sentence by the inversion model, and (b) LLaMA2's next-token prediction at each position. These heatmaps illustrate the model's attention during inference.
    }
\end{figure}
To further investigate the GRU model's recovery behavior on \gls{llm} intermediate outputs, we employed gradient-based feature attribution methods. We generated heatmaps comparing the GRU model's output relative to its input smashed data with the \gls{llm}‘s decoder output (next-token prediction at each step) relative to the same smashed data, as shown in Figure \ref{fig:sal}. It is evident that the unidirectional GRU's “preceding sequence-only” focus aligns with the attention mechanism of causal language models. However, during text reconstruction, the GRU focuses on the current token and the preceding few token positions, demonstrating a much narrower attention span compared to \glspl{llm}. The GRU's ability to achieve high reconstruction performance despite its narrower focus suggests that it performs inversion primarily based on the intermediate representation of individual tokens rather than on the interaction between different tokens' intermediate representations. This also indicates that the high-dimensional intermediate representations (hidden states) of \glspl{llm} contain rich information about the raw input tokens. This observation aligns with the hypothesis drawn from the mutual attack experiments on the Sensi-Series Datasets discussed in Section \ref{sec:exp-m1}.

\subsection{Impact of LLM's Hidden Dimensionality}

\begin{figure}
    \centering
    \includegraphics[width=1\linewidth]{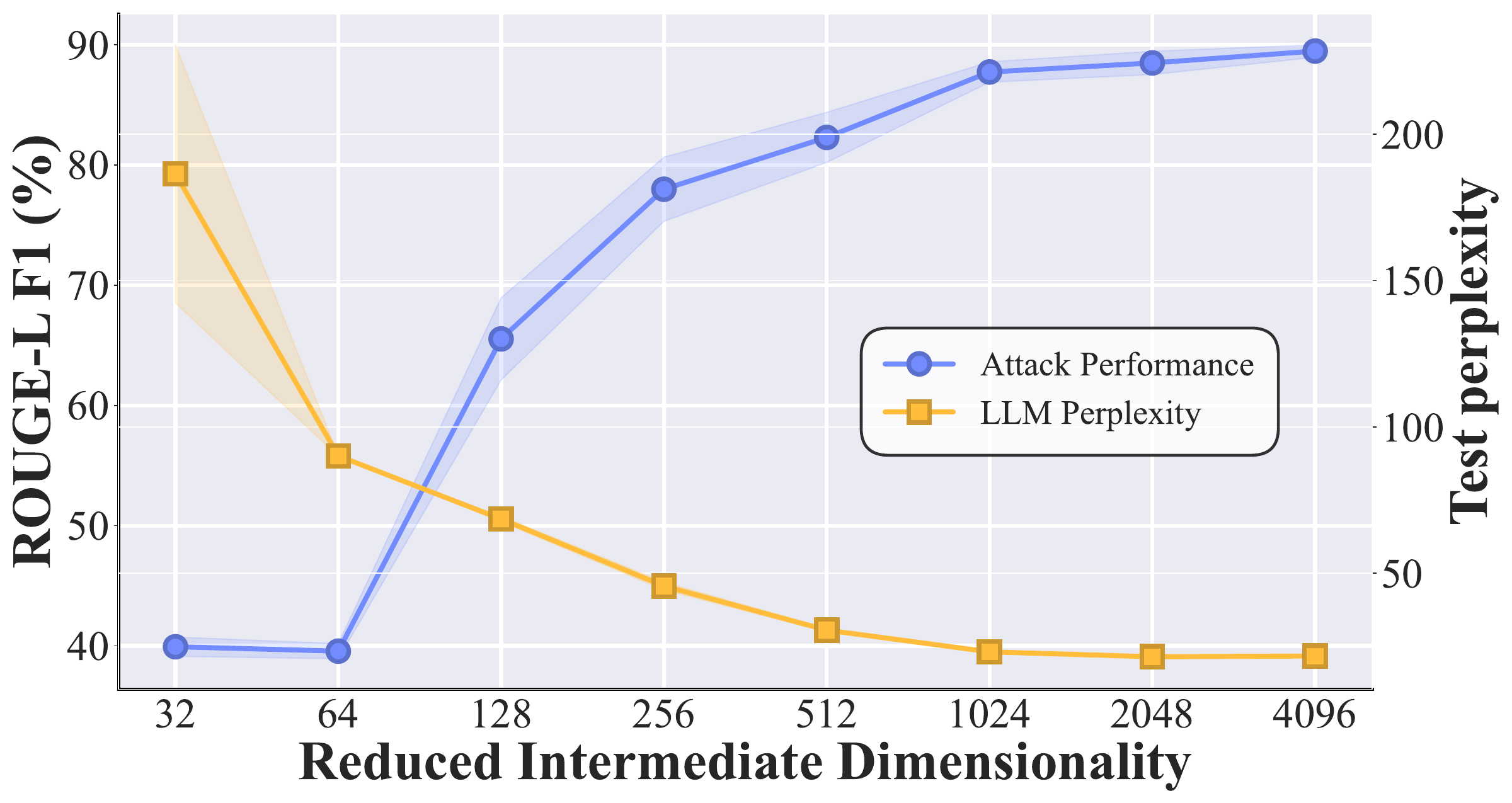}
    \caption{The impact of intermediate representation dimensionality (simulated via dimensionality reduction) in \gls{llm} (LLaMA2-chat-7B) on its own performance (test perplexity $\downarrow$ on PIQA) and the attack performance (ROUGE-L F1 Score \%) of an \gls{m1} model trained on SensiReplaced.}
    \label{fig:dim}
    \Description{
    The impact of intermediate representation dimensionality (simulated via dimensionality reduction) in \gls{llm} (LLaMA2-chat-7B) on its own performance (test perplexity $\downarrow$ on PIQA) and the attack performance (ROUGE-L F1 Score \%) of an \gls{m1} model trained on SensiReplaced.
    }
\end{figure}
The hypothesis that the high-dimensional intermediate representations of tokens support \gls{m1} reconstruction attacks prompted us to investigate: \emph{What is the minimum dimensionality of intermediate representations that can sustain high-performance \gls{m1} attacks?} To explore this, we conducted experiments on LLaMA2-chat-7B, focusing on hidden dimensionality. Specifically, we progressively reduced the intermediate dimensions of the \gls{llm} and observed the changes in attack performance (measured by ROUGE-L \% $\uparrow$ on PIQA) and the \gls{llm}'s own performance (measured by Testing Perplexity $\downarrow$ on PIQA). Assuming the original hidden dimensionality of the \gls{llm} is $D$, we simulated reduced intermediate dimensions by training a pair of dimensionality reduction-expansion matrices, $A^{D\times\alpha}$ and $B^{\alpha \times D}$. The output $h$ of a specific layer (in this case, the 6th layer) is first processed by these matrices, resulting in $h A B$, which is then fed into the next layer. We pre-trained these matrices on SensiReplaced to ensure that their output mirrors the input as closely as possible. We then applied a pre-trained $AB$ to the target \gls{llm}, allowing the \gls{m1} inversion model to attack the intermediate output $h_p^{S\times D}$ of a sensitive sample after dimensionality reduction, resulting in $(h_p A)^{S\times \alpha}$.

As shown in Figure \ref{fig:dim}, both model performance and attack performance improve as the intermediate dimensionality $\alpha$ increases. However, even with an intermediate dimensionality of just 32, \gls{m1} is still able to recover 40\% of the data, with the increase in attack performance significantly outpacing the improvement in model performance. For example, when $\alpha$ reaches 128, the attack performance sharply increases to over 60\%, while the model's performance sees only a modest gain. This suggests that while the intermediate representations of \glspl{llm} are low-rank, the information related to the original input tokens is more easily retained in low-rank representations compared to the information needed to support language modeling.

\section{Comparison with Vision Models}
\label{app:vis}
\begin{figure}
    \centering
    \includegraphics[width=1\linewidth]{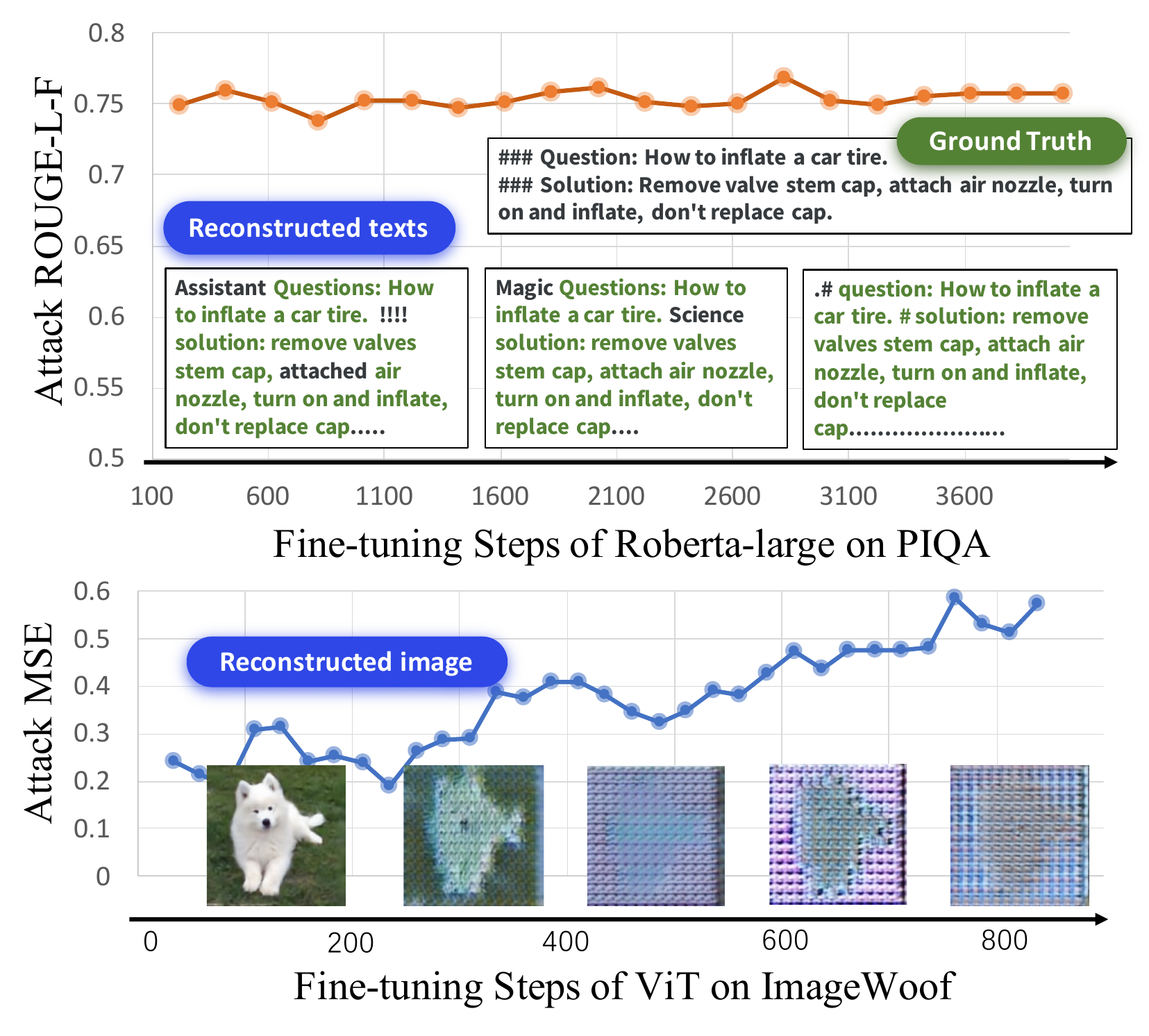}
    \caption{A comparison of the attack performance on language models (Top) and image models (Bottom) as the number of fine-tuning steps on the target model changes. For language models, the attack targets the sixth layer output of the Roberta-large model, fine-tuned for a binary classification task on the PIQA dataset, with attack performance evaluated using RougeL-F1 (higher is better). For image models, the attack targets the sixth layer output of the ViT-large model, fine-tuned for a multi-classification task on the ImageWoof dataset, with attack performance evaluated using MSE (lower is better).}
    \label{fig:image-steps}
    \Description{
    A comparison of the attack performance on language models (Top) and image models (Bottom) as the number of fine-tuning steps on the target model changes. For language models, the attack targets the sixth layer output of the Roberta-large model, fine-tuned for a binary classification task on the PIQA dataset, with attack performance evaluated using RougeL-F1 (higher is better). For image models, the attack targets the sixth layer output of the ViT-large model, fine-tuned for a multi-classification task on the ImageWoof dataset, with attack performance evaluated using MSE (lower is better).
    }
\end{figure}

To investigate whether Assumption \ref{as:nottoofar} holds for other models following the pretrain and fine-tune paradigm, we adapted the \gls{m1} attack method to the Transformer-based vision model ViT~\citep{DBLP:conf/iclr/DosovitskiyB0WZ21}. For ViT's intermediate outputs, we first captured their sequence information using a unidirectional GRU, followed by reconstructing the images through multiple deconvolution layers. Using MSE as the loss function, we trained this model on the 6th layer output of a pretrained ViT-large using the ImageWoof test set. The model successfully converged on the training dataset and exhibited good recovery performance on the test set. We then applied it to attack a ViT model undergoing split fine-tuning on the ImageWoof train set for a multi-classification downstream task. However, unlike text models, as fine-tuning of the target system progressed, the performance of the attack model trained in this manner significantly deteriorated, as shown in Figure \ref{fig:image-steps}.

In contrast, an attack on the language model Roberta, performing a text classification task with a similar parameter scale, demonstrated robustness against the target system's fine-tuning, as depicted in Figure \ref{fig:image-steps}. After two epochs of fine-tuning on PIQA, the attack model's performance was not notably affected. Based on this observation, we preliminarily conclude that semi-white-box settings pose a greater privacy risk in the context of language models, as the features of intermediate outputs learned by learning-based methods may not change significantly with \gls{llm} fine-tuning on downstream tasks. In other words, Assumption \ref{as:nottoofar}, i.e., Not-too-far, holds true. In contrast, for image models, a semi-white-box setting leans more towards a black-box scenario, indicating that the effectiveness of attacks using pre-trained parameters is more limited.

%% file: images/tables/dataset_detail.tex
\begin{table*}[]
\caption{Details of the datasets used in our experiments.}
\label{tab:dataset_detail}
\resizebox{\textwidth}{!}{%
\begin{tabular}{lllll}
\hline
Dataset & \# Samples & Task & Splits & Example \\ \hline
PIQA & 21,035 & \begin{tabular}[c]{@{}l@{}}Commonsense QA;\\ Text classification\end{tabular} & \begin{tabular}[c]{@{}l@{}}Train (77\%)\\ Test (14\%)\\ Validation (9\%)\end{tabular} & \begin{tabular}[c]{@{}l@{}}\textbf{Goal:} How to prevent pain while cutting your nails?\\ \textbf{Sol1:} Dip your nails inside lukewarm water for 5 minutes before cutting.\\ \textbf{Sol2:} Cutting nails doesn’t cause pain as nails don’t have sensation.\\ \textbf{label:} 1\end{tabular} \\
WikiText2-v1 & 44,836 & Text Generation & \begin{tabular}[c]{@{}l@{}}Train (82\%)\\ Test (8\%)\\ Validation (10\%)\end{tabular} & \begin{tabular}[c]{@{}l@{}}\textbf{Text: } In the late 1920s , Barker began to doubt she was doing enough \\ for the church and considered focusing solely on sacred works. Family \\ and friends recommended she continue secular and sacred works , which \\ she did .\end{tabular} \\
GSM8K & 8,762 & Math Problem QA & \begin{tabular}[c]{@{}l@{}}Train (85\%)\\ Validation (15\%)\end{tabular} & \begin{tabular}[c]{@{}l@{}}\textbf{Question:} James buys 5 packs of beef that are 4 pounds each. The price of\\  beef is \$5.50 per pound. How much did he pay?\\\textbf{Answer: } He bought 5*4=«5*4=20»20 pounds of beef So he paid 20*5.5=\\ \$«20*5.5=110»110 \#\#\#\# 110\end{tabular} \\
CodeAlpaca-20K & 20,022 & \begin{tabular}[c]{@{}l@{}}Instructed Code \\ Generation\end{tabular} & \begin{tabular}[c]{@{}l@{}}Train (90\%)\\ Validation (10\%)\end{tabular} & \begin{tabular}[c]{@{}l@{}}\textbf{Prompt:} Create a SQL query to find the highest grossing movie. \\ Table Name: "Movies" Columns: "MovieName", "Box- OfficeGross"\\ \textbf{Completion: }SELECT MovieName FROM Movies ORDER BY \\ BoxOfficeGross DESC LIMIT 1;\end{tabular} \\
Sensi-Series & 19,665 & Text Generation & \begin{tabular}[c]{@{}l@{}}Train (60\%)\\ Test (25\%)\\ Validation (15\%)\end{tabular} & \begin{tabular}[c]{@{}l@{}}\textbf{Text: }\textless{}P\textgreater{}McMillan\textless{}\textbackslash{}P\textgreater clinched a bronze in the boys’ \textless{}P\textgreater{}200m freestyle\\ \textless{}\textbackslash{}P\textgreater final. \textless{}P\textgreater{}Rachel Bethel\textless{}\textbackslash{}P\textgreater missed out on earning another swimming \\ medal by \textless{}P\textgreater{}0.15 seconds\textless{}\textbackslash{}P\textgreater as she took fourth in the girls \textless{}P\textgreater 200m \\ freestyle\textless{}\textbackslash{}P\textgreater final.\end{tabular} \\ \hline
\end{tabular}%
}
\end{table*}

%% file: images/tables/label_type.tex
\begin{table}[htbp]
\caption{Named entity types of sensitive entities}
\label{tab:label_type}
\small
\begin{tabular*}{0.42\textwidth}{@{\extracolsep{\fill}}lll|llllll}
\hline
\multicolumn{3}{c|}{Type} & \multicolumn{6}{c}{Description}                                \\ \hline
\multicolumn{3}{c|}{<PERSON>}                                         & \multicolumn{6}{l}{People, including fictional} \\
\multicolumn{3}{c|}{<GPE>}                                        & \multicolumn{6}{l}{Countries, cities, states} \\ 
\multicolumn{3}{c|}{<LOC>}                                        & \multicolumn{6}{l}{Non-GPE locations} \\ 
\multicolumn{3}{c|}{<DATE>}                                        & \multicolumn{6}{l}{Absolute or relative dates or periods} \\ 
\multicolumn{3}{c|}{<QUANTITY>}                                        & \multicolumn{6}{l}{Measurements, as of weight or distance} \\ 
\multicolumn{3}{c|}{<TIME>}                                        & \multicolumn{6}{l}{Times smaller than a day} \\ 
\multicolumn{3}{c|}{<PERCENT>}                                        & \multicolumn{6}{l}{Percentage} \\ 
\multicolumn{3}{c|}{<ORG>}                                        & \multicolumn{6}{l}{Companies, agencies, institutions, etc.} \\ 
\multicolumn{3}{c|}{<NORP>}                                        & \multicolumn{6}{l}{Nationalities or religious} \\ 
\multicolumn{3}{c|}{<MONEY>}                                        & \multicolumn{6}{l}{Monetary values, including unit} \\
\multicolumn{3}{c|}{<LAW>}                                        & \multicolumn{6}{l}{Named documents made into laws} \\ 
\multicolumn{3}{c|}{<WORK\_OF\_ART>}                                        & \multicolumn{6}{l}{Titles of books, songs, etc.} \\ \hline
\end{tabular*}
\end{table}

%% file: images/tables/big_table_left.tex
\begin{table*}[]
\caption{Performance of \gls{ours} (ROUGE-1 F1 \% and Token Accuracy) on various \glspl{llm} fine-tuned on different datasets compared to alternative methods. Experiments utilized split points set at 6-26, with SensiReplaced serving as the auxiliary dataset $D_a$. The best results for each configuration are highlighted in \textbf{bold}, with the second best results \underline{underlined}.}
\label{tab:big-table-left}
\resizebox{\textwidth}{!}{%
\begin{tabular}{l|l|ccccllllll}
\hline
\multicolumn{1}{c|}{} &
  \multicolumn{1}{c|}{} &
  \multicolumn{10}{c}{Split Fine-tuning Datasets} \\ \cline{3-12} 
\multicolumn{1}{c|}{} &
  \multicolumn{1}{c|}{} &
  \multicolumn{2}{c}{SensiMkd} &
  \multicolumn{2}{c}{Codealpaca} &
  \multicolumn{2}{c}{Gsm8k} &
  \multicolumn{2}{c}{PIQA} &
  \multicolumn{2}{c}{WikiText} \\ \cline{3-12} 
\multicolumn{1}{c|}{\multirow{-3}{*}{Model}} &
  \multicolumn{1}{c|}{\multirow{-3}{*}{Methods}} &
  ROUGE-1-F &
  TokACC &
  ROUGE-1-F &
  Meteor &
  \multicolumn{1}{c}{ROUGE-1-F} &
  \multicolumn{1}{c}{Meteor} &
  \multicolumn{1}{c}{ROUGE-1-F} &
  \multicolumn{1}{c}{Meteor} &
  \multicolumn{1}{c}{ROUGE-1-F} &
  \multicolumn{1}{c}{Meteor} \\ \hline
 &
  AE &
  {\color[HTML]{333333} 50.71$\pm$0.42} &
  {\color[HTML]{333333} 52.14$\pm$0.47} &
  {\color[HTML]{333333} 60.66$\pm$0.53} &
  {\color[HTML]{333333} 60.47$\pm$0.57} &
  {\color[HTML]{333333} 42.83$\pm$0.22} &
  {\color[HTML]{333333} 45.28$\pm$0.27} &
  {\color[HTML]{333333} 33.27$\pm$1.23} &
  {\color[HTML]{333333} 36.21$\pm$0.87} &
  {\color[HTML]{333333} 47.77$\pm$0.52} &
  {\color[HTML]{333333} 58.12$\pm$0.62} \\
 &
  EIA* &
  {\color[HTML]{333333} 79.96$\pm$4.50} &
  {\color[HTML]{333333} 81.00$\pm$5.92} &
  {\color[HTML]{333333} 56.66$\pm$2.93} &
  {\color[HTML]{333333} 56.38$\pm$2.79} &
  \multicolumn{1}{c}{{\color[HTML]{333333} 63.36$\pm$3.87}} &
  \multicolumn{1}{c}{{\color[HTML]{333333} 63.54$\pm$3.37}} &
  \multicolumn{1}{c}{{\color[HTML]{333333} 57.74$\pm$0.35}} &
  \multicolumn{1}{c}{{\color[HTML]{333333} 57.12$\pm$0.83}} &
  \multicolumn{1}{c}{{\color[HTML]{333333} 84.32$\pm$1.61}} &
  \multicolumn{1}{c}{{\color[HTML]{333333} 83.06$\pm$1.84}} \\
 &
  TAG* &
  {\color[HTML]{333333} 81.60$\pm$1.28} &
  {\color[HTML]{333333} 85.81$\pm$1.18} &
  {\color[HTML]{333333} 85.68$\pm$0.26} &
  {\color[HTML]{333333} 86.32$\pm$0.48} &
  \multicolumn{1}{c}{{\color[HTML]{333333} 84.39$\pm$1.36}} &
  \multicolumn{1}{c}{{\color[HTML]{333333} 84.77$\pm$1.26}} &
  \multicolumn{1}{c}{{\color[HTML]{333333} 79.30$\pm$2.71}} &
  \multicolumn{1}{c}{{\color[HTML]{333333} 80.63$\pm$2.42}} &
  \multicolumn{1}{c}{{\color[HTML]{333333} 76.54$\pm$1.19}} &
  \multicolumn{1}{c}{{\color[HTML]{333333} 77.66$\pm$0.87}} \\
 &
  LAMP* &
  {\color[HTML]{333333} 80.83$\pm$0.48} &
  {\color[HTML]{333333} 84.40$\pm$0.21} &
  {\color[HTML]{333333} 84.39$\pm$0.29} &
  {\color[HTML]{333333} 85.21$\pm$0.31} &
  \multicolumn{1}{c}{{\color[HTML]{333333} 84.85$\pm$0.33}} &
  \multicolumn{1}{c}{{\color[HTML]{333333} 85.03$\pm$0.39}} &
  \multicolumn{1}{c}{{\color[HTML]{333333} 77.67$\pm$0.99}} &
  \multicolumn{1}{c}{{\color[HTML]{333333} 78.86$\pm$0.98}} &
  \multicolumn{1}{c}{{\color[HTML]{333333} 76.33$\pm$0.86}} &
  \multicolumn{1}{c}{{\color[HTML]{333333} 77.33$\pm$0.81}} \\ \cline{2-12} 
 &
  SIP-only &
  {\color[HTML]{333333} 83.07$\pm$0.97} &
  83.53$\pm$0.85 &
  {\color[HTML]{333333} 85.33$\pm$0.54} &
  86.53$\pm$0.34 &
  \multicolumn{1}{c}{{\color[HTML]{333333} 95.91$\pm$0.38}} &
  95.30$\pm$0.37 &
  \multicolumn{1}{c}{{\color[HTML]{333333} 65.43$\pm$0.87}} &
  67.57$\pm$1.11 &
  \multicolumn{1}{c}{{\color[HTML]{333333} 95.66$\pm$0.98}} &
  96.20$\pm$0.73 \\
 &
  BiSR(b) &
  {\color[HTML]{333333} 90.33$\pm$1.81} &
  90.94$\pm$1.60 &
  {\color[HTML]{333333} 90.82$\pm$0.70} &
  90.40$\pm$0.80 &
  \multicolumn{1}{c}{{\color[HTML]{333333} 96.40$\pm$0.36}} &
  95.68$\pm$0.39 &
  \multicolumn{1}{c}{{\color[HTML]{333333} 83.24$\pm$1.14}} &
  82.73$\pm$1.29 &
  \multicolumn{1}{c}{{\color[HTML]{333333} 96.80$\pm$0.78}} &
  97.49$\pm$0.66 \\
 &
  BiSR(f) &
  {\color[HTML]{333333} {\ul 92.87$\pm$1.95}} &
  {\ul 90.46$\pm$1.85} &
  {\color[HTML]{333333} {\ul 95.33$\pm$0.17}} &
  {\ul 93.35$\pm$0.28} &
  \multicolumn{1}{c}{{\color[HTML]{333333} {\ul 98.12$\pm$0.36}}} &
  {\ul 97.17$\pm$0.48} &
  \multicolumn{1}{c}{{\color[HTML]{333333} {\ul 88.24$\pm$1.52}}} &
  {\ul 85.08$\pm$1.45} &
  \multicolumn{1}{c}{{\color[HTML]{333333} {\ul 98.70$\pm$0.58}}} &
  {\ul 98.69$\pm$0.44} \\
\multirow{-8}{*}{\begin{tabular}[c]{@{}l@{}}LLa-\\ MA2\end{tabular}} &
  BiSR &
  {\color[HTML]{333333} \textbf{95.62$\pm$0.23}} &
  \textbf{92.28$\pm$0.28} &
  {\color[HTML]{333333} \textbf{95.99$\pm$0.18}} &
  \textbf{93.53$\pm$0.07} &
  \multicolumn{1}{c}{{\color[HTML]{333333} \textbf{98.48$\pm$0.25}}} &
  \textbf{97.57$\pm$0.42} &
  \multicolumn{1}{c}{{\color[HTML]{333333} \textbf{92.72$\pm$0.69}}} &
  \textbf{88.50$\pm$0.77} &
  \multicolumn{1}{c}{{\color[HTML]{333333} \textbf{99.64$\pm$0.26}}} &
  \textbf{99.43$\pm$0.32} \\ \hline
 &
  AE &
  {\color[HTML]{333333} 75.91$\pm$0.32} &
  75.77$\pm$0.15 &
  {\color[HTML]{333333} 85.67$\pm$0.35} &
  85.39$\pm$0.40 &
  {\color[HTML]{333333} 73.25$\pm$1.02} &
  72.06$\pm$1.00 &
  {\color[HTML]{333333} 64.49$\pm$0.86} &
  64.39$\pm$0.76 &
  {\color[HTML]{333333} 82.55$\pm$4.96} &
  85.83$\pm$3.42 \\
 &
  EIA* &
  {\color[HTML]{333333} 85.15$\pm$3.11} &
  87.07$\pm$2.41 &
  {\color[HTML]{333333} 53.17$\pm$6.27} &
  54.27$\pm$6.30 &
  {\color[HTML]{333333} 65.58$\pm$3.57} &
  66.71$\pm$4.12 &
  {\color[HTML]{333333} 64.02$\pm$1.92} &
  64.06$\pm$1.68 &
  {\color[HTML]{333333} 82.75$\pm$0.59} &
  82.45$\pm$0.41 \\
 &
  TAG* &
  {\color[HTML]{333333} 62.39$\pm$5.57} &
  69.35$\pm$5.32 &
  {\color[HTML]{333333} 79.52$\pm$1.16} &
  80.10$\pm$0.95 &
  {\color[HTML]{333333} 81.66$\pm$1.09} &
  82.35$\pm$1.16 &
  {\color[HTML]{333333} 72.72$\pm$1.77} &
  74.17$\pm$1.11 &
  {\color[HTML]{333333} 70.78$\pm$1.02} &
  72.10$\pm$0.98 \\
 &
  LAMP* &
  {\color[HTML]{333333} 60.51$\pm$0.31} &
  69.91$\pm$0.61 &
  {\color[HTML]{333333} 78.65$\pm$0.16} &
  79.02$\pm$0.20 &
  {\color[HTML]{333333} 78.29$\pm$0.53} &
  78.17$\pm$0.11 &
  {\color[HTML]{333333} 70.16$\pm$0.36} &
  71.40$\pm$0.13 &
  {\color[HTML]{333333} 70.49$\pm$0.14} &
  71.87$\pm$0.14 \\ \cline{2-12} 
 &
  SIP-only &
  {\color[HTML]{333333} 80.84$\pm$0.75} &
  80.95$\pm$0.73 &
  {\color[HTML]{333333} 88.77$\pm$0.32} &
  88.42$\pm$0.30 &
  {\color[HTML]{333333} 93.19$\pm$0.17} &
  92.84$\pm$0.22 &
  {\color[HTML]{333333} 78.46$\pm$1.21} &
  78.10$\pm$1.12 &
  {\color[HTML]{333333} 93.53$\pm$0.34} &
  93.47$\pm$0.76 \\
 &
  BiSR(b) &
  {\color[HTML]{333333} 88.33$\pm$0.69} &
  87.77$\pm$0.73 &
  {\color[HTML]{333333} 94.87$\pm$0.31} &
  94.60$\pm$0.30 &
  {\color[HTML]{333333} 95.80$\pm$0.18} &
  95.74$\pm$0.26 &
  {\color[HTML]{333333} 91.23$\pm$0.31} &
  90.74$\pm$0.29 &
  {\color[HTML]{333333} 97.28$\pm$0.75} &
  96.80$\pm$0.28 \\
 &
  BiSR(f) &
  {\color[HTML]{333333} {\ul 95.07$\pm$0.60}} &
  {\ul 94.46$\pm$0.94} &
  {\color[HTML]{333333} \textbf{97.22$\pm$0.81}} &
  \textbf{96.75$\pm$0.90} &
  {\color[HTML]{333333} {\ul 99.06$\pm$0.29}} &
  {\ul 99.07$\pm$0.34} &
  {\color[HTML]{333333} {\ul 92.36$\pm$1.15}} &
  {\ul 90.84$\pm$1.47} &
  {\color[HTML]{333333} {\ul 99.71$\pm$0.26}} &
  {\ul 99.56$\pm$0.51} \\
\multirow{-8}{*}{\begin{tabular}[c]{@{}l@{}}GPT2\\ -large\end{tabular}} &
  BiSR &
  {\color[HTML]{333333} \textbf{95.20$\pm$0.83}} &
  \textbf{94.47$\pm$0.98} &
  {\color[HTML]{333333} {\ul 97.16$\pm$0.55}} &
  {\ul 96.63$\pm$0.79} &
  {\color[HTML]{333333} \textbf{99.14$\pm$0.11}} &
  \textbf{99.17$\pm$0.13} &
  {\color[HTML]{333333} \textbf{92.64$\pm$0.70}} &
  \textbf{91.23$\pm$0.98} &
  {\color[HTML]{333333} \textbf{99.79$\pm$0.29}} &
  \textbf{99.62$\pm$0.54} \\ \hline
 &
  AE &
  66.50$\pm$0.94 &
  68.77$\pm$0.53 &
  68.89$\pm$0.17 &
  69.35$\pm$0.19 &
  \multicolumn{1}{c}{62.98$\pm$0.88} &
  67.62$\pm$0.58 &
  \multicolumn{1}{c}{49.52$\pm$0.85} &
  55.91$\pm$1.04 &
  65.98$\pm$1.06 &
  72.02$\pm$1.30 \\
 &
  EIA* &
  65.78$\pm$9.52 &
  70.82$\pm$7.45 &
  9.87$\pm$3.94 &
  7.68$\pm$5.63 &
  \multicolumn{1}{c}{11.68$\pm$0.32} &
  11.68$\pm$0.14 &
  \multicolumn{1}{c}{30.38$\pm$6.65} &
  27.29$\pm$6.83 &
  26.78$\pm$0.23 &
  25.86$\pm$1.00 \\
 &
  TAG* &
  61.68$\pm$2.16 &
  67.37$\pm$1.40 &
  79.49$\pm$0.75 &
  78.26$\pm$1.09 &
  \multicolumn{1}{c}{81.69$\pm$2.08} &
  80.49$\pm$2.63 &
  \multicolumn{1}{c}{70.04$\pm$0.58} &
  67.19$\pm$2.25 &
  \multicolumn{1}{c}{68.90$\pm$1.79} &
  69.87$\pm$1.56 \\
 &
  LAMP* &
  39.21$\pm$0.13 &
  50.50$\pm$0.18 &
  78.25$\pm$0.39 &
  77.58$\pm$0.46 &
  79.48$\pm$0.32 &
  79.58$\pm$0.35 &
  70.31$\pm$0.34 &
  66.97$\pm$0.10 &
  \multicolumn{1}{c}{69.03$\pm$0.52} &
  69.62$\pm$0.52 \\ \cline{2-12} 
 &
  SIP-only &
  77.13$\pm$0.69 &
  74.70$\pm$0.38 &
  82.37$\pm$0.69 &
  81.32$\pm$0.68 &
  92.59$\pm$0.31 &
  92.20$\pm$0.44 &
  64.14$\pm$1.32 &
  66.48$\pm$1.11 &
  93.78$\pm$1.14 &
  94.89$\pm$0.89 \\
 &
  BiSR(b) &
  \textbf{91.33$\pm$0.77} &
  \textbf{91.58$\pm$0.39} &
  \textbf{93.21$\pm$0.76} &
  \textbf{92.51$\pm$0.63} &
  {\ul 94.50$\pm$0.21} &
  94.22$\pm$0.31 &
  \textbf{87.42$\pm$0.45} &
  \textbf{88.77$\pm$0.49} &
  94.96$\pm$1.29 &
  95.85$\pm$0.76 \\
 &
  BiSR(f) &
  89.92$\pm$1.07 &
  83.87$\pm$1.16 &
  86.12$\pm$1.72 &
  83.95$\pm$1.85 &
  \textbf{99.72$\pm$0.10} &
  \textbf{99.54$\pm$0.15} &
  86.64$\pm$1.48 &
  84.74$\pm$1.28 &
  {\ul 100.00$\pm$0.00} &
  {\ul 100.00$\pm$0.00} \\
\multirow{-8}{*}{\begin{tabular}[c]{@{}l@{}}Chat-\\ GLM3\end{tabular}} &
  BiSR &
  {\ul 90.90$\pm$0.31} &
  {\ul 90.08$\pm$0.52} &
  {\ul 86.54$\pm$1.96} &
  {\ul 85.38$\pm$2.12} &
  \multicolumn{1}{c}{99.71$\pm$0.09} &
  {\ul 99.53$\pm$0.17} &
  \multicolumn{1}{c}{{\ul 86.68$\pm$1.52}} &
  {\ul 85.57$\pm$1.27} &
  \textbf{100.00$\pm$0.00} &
  \textbf{100.00$\pm$0.00} \\ \hline
\end{tabular}%
}
\end{table*}

%% file: images/tables/quant.tex
\begin{table}[]
\caption{Impact of \gls{llm} Quantization on \gls{ours}'s Attack Performance (ROUGE-L F1 \% $\uparrow$) and LLM's fine-tuning Performance (Perplexity on PIQA-test $\downarrow$), tested on LLaMA2-chat-7B.}
\label{tab:quant}
\resizebox{\columnwidth}{!}{%
\begin{tabular}{l|ccccccccc}
\hline
 & \multicolumn{9}{c}{Quantization} \\ \cline{2-10} 
 & \multicolumn{3}{c}{Raw} & \multicolumn{3}{c}{8-bit} & \multicolumn{3}{c}{4-bit} \\ \cline{2-10} 
\multirow{-3}{*}{Methods} & RgLF & Mtor & TRR & RgLF & Mtor & TRR & RgLF & Mtor & TRR \\ \hline
SIP & \begin{tabular}[c]{@{}c@{}}82.83\\ $\pm$0.15\end{tabular} & \begin{tabular}[c]{@{}c@{}}88.57\\ $\pm$0.13\end{tabular} & \begin{tabular}[c]{@{}c@{}}84.10\\ $\pm$0.25\end{tabular} & \begin{tabular}[c]{@{}c@{}}82.44\\ $\pm$0.08\end{tabular} & \begin{tabular}[c]{@{}c@{}}87.84\\ $\pm$0.02\end{tabular} & \begin{tabular}[c]{@{}c@{}}82.99\\ $\pm$0.15\end{tabular} & \begin{tabular}[c]{@{}c@{}}82.07\\ $\pm$0.22\end{tabular} & \begin{tabular}[c]{@{}c@{}}88.77\\ $\pm$0.19\end{tabular} & \begin{tabular}[c]{@{}c@{}}83.56\\ $\pm$0.03\end{tabular} \\
BiSR(b) & \begin{tabular}[c]{@{}c@{}}89.66\\ $\pm$0.27\end{tabular} & \begin{tabular}[c]{@{}c@{}}93.68\\ $\pm$0.27\end{tabular} & \begin{tabular}[c]{@{}c@{}}90.64\\ $\pm$0.03\end{tabular} & \begin{tabular}[c]{@{}c@{}}89.46\\ $\pm$0.99\end{tabular} & \begin{tabular}[c]{@{}c@{}}93.58\\ $\pm$1.21\end{tabular} & \begin{tabular}[c]{@{}c@{}}90.15\\ $\pm$1.03\end{tabular} & \begin{tabular}[c]{@{}c@{}}86.85\\ $\pm$0.47\end{tabular} & \begin{tabular}[c]{@{}c@{}}91.50\\ $\pm$0.06\end{tabular} & \begin{tabular}[c]{@{}c@{}}87.08\\ $\pm$0.31\end{tabular} \\
BiSR & \begin{tabular}[c]{@{}c@{}}95.97\\ $\pm$0.09\end{tabular} & \begin{tabular}[c]{@{}c@{}}97.97\\ $\pm$0.03\end{tabular} & \begin{tabular}[c]{@{}c@{}}92.29\\ $\pm$0.43\end{tabular} & \begin{tabular}[c]{@{}c@{}}95.54\\ $\pm$0.20\end{tabular} & \begin{tabular}[c]{@{}c@{}}97.51\\ $\pm$0.17\end{tabular} & \begin{tabular}[c]{@{}c@{}}92.12\\ $\pm$0.29\end{tabular} & \begin{tabular}[c]{@{}c@{}}96.80\\ $\pm$0.37\end{tabular} & \begin{tabular}[c]{@{}c@{}}98.65\\ $\pm$0.05\end{tabular} & \begin{tabular}[c]{@{}c@{}}93.20\\ $\pm$0.15\end{tabular} \\ \hline
\rowcolor[HTML]{EFEFEF} 
TestPPL & \multicolumn{3}{c}{\cellcolor[HTML]{EFEFEF}14.94$\pm$0.04} & \multicolumn{3}{c}{\cellcolor[HTML]{EFEFEF}15.67$\pm$0.01} & \multicolumn{3}{c}{\cellcolor[HTML]{EFEFEF}20.28$\pm$0.14} \\ \hline
\end{tabular}%
}
\end{table}

%% file: images/tables/diff_model.tex
\begin{table}[]
\caption{Attack performance (ROUGE-L F1\%, Meteor\%, TokACC) of different inversion models (trained on SensiReplaced) on LLaMA2-chat-7B (attack on the 6-th layer) and PIQA dataset.}
\label{tab:diff_models}
\resizebox{\columnwidth}{!}{%
\begin{tabular}{l|ccccc}
\hline
\multicolumn{1}{c|}{} & Linear & LSTM & SelfAttn & GRU(bi) & GRU \\ \hline
RougeLF & 77.15$\pm$0.93 & 80.72$\pm$0.64 & 76.32$\pm$1.53 & 82.91$\pm$0.72 & \textbf{83.54$\pm$1.16} \\
Meteor & 86.88$\pm$0.27 & 86.76$\pm$0.20 & 85.46$\pm$1.41 & 88.03$\pm$0.18 & \textbf{88.58$\pm$0.69} \\
TokACC & 77.69$\pm$0.72 & 80.92$\pm$0.65 & 76.83$\pm$1.62 & 83.21$\pm$0.55 & \textbf{83.92$\pm$0.78} \\ \hline
\end{tabular}%
}
\end{table}